\documentclass[pra, aps, floats, showpacs, floatfix, showkeys, reprint]{revtex4-2}

\usepackage[a4paper, left=1.5cm, right=1.5cm, top=2.5cm, bottom=2.5cm]{geometry}
\usepackage{amssymb}
\usepackage{amsmath}
\usepackage{graphicx}
\usepackage[cp1251]{inputenc}
\usepackage[english]{babel}
\usepackage{float}
\usepackage{multirow}
\usepackage{array}
\usepackage[usenames,dvipsnames]{color}
\usepackage[hidelinks]{hyperref}
\usepackage{fancyhdr}

\makeatletter
\def\p@subsection{}
\def\p@subsubsection{}
\makeatother

\begin{document}

\setcounter{MaxMatrixCols}{10}

\title{Optics of Plasmon-Exciton Nanostructures: Theoretical Models and Physical Phenomena in Metal/J-aggregate Systems}


\author{V.S.~Lebedev}  
\email{vlebedev@lebedev.ru}  

\author{A.D.~Kondorskiy}
\email{kondorskiy@lebedev.ru}

\affiliation{\medskip P.N. Lebedev Physical Institute of the Russian Academy of Sciences,\\
	Leninskiy prosp. 53, 119991 Moscow, Russia}

\begin{abstract}
We review the studies of a wide range of optical phenomena resulting from near-field coupling between excitons and localized surface plasmon-polaritons in hybrid nanostructures. Modern physical approaches and theoretical models reported here for the description of light absorption, scattering, and extinction spectra are appropriate for interpreting physical effects in nanosystems containing metals and various excitonic materials, such as molecular aggregates of organic dyes or inorganic quantum-confined semiconductor structures. Using the example of hybrid nanosystems composed of a metal core and an outer shell of dye J-aggregate, we perform a theoretical analysis of the optical spectra behavior in the regimes of weak, strong, and ultrastrong plasmon--exciton coupling. We consider resonance and antiresonance phenomena induced by the coupling of an exciton with dipole and multipole plasmons, including a pronounced dip in light absorption, as well as the spectral band replication effect of plexcitonic nanoparticles and their dimers. We discuss the significant roles of the size-dependent permittivity of the metal core, the effects of anisotropy and chirality of the excitonic J-aggregate shell, and the influence of an intermediate passive layer on the formation of the optical spectra of bilayer, trilayer, and multilayer nanoparticles. The review outlines the experimental and theoretical results for hybrid nanosystems of various geometrical shapes, sizes, and compositions, broadens our understanding of the physical phenomena caused by the plasmon--exciton coupling, and represents the current state of research in the optics of metalorganic nanostructures.
\end{abstract}


\maketitle

\thispagestyle{fancy}

\tableofcontents

\section{Introduction}\label{sect:intro}

Over the past two decades, optical properties of various hybrid nanoparticles and nanostructures, as well as effects of their interaction with light fields, have been comprehensively studied by leading scientific centers. Fundamental interest in hybrid nanosystems is motivated to a considerable extent by the intense development of nanophotonics, nanoplasmonics and physics of quantum-confined structures. Besides, hybrid nanomaterials are extensively used in the development of future generations of photonic, optoelectronic and light-emitting devices  \cite{Gallop2024, Perego2022, Koduru2023}. Of note are advances in the field of optical switches \cite{Beddoes2023, Xie2023}, memory elements \cite{Vats2023, Youngblood2023}, photodetectors \cite{Kim2022}, photovoltaic cells \cite{Reus2024, Manzhos2021, Milichko2016}, light-emitting diodes \cite{Yang2024, Trapani2022}, and nanosensors \cite{Narayan2023, Firoozi2023, Ates2022}. Much attention is being paid to the development of the element base for nanophotonic integrated circuits \cite{Rodrigues2023, Chandrasekar2022} and plasmonic networks controlling optical information \cite{Davis2017}, nanolasers and spasers \cite{Fernandez-Bravo2019, Shalaev2017, Balykin2018}, superlenses, nanowaveguides, and near-field optical probes \cite{Park2014, Kazantsev2017, Khodadadi2020}, as well as a number of other devices operating on the basis of the effects of subwavelength optics, quantum-confinement, optical nonlinearity, and plasmon-induced phenomena. In this context, studies on hybrid organic/inorganic photonics and optoelectronics are of significant importance \cite{Agranovich2011, Will2019, Zhou2021}. 

A particular focus has been given to the study of the optical properties of hybrid structures consisting of metal nanoparticles and complex molecular systems, including dye aggregates. The physical properties of metal nanostructures and the effects of their interaction with light fields have been studied in detail \cite{Klimov2009, Lindquist2012, Qazi2016, H-Chang2021, Kond-Leb_JRLR2021, Miroshnichenko2022, Khlebtsov2022}. Metal structures and nanoantennas created on their basis \cite{Krasnok2013, Lepeshov2018} are capable of converting light into a highly localized electromagnetic field associated with collective oscillations of free electrons in metals. Such structures and devices make it possible to control and manipulate light fields on nanometer scales \cite{Barbillon2017, Klimov2021}, raising significantly the efficiency of photodetection \cite{Diedenhofen2015}, the intensity of photoluminescence \cite{Tam2007, Liaw2009, Ming2012, Dong2015, LeeLee2020} and light scattering \cite{Kneipp1997, Ru2008, Chen_2015, Khlebtsov2023}, and the sensitivity of biological and chemical sensors \cite{Arslanagic2015, Smirnov2017, LeeLawrie2021}.

The use of the ordered dye aggregates as an organic subsystem of metalorganic nanostructures has advantages over usual monomer molecules. Characteristically for J-aggregates, due to the translational order the electronic excitations of individual molecules are collectivized, forming Frenkel excitons. J-aggregates of polymethine (cyanine) dyes have a very narrow optical absorption peak (J-band), shifted to the red region of the spectrum relative to the peak of the monomer of the same dye, showing resonant fluorescence with a small Stokes shift, an abnormally high transition oscillator strength in the J-band, and giant nonlinear optical susceptibility. These unique optical properties arise due to the coherence of the states of the molecules constituting J-aggregates. A number of reviews \cite{Shapiro2006, Wurthner2011, J-Aggregates2012, Bricks2018, Hestand2018} have described the structure, methods of synthesis and physicochemical properties of aggregates of cyanine dyes. Intensive research has been carried out recently on the optical properties and effects of the excitation energy transfer in aggregates of more complex types, such as tubular \cite{Otsuki2018}, columnar \cite{Hecht-Wurthner2021}, X-aggregates \cite{Ma2021}, as well as multichromic multilayer \cite{Shapiro_OE2018, Shapiro_QE2018} aggregates.

As has been shown in \cite{Wiederrecht2008, Fofang2008, Lebedev2008, Lebedev2010, Antosiewicz2014, Shapiro2015, Todisco2018, Song2019}, hybrid metalorganic nanoparticles containing J-aggregates have unique optical properties. In certain regimes of their interaction with light, they combine the advantages of the excitonic subsystem, associated with the high oscillator strength of the radiative transition and the small width of the absorption J-band, and the plasmonic subsystem, capable of strongly enhancing the local field near the surface of a metal nanoparticle as compared to the external electromagnetic radiation. Strong local light fields can lead to modifications in the electronic structure and properties of J-aggregates \cite{Watanabe2006}. They can, as a result, have a significant impact on the photoluminescence of dyes deposited on plasmonic nanostructures \cite{Kravets2010a, Kravets2010b, Singh2024}. Many interesting effects have been shown to appear in a variety of metalorganic systems, such as giant Raman scattering \cite{A-Wang2015,Walters2018,Ralevich2018}, plasmon-enhanced fluorescence of J-aggregates \cite{Sorokin2015,Sorokin2020}, superquenching (sharp decrease in fluorescence intensity) \cite{L-Lu2002}, photoinduced charge separation reactions in J-aggregates near metal surfaces or particles \cite{Hranisavljevic2002}, plasmon-enhanced resonance energy transfer from donor molecules to acceptors, as well as energy transfer from molecular aggregates to surface plasmons \cite{Jian-Zhang2007, Akhavan2017, Kabbash2016, Weeraddana2017}.

In the physics of composite nanosystems containing metal particles of various shapes and sizes and molecular dye aggregates deposited on them, most of the focus is on the effects caused by the coupling of Frenkel excitons with localized surface plasmons induced in the metal subsystem upon its interaction with light \cite{Wiederrecht2004, Bellessa2009, Yoshida2010, Leb-Medv2012, Balci2013, Salomon2013, DeLacy2015, Zengin2015, Kondorskiy2015, Melnikau2022}. This area of research also includes studies of the effects of near-field coupling between Frenkel excitons and propagating surface plasmons in planar systems containing J-aggregates of dyes deposited on a flat metal substrate \cite{Bellessa2004,Symonds2008,Cade2009,Bellessa2014,Chmereva2016}. To a considerable extent these physical effects are similar to the extensively studied effects resulting from the plasmon--exciton coupling in hybrid systems containing a quantum-confined semiconductor compound and metal nanostructures \cite{Matsui2017, Forn-Diaz2019, Bitton2019, Liu2021, Kucherenko2022, Kim-Barulin2023}, including nanoparticles and thin films. Other similar optical phenomena are caused by the electromagnetic coupling of excitons with polaritons or plasmon-polaritons in microcavities \cite{Hirai2023, Jiang2019, Tserkezis2020, Deng2023, Tserkezis2023}.

The plasmon--exciton coupling results in some new hybrid states of the system with optical properties different from those of each of its constituents. In a general case, the spectrum of a composite system cannot be described as a simple superposition of the spectra of individual components. The effects that arise in this case are termed plasmon--exciton or plexcitonic, for short \cite{Nordlander2011, DeLacy2013, Schlather2013}. It has been shown that, due to the plexcitonic coupling, it is possible to effectively control the luminescence, absorption and light scattering spectra of metalorganic nanostructures by varying the geometrical parameters and shapes of their constituents \cite{Wurtz2007,Leb-Medv2013a,Leb-Medv2013b,Moritaka2020,Moritaka2023}. The behavior of these spectra depends significantly on the specific magnitudes of the optical constants of metals and molecular aggregates constituting hybrid nanostructures, including the oscillator strength of the J-band transition in the dye \cite{Leb-Medv2012, H-Chen2012, Thomas2018} and on the mutual positions of plasmon resonance peaks and the center of the J-absorption band transition \cite{Leb-Medv2012, Zengin2013}. All these factors make it possible to affect drastically the spectral-kinetic \cite{Wiederrecht2008,Fofang2008,Lebedev2008,Lebedev2010,Antosiewicz2014,Shapiro2015,Todisco2018} and nonlinear-optical \cite{Simon2016,Nan2015} properties of the composite plexcitonic systems. The most vivid phenomena occur in the strong and ultrastrong coupling regimes between plasmons and Frenkel excitons of molecular J-aggregates \cite{Wiederrecht2004, Bellessa2009, Yoshida2010, Leb-Medv2012, Balci2013, Salomon2013, DeLacy2015, Zengin2015, Kondorskiy2015, Lekeufack2010, Leb-Medv2013b, Ni2010, Zengin2013, H-Chen2012, Simon2016, Nan2015, Thomas2018, Melnikau2013, Vasa2013, Vasa2013Rabi, Vasa2018, Fain2019, Ates2020, Guo-Wu2021, Dey2023, Moritaka2020, Moritaka2023, Melnikau2022,Bellessa2004,Symonds2008,Cade2009,Bellessa2014,Chmereva2016,Matsui2017,Forn-Diaz2019,Bitton2019,Liu2021,Kucherenko2022,Kim-Barulin2023}.

To date, several review articles on various aspects of plasmon--exciton interaction have been published \cite{Barnes2015, Sukharev2017, Cao2018, Manuel2019, Kholmicheva2019, Vasa2020, He-Li2020, Wei2021}. They discuss the general principles of plexcitonic coupling and some applications in photonics and optoelectronics, describe methods for creating plexcitonic nanosystems, consider the effects of plasmonic amplification and quenching of spontaneous emission, exciton energy transfer between quantum emitters, as well as a number of other topical issues. However, so far there have not been any reviews that would present a clear and sufficiently complete physical picture of plexcitonic phenomena in the optics and spectroscopy of metal nanoparticles and nanostructures of various shapes and sizes, coated with an outer shell of molecular dye aggregates. Moreover, a number of physical approaches and analytical models of plexcitonic coupling have not currently received a proper theoretical consideration in the literature. Nor do the existing reviews reflect many important experimental results and theoretical calculations of light absorption, scattering and extinction processes, as well as  photoluminescence in such composite nanosystems.

In this review we present the current state of research in the optics of metalorganic nanosystems and provide a detailed description of the effects of the near-field coupling of Frenkel excitons with localized surface plasmons in such systems. The analysis of plexcitonic coupling effects is performed for the exciton interacting not only with dipole plasmon, but also with higher-order multipole plasmons (quadrupole, octupole, etc.), whose relative contribution grows with the increase in the particle size. We report the results of numerical calculations and experimental data for hybrid spherical particles and for more complex geometrical shapes including nanorods, nanodisks, and nanoprisms. The contributions of light absorption and scattering processes to the total extinction cross sections for different particle sizes are discussed. We clarify the influence of the size-dependent dielectric function of the metal core on the resulting spectral distributions of light absorbed by a core-shell nanoparticle. Special attention is paid to the investigation of the behavior and specific features of the optical spectra for two-layer (Metal/J-aggregate), three-layer (Metal/Spacer/J-aggregate) and multilayer nanoparticles. In addition to the conventional isotropic model of the excitonic outer shell, we consider its tensor nature and demonstrate the essential role of orientational and anisotropic effects of the J-aggregated shell for the formation of the optical spectra of hybrid plexcitonic nanoparticles with a metal core. We provide a detailed discussion of various plasmon--exciton coupling regimes and consider a number of resonance and antiresonance phenomena observed when light interacts with plexcitonic nanoparticles. They include a pronounced dip in light absorption (or induced transparency in transmission spectra), as well as spectral band replication in hybrid metalorganic dimers, consisting of metalorganic nanospheres and nanodisks placed close to each other. We also briefly indicate possible applications of some fundamental results in the topical field of plexcitonic physics for the development of new materials with unique optical properties potentially useful in fabrication of efficient photonic and optoelectronic devices.

\section{\label{Sect-SPP}Surface Plasmon-Polaritons}

\begin{figure*}[ht]
\centering\includegraphics[width=0.9\linewidth]{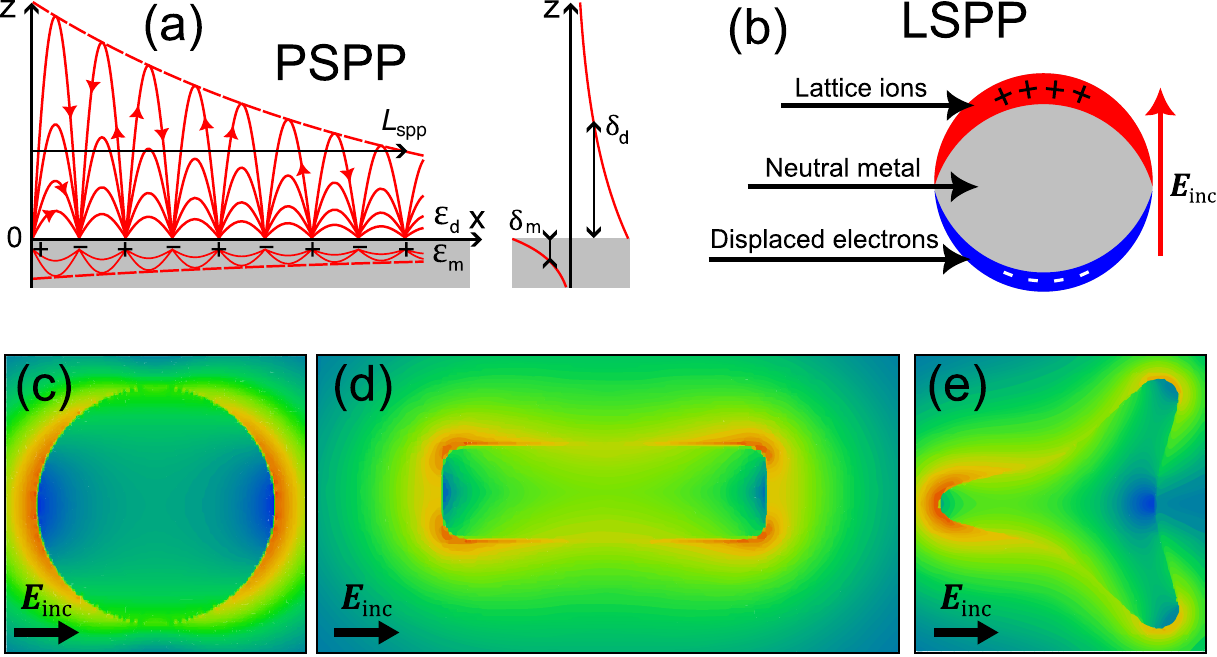}
\caption{(a) Schematic view of a propagating surface plasmon-polariton (PSPP) along a planar metal/dielectric interface. (b) Illustration of the mechanism of the charge separation on the surface of a metal nanosphere which leads to the formation of localized surface plasmon-polariton (LSPP) induced by the electric field, $\mathbf{E}_{\text{inc}}$, of the incident light. (c), (d), (e) Spatial distributions of the squared modulus of the electric field, $\left|E\right|^2$, in the longitudinal section of nanoparticles: (c) disk, (d) rod and (e) three-pointed star. The calculations have been performed using the FDTD method for light wavelengths, $\lambda$, that correspond to the positions of spectral peak maxima of longitudinal plasmon resonances, namely, (c) $\lambda = 425$ nm for the disk; (d) $\lambda = 690$ nm for the rod; (e) $\lambda = 510$ nm for the three-pointed star.
The following geometrical parameters of nanoparticles have been used: (c) $D = 27$ nm is the diameter of the disk, $h = 15$ nm is its thickness; (d) $D = 15$ nm is the diameter of the rod, $L = 49$ nm is its length; (e) $L=35$ nm is the distance between the tips of the star, $h=15$ nm is its thickness. In panels (c), (d), and (e) the directions of the electric field, $\mathbf{E}_{\text{inc}}$, of light incident normal to the plane of these panels are indicated by black arrows.}
\label{fig:spp}
\end{figure*}

\subsection{Introductory remarks}

Below we offer some preliminary remarks on surface plasmon-polaritons which are believed necessary to understand the effects of the near-field plexcitonic coupling. Surface plasmon-polaritons are formed under certain conditions at the metal/dielectric interface as a result of the interaction of free electrons of a conductor with the electromagnetic radiation. The physics of these surface waves of the combined nature is such that oscillations of the electron density in the metal are coherent with oscillations of the electromagnetic field in the surface layers of both the metal and the dielectric. There are two types of surface plasmon-polaritons: propagating surface plasmon-polaritons (PSPP) and localized surface plasmon-polaritons (LSPP) \cite{Barnes2015, Cao2018}. Such waves can be excited not only in systems with flat or extended curved surfaces (for example, in metal films, nanowires and nanowaveguides), but also in nanoparticles of various geometrical shapes and sizes (nanospheres, nanorods, nanodisks, nanoprisms, nanocubes, nanostars) and in their arrays, as well as in nano-sized pores of three-dimensional structures and holes in metal films. The necessary condition for the excitation of surface plasmon-polaritons, both propagating along the flat boundary of a two-component metal/dielectric system and localized in a metal nanoparticle surrounded by a dielectric medium, is the difference of signs of the real parts of the metal,
$\operatorname{Re}\!\left\{\varepsilon_{\text{m}}\right\}<0$, and the dielectric, $\operatorname{Re}\!\left\{\varepsilon_{\text{d}}\right\}>0$, permittivities. We shall start by listing here the main concepts of the theory of propagating surface plasmon-polaritons, highlighting the common physical nature of these two types of plasmon-polaritons, PSPP and LSPP, and pointing to the differences in their properties. Note that the words ''surface'' and ''polariton'' in the terms of PSPP and LSPP are often omitted for brevity.

\subsection{Propagating Surface Plasmon-Polaritons}

\begin{table*}[ht]
	\centering
	\caption{
		Results of calculations using formulas \eqref{eq:pspp-penetration-depth}--\eqref{eq:pspp-propagation-length} of the field penetration depth, $\delta_{\text{m}}$ and $\delta_{\text{d}}$, and the propagation length, $L_{\text{SPP}}$, of the surface plasmon-polariton at flat interfaces between silver or gold and quartz ($\mathrm{Ag}/\mathrm{SiO}_{2}$ and $\mathrm{Au}/\mathrm{SiO}_{2}$) at different wavelengths, $\lambda$, of light in vacuum.
	}
	\begin{tabular}{|c|c|c|c|c|c|c|c|c|}
		\hline
		&& \multicolumn{3}{c|}{$\mathrm{Ag}/\mathrm{SiO}_{2}$} 
		&& \multicolumn{3}{c|}{$\mathrm{Au}/\mathrm{SiO}_{2}$}  \\
		\hline
		\hspace{12pt} $\lambda$, $\mu$m \hspace{12pt}
		&& \hspace{12pt} $\delta_{\text{Ag}}$, nm \hspace{12pt} & \hspace{12pt} $\delta_{\text{SiO}_{2}}$, nm \hspace{12pt} & \hspace{12pt} $L_{\text{SPP}}$, $\mu$m \hspace{12pt} 
		&& \hspace{12pt} $\delta_{\text{Au}}$, nm \hspace{12pt} & \hspace{12pt} $\delta_{\text{SiO}_{2}}$, nm \hspace{12pt} & \hspace{12pt} $L_{\text{SPP}}$, $\mu$m \hspace{12pt} \\
		\hline
		1.5	&& 21.7 & 1236 & 350 && 22.8 & 1174 & 87.7 \\
		\hline
		1.0 && 21.9 & 527  & 225 && 24.0 & 478  & 28.7 \\
		\hline
		0.7 && 22.1 & 241  & 42.3 && 25.6 & 200  & 7.62 \\
		\hline
		0.5 && 22.5 & 103  & 5.42 && 34.5 & 71   & 0.003  \\
		\hline
		0.4 && 21.7 & 44   & 0.68 && \multicolumn{3}{c|}{no propagation}  \\
		\hline
	\end{tabular}
	\label{tab:PSPP-data}
\end{table*}

Under appropriate conditions, surface plasmon-polaritons can be excited by light and propagate within a thin layer along the flat interface between a metal and a dielectric. This is shown in Fig.~\ref{fig:spp}, which demonstrates a schematic view of the electric field of a surface wave propagating along the $X$ axis directed along this interface. The $Z$ axis is directed normal to the surface, such that at $z > 0$ the material is a dielectric, and the region $z < 0$ is filled with a metal. One can also see in Fig. \ref{fig:spp} the alternating signs of charges in the near-surface layer of the metal, which means that the oscillations of the electron density are predominantly longitudinal along the $x$ direction of the electromagnetic field propagation \cite{Bozhevolnyi2013}. Analysis of the boundary conditions shows \cite{Bozhevolnyi2013} that the electromagnetic field here is a transverse magnetic (TM) wave in which the magnetic field, $\mathbf{H}$, is directed perpendicularly to the plane of the figure ($H_{\mathrm{ x}} = H_{\mathrm{z}} = 0$), and the electric field, $\mathbf{E}$, has both longitudinal, $E_{\mathrm{x}}$, and transverse, $ E_{\mathrm{z}}$, components. For such plasmon-polaritons, the dispersion relation can be written as
\begin{equation}
\begin{split}
	&k^{\text{SPP}}_{\mathrm{x}} = \frac{2\pi}{\lambda} \sqrt{\frac{\varepsilon_{\text{d}}\varepsilon_{\text{m}}}{\varepsilon_{\text{d}} + \varepsilon_{\text{m}}}},\\
	&k^{\text{SPP}}_{\mathrm{z}} = \begin{cases}
		\frac{2\pi}{\lambda}
		\sqrt{\frac{\varepsilon^{2}_{\text{d}}}{\varepsilon_{\text{d}} + \varepsilon_{\text{m}}}}, \quad \mathrm{z} > 0, \\
		\frac{2\pi}{\lambda}
		\sqrt{\frac{\varepsilon^{2}_{\text{m}}}{\varepsilon_{\text{d}} + \varepsilon_{\text{m}}}} , \quad \mathrm{z} < 0.
	\end{cases}
\end{split}	
	\label{eq:PSPP-quasimomentum}
\end{equation}
As is evident from the expressions \eqref{eq:PSPP-quasimomentum}, when the following conditions are satisfied
\begin{equation}
	\operatorname{Re}\!\left\{\varepsilon_{\text{m}} + \varepsilon_{\text{d}}\right\} < 0, \quad
	\operatorname{Re}\!\left\{\varepsilon_{\text{m}} \varepsilon_{\text{d}}\right\} < 0,
	\label{eq:PSPP-eps-conditions}
\end{equation}
then the surface wave propagates along the interface, and the electric field, $\mathbf{E}$, decreases exponentially along the normal direction both into the metal ($z < 0$) and into the dielectric ($z > 0$).
The field penetration depths into a metal ($\delta_{\text{m}}$) and a dielectric ($\delta_{\text{d}}$) are given by  \cite{Barnes2003, Bozhevolnyi2013, Zhang2012}
\begin{equation}
	\delta_{\text{d}} = \frac{\lambda}{2\pi} \sqrt{\left|\frac{\varepsilon_{\text{d}}+\varepsilon_{\text{m}}}{\varepsilon^{2}_{\text{d}}}\right|},\quad
	\delta_{\text{m}} = \frac{\lambda}{2\pi} \sqrt{\left|\frac{\varepsilon_{\text{d}}+\varepsilon_{\text{m}}}{\varepsilon^{2}_{\text{m}}}\right|},
	\label{eq:pspp-penetration-depth}
\end{equation}
where $\lambda$ is the wavelength of light in vacuum; $\varepsilon_{\text{m}}$ and $\varepsilon_{\text{d}}$ are the permittivities of the metal and dielectric, respectively.

Table \ref{tab:PSPP-data} shows the results of calculations using formula \eqref{eq:pspp-penetration-depth} of the penetration depth into metal ($\delta_{\text{m}}$) and dielectric ($\delta_{\text{d}}$) for different wavelengths $\lambda$ of light in vacuum.
Quartz ($\mathrm{Si} \mathrm{O}_{2}$) was chosen as a dielectric in our calculations, with silver ($\mathrm{Ag}$) or gold ($\mathrm{Au}$) as a metal. It is clear from the data presented in Table \ref{tab:PSPP-data} that a wave traveling along the interface penetrates the metal to a much shorter distance than the dielectric.

The surface plasmon-polariton exhibits attenuation as it propagates along the $x$ axis due to the energy dissipation in the metal (see the dashed envelope in Fig. \ref{fig:spp}).
The corresponding propagation length, $L_{\text{SPP}}$, can be expressed in terms of the imaginary part, $\operatorname{Im}\! \left\{k^{\text{SPP}}_{\mathrm{x}}\right\}$, of the complex wave vector, $\mathbf{k}^{\text{SPP}}$, of the surface plasmon determined by the dispersion law \eqref{eq:PSPP-quasimomentum}. According to \cite{Barnes2003, Zhang2012}, the value of $L_{\text{SPP}}$ primarily depends on the metal permittivity, $\varepsilon_{\text{m}}$, at the surface plasmon oscillation frequency, $\omega = 2 \pi c/\lambda$. It is determined by the expression \cite{Barnes2003, Zhang2012}
\begin{equation}
	L_{\text{SPP}} = \frac{\lambda}{2\pi} \left(
	\frac{\varepsilon_{\text{d}} + \operatorname{Re}\!\left\{\varepsilon_{\text{m}}\right\} } {\varepsilon_{\text{d}}\operatorname{Re}\!\left\{\varepsilon_{\text{m}}\right\}}
	\right)^{\frac{3}{2}}
	\frac{\left(\operatorname{Re}\!\left\{\varepsilon_{\text{m}}\right\}\right)^{2}}{\operatorname{Im}\!\left\{\varepsilon_{\text{m}}\right\}}.
	\label{eq:pspp-propagation-length}
\end{equation}

Table \ref{tab:PSPP-data} shows the results of calculations using formula \eqref{eq:pspp-propagation-length} of the propagation length, $L_{\text{SPP}}$, of the surface plasmon-polariton along a flat metal/dielectric interface in the direction of the $X$ axis.
The data presented for quartz on silver and gold show that the value of $L_{\text{SPP}}$ depends significantly on the wavelength, $\lambda$, of incident light in a vacuum and decreases significantly in the short-wavelength part of the visible spectrum, so in the case of silver at $\lambda = 0.5$ $\mu$m the length is $L_{\text{SPP}} \approx 5$ $\mu$m, and there is a cutoff effect for gold near this wavelength, i.e. the wave stops propagating along the gold/quartz interface along the $X$ axis.
It also follows from Table \ref{tab:PSPP-data} that the longest propagation lengths are realized for plasmon-polaritons in the near-infrared (IR) region. For example, in the vicinity of the telecommunication wavelength $\lambda \approx 1.5$ $\mu m$, the value of $L_{\text{SPP}}$ is several hundred micrometers for $\mathrm{Ag}/\mathrm{SiO}_{2}$ interface and approaches approximately one hundred micrometers in $\mathrm{Au}/\mathrm{SiO}_{2}$.

It is important to recall one well-known feature of the propagating plasmon-polariton dispersion law. According to \eqref{eq:PSPP-quasimomentum}, provided that conditions \eqref{eq:PSPP-eps-conditions} are met, the component $k^{\text{SPP}}_{\mathrm{x}}$ of the plasmon-polariton wave vector along the direction of its propagation $x$ always turns out to be greater than the wave number $k_{\text{d}} = 2 \pi \sqrt{\varepsilon_{\text{d}}}/\lambda$ of light in a dielectric medium at the same frequency $\omega = 2 \pi c/\lambda$.
Therefore, due to the momentum conservation law, light propagating freely in a dielectric cannot directly excite a surface plasmon-polariton (PSPP). Thus, a huge number of works have been devoted to diverse methods of excitation of propagating surface plasmon-polaritons to ensure the matching of momenta. Most common excitation schemes include as follows: Kretschmann and Otto schemes based on the frustrated total internal reflection; excitation with a near-field microscope optical probe, as well as diffraction on a grating, on a corrugated structure, and on inhomogeneities and roughness of the metal/dielectric interface (see, e.g., \cite{Raether1988, Zhang2012}).

\subsection{Localized Surface Plasmon-Polaritons}

\begin{figure}[ht]
	\centering\includegraphics[width=0.98\linewidth]{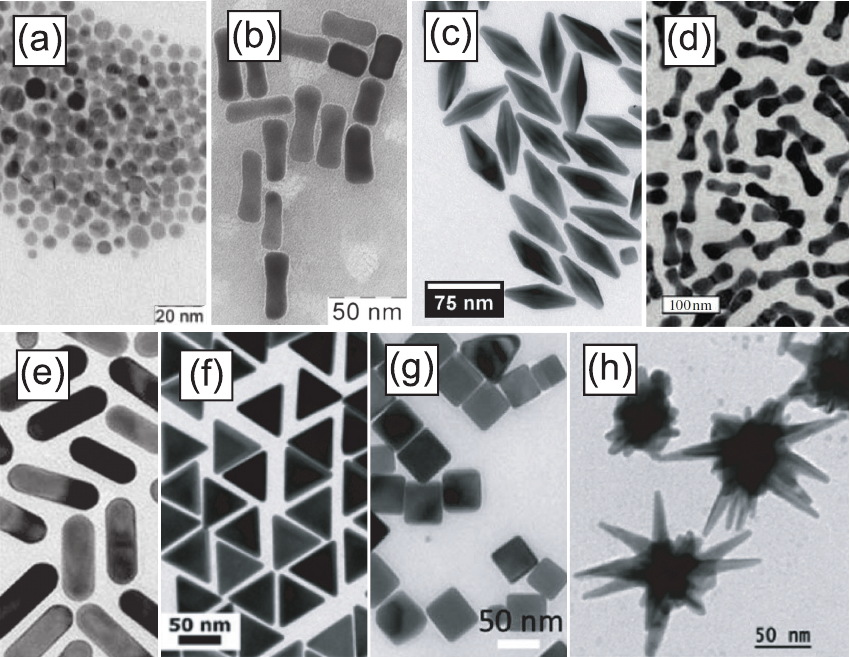}
	\caption{TEM images of metal nanoparticles of various shapes: (a) nanospheres~\cite{Lebedev2010}; (b) nanorods~\cite{Shapiro2015}; (c) bipyramids~\cite{Chateau2015}; (d) nanodumbbells~\cite{Cardinal2010}; (e) nanorods with round ends~\cite{Melnikau2016}; (f) triangular nanoprisms \cite{Scarabelli2021}; (g) cubes \cite{Yin2016}; (h) nanostars~\cite{Swarnapali2015}.}
	\label{fig:tem_au}
\end{figure}

Localized surface plasmon-polaritons are excited in metal nanoparticles and in metal-containing nanostructures, the size of which may be less than the light wavelength, $\lambda$, and even less than the penetration depth, $\delta_{\text{m}}$, of the electromagnetic field given in Table \ref{tab:PSPP-data} for a bulk metal. Modern synthesis techniques make it possible to obtain nanoparticles of various shapes, including rather complex ones (see Fig. \ref{fig:tem_au}). Similarly to the case of propagating surface plasmon-polaritons, the electromagnetic field of these waves of combined nature is evanescent \cite{Girard2000}, and its intensity exhibits a strong local increase in the regions adjacent to the nanoparticle surface inside and outside the metal. When localized plasmon-polaritons appear in metal nanostructures, collective coherent oscillations in the charge density in the near-surface layer are excited, as is the case for extended geometry. However, in contrast to propagating plasmon-polaritons, the oscillations of electrons are confined by the closed surface of the nanostructure while the evanescent field is concentrated in the vicinity of the interface. Therefore, the properties of the localized surface plasmon-polaritons are primarily determined by the geometrical shape and size of the nanoparticle, in addition to the  permittivities of the particle material and the environment.

The formation mechanism of a localized surface plasmon-polariton is schematically shown in Fig. \ref{fig:spp}b using a spherical particle as an example. An electromagnetic wave in metal nanoparticles displaces free conduction electrons relative to the ionic core of the crystal lattice. As a result, surface charges of different signs at the opposite ends of the particle create a restoring field, the strength of which is proportional to the displacement of electrons relative to the ionic core. The signs of positive and negative charges shown in Fig. \ref{fig:spp}b change periodically with time, following the oscillations of the external field, $\mathbf{{E}}_{\text{inc}}$. Thus, a metal particle is a system with a set of eigen-frequencies, the values of which are determined by the shape, size and optical constants of the particle material and its environment \cite{Bohren1998, Hohenau2007}.

The shape of the nanoparticle affects radically the spatial distribution of the electric field. This is clearly demonstrated in Fig. \ref{fig:spp}c--\ref{fig:spp}e, which shows our calculations of the $\left|E\right|^{2}$ value for the disk, rod and three-point star. The calculations were performed using the Finite-Difference Time-Domain (FDTD) method. As is evident from the figure, for particles of complex geometrical shape the field is strongly localized near the curved surface of the metal nanoparticle, and its strength is maximum where the curvature of the particle surface is the highest, i.e. where the radius of curvature is particularly small. Thus, the field distributions shown in Fig. \ref{fig:spp}c--\ref{fig:spp}e confirm the justification for using the term ''localized surface plasmon-polariton'', since most of the volume of the depicted particles the field strength is extremely small, existing actually only in the vicinity of the surface inside and outside the metal.

Note another fundamental difference between a localized plasmon-polariton and a propagating one. The excitation of the latter requires the matching of the photon and plasmon-polariton wave vectors. In the case of localized plasmon-polaritons, due to the lack of propagation there is no excitation criteria based on the wave vector matching. So there is no difficulty exciting plasmon-polaritons, as is the case of flat extended metal/dielectric interfaces. 

When the radius of curvature, $r$, of the nanoparticle surface is less than the thickness of the penetration depth, $\delta_{\text{m}}$, in a bulk metal, which lies in the range of $20$--$35$ nm in the visible and near-infrared spectral regions, then the fraction of the particle volume into which the field effectively penetrates gradually increases \cite{Hohenau2007}. The specific nature of the field localization inside and outside the particle and the depth of its penetration into the metal ($\delta_{\text{m}}$) and the environment surrounding the particle ($\delta_{\text{d}}$) substantially depend on the value of $r$. Correspondingly, the greatest increase in the local field compared to the incident field occurs near the vertices and tips of nanoparticles, for example, bipyramids, prisms, cubes and stars (see Fig. \ref{fig:tem_au}c, \ref{fig:tem_au}f, \ref{fig:tem_au}g and \ref{fig:tem_au}h, respectively). Note that in the interaction of optical radiation with small nanoparticles of size $a$, a quasistatic approximation is applicable if the condition $ka\ll 1$ is satisfied. In this approximation the contribution of the magnetic field into the total energy of the evanescent electromagnetic field is small and the role of electric field is predominant.

In the opposite limiting case, when the radius of curvature $r$ of the particle surface becomes much greater than the characteristic values of the field penetration depths into the metal and dielectric, $r \gg \delta_{\text{m}},\,\delta_{\text{d}}$, the situation becomes similar to that of a flat surface of these materials. Here, the field is concentrated inside and outside the particle near its surface in a region whose volume is significantly less than that of the particle. In this limit, the depth of field penetration into the metal is practically independent of the shape and size of the particle and is determined only by the permittivity of the metal, $\varepsilon_{\text{m}}$, and the surrounding medium, $\varepsilon_{\text{d}}$, while their characteristic values correspond to the case of propagating plasmon-polaritons (see Table \ref{tab:PSPP-data}).

\subsection{Size-dependent dielectric functions of nanoparticles}

For metal particles two ranges of sizes are usually distinguished, showing qualitatively different electronic structure. The characteristic size separating these regions is the Fermi wavelength of the electron, which is equal to $\lambdabar_{\text{F}} \sim 1$ nm for gold, silver and copper. In the range of $r\lesssim \lambdabar_{\text{F}}$, quantum confinement effects can be observed in the energy spectrum and optical properties of metal clusters, while the role of these effects increases with their sizes decreasing. The properties of subnanometer-sized metal clusters can be described by a variety of quantum mechanical methods most notably the time-dependent density functional theory \cite{Koch2001}. If the size of a noble metal particle exceeds $\lambdabar_{\text{F}}\sim 1$ nm, then its energy spectrum becomes quasicontinuous within the allowed bands. Therefore, starting with sizes of several nanometers, it is possible to introduce the concept of the size-dependent dielectric function of a metal particle, $\varepsilon_{\text{m}}$, and describe its optical properties using the Maxwell's equations within the framework of classical electrodynamics of continuous media.

The local permittivity of a noble metal in the visible and near-IR spectral ranges can be represented as the sum of the intraband, $\varepsilon_{\text{intra}}$, and interband, $\varepsilon_{\text{inter}}$, transitions:
\begin{equation}
	\varepsilon_{\text{m}} \left( \omega \right) = \varepsilon_{\text{intra}} \left( \omega \right) + \varepsilon_{\text{inter}} \left( \omega \right).
\label{eps-metal}
\end{equation}
\begin{equation}
	\varepsilon_{\text{intra}} \left( \omega \right) =\varepsilon^{\infty}_{\text{m}} - \frac{\omega _{\text{p}}^{2}}{\omega ^{2}+i\omega \gamma_{\text{intra}} }\,, \quad 
	\omega _{\text{p}}=\left(\frac{4 \pi n_e e^2}{m_e}\right)^{1/2}.
	\label{eps-intra} 
\end{equation}
\begin{equation}
\begin{split}
	\varepsilon_{\text{inter}}\left( \omega \right) &=K\int\limits_{\omega_{%
			\text{g}}}^{\infty} dx\frac{\sqrt{x-\omega _{\text{g}}}}{x}\left[ 1-F\left(
	x,\Theta \right) \right]  \times\\
	&\times \frac{\left( x^{2}-\omega ^{2}+\gamma _{\text{inter}}^{2}-2i\omega
		\gamma _{\text{inter}}\right) }{\left( x^{2}-\omega ^{2}+\gamma _{\text{inter%
		}}^{2}\right) ^{2}+4\omega ^{2}\gamma _{\text{inter}}^{2}}\,.
\end{split}
	\label{eps_inter}
\end{equation}
The quantity of $\varepsilon_{\text{intra}}$ in \eqref{eps-metal} is described by the modified Drude formula \eqref{eps-intra} and includes a frequency-dependent part (the second term in \eqref{eps-intra}), responsible for collective oscillations of free electrons in a metal with a plasma frequency $\omega_{\text{p}}$ and a damping coefficient $\gamma_{\text{intra}}$ ($1/\gamma_{\text{intra}}$ is the corresponding relaxation time of free electrons). In addition, following \cite{Gaponenko2010, Novotny2012}, in formula \eqref{eps-intra} a frequency-independent term is introduced. It is the constant $\varepsilon_{\text{m}}^{\infty}$, which is generally not equal to unity.

The quantity $\varepsilon_{\text{inter}}$, described by formula \eqref{eps_inter}, determines the frequency-dependent contribution of bound electrons, i.e. electronic transitions between the valence d-band and the conduction sp-band of noble metals; $F(x,\Theta)$ is a distribution of electron energy, $\hbar x$, at temperature of $\Theta$; $\gamma_{\text{inter}}$ is a damping coefficient for the specified interband transitions. In accordance with \cite{Bigot1995, Inouye1998}, the d-band is dispersionless, while the sp-band is parabolic with a minimal energy of $\hbar \omega_{\text{g}}$ relative to the d-band. Constant $K \propto D^2$, where $D$ is the transition dipole moment. 

Taking into account only the contribution of free electrons, we can put $\varepsilon_{\text{m}}^{\infty} = 1$ in formula \eqref{eps-intra}. However, it is convenient \cite{Gaponenko2010, Novotny2012} to add there a term, $\varepsilon_{\text{m}}^{\infty}-1$, which describes the contribution of the interband transitions that do not belong to the visible and near-IR spectral regions, thus neglecting their frequency dispersion. According to \cite{Gaponenko2010}, the typical values of $\varepsilon_{\text{m}}^{\infty}$ constant vary in the range of $1 \leq \varepsilon_{\text{m}}^{\infty} \leq 10$ for different metals.

When the particle size is small compared to the electron mean free path, $l_{\infty}$, in a bulk metal, in the calculation of $\varepsilon_{\text{intra}}\left(\omega \right)$ one should additionally take into account the size effect associated with the free-electron scattering from the interface between a metal particle and its external medium (e.g., solution or J-aggregate shell). This leads to the dependence of the damping coefficient $\gamma_{\text{intra}}$, and hence the permittivity, $\varepsilon_{\text{intra}}\left(\omega,r \right)$, on both frequency $\omega$ and the particle radius $r$. For the effective damping coefficient we will use the well-known phenomenological expression \cite{Kreibig1995}
\begin{equation}
\gamma_{\text{intra}}^{\left(r\right)} = \gamma _{\text{intra}}^{\text{bulk}} + \xi \frac{v_{\text{F}}}{r}, \quad
\gamma_{\text{intra}}^{\text{bulk}}=\frac{1}{\tau}=\frac{1}{\tau_{\text{e-e}}} + \frac{1}{\tau_{\text{e-ph}}}+\frac{1}{\tau_{\text{e-defect}}},
\label{gamma-free-size-dep}
\end{equation}
\noindent which was obtained for spherical particles. Here $v_{\text{F}}$ is the Fermi velocity; $\xi$ is a dimensionless constant ($\xi\sim 1$), which is determined by the specifics of the scattering process of free electrons from the interface between a metal nanoparticle and its environment and can be deduced from experimental data on photoabsorption. The damping coefficient in bulk metal, $\gamma_{\text{intra}}^{\text{bulk}}$, is determined by the relaxation times $\tau_{\text{e-e}}$ and $\tau_{\text{e-ph}}$ in electron-electron and electron-phonon collisions, as well as the relaxation time $\tau_{\text{e-defect}}$ in electron scattering from defects, respectively.

Since the size effect has a weak influence on the contribution of interband transitions, the following expression is often used to reliably describe both contributions to the permittivity of a metal nanoparticle, taking into account the size effect
\begin{equation}
\varepsilon_{\text{m}}\left( \omega ,r\right) =\varepsilon _{\text{bulk}}\left( \omega \right) + \frac{\omega _{\text{p}}^{2}}{\omega^{2}+i\omega \gamma _{\text{intra}}^{\text{bulk}} }
- \frac{\omega _{\text{p}}^{2}}{\omega ^{2} + i \omega \gamma_{\text{intra}}^{\left(r\right)}}
\label{eps-size-dep}
\end{equation}
\noindent following a semiempirical approach \cite{Kreibig1995}. Here $\varepsilon_{\text{bulk}}$ is the permittivity of a bulk metal obtained from experimental data (see Refs. \cite{Johnson1972,Babar2015, Babar2015,Palik1991, Rakic1995} for Ag, Au, Cu, Al); $\gamma_{\text{intra}}^{\text{bulk}}$ is the frequency of electron scattering in a bulk metal, and $\gamma_{\text{intra}}^{\left(r\right)}$ is the frequency of electron scattering (\ref{gamma-free-size-dep}) from the surface of a metal particle.

\begin{figure}[b]
	\centering\includegraphics[width=0.98\linewidth]{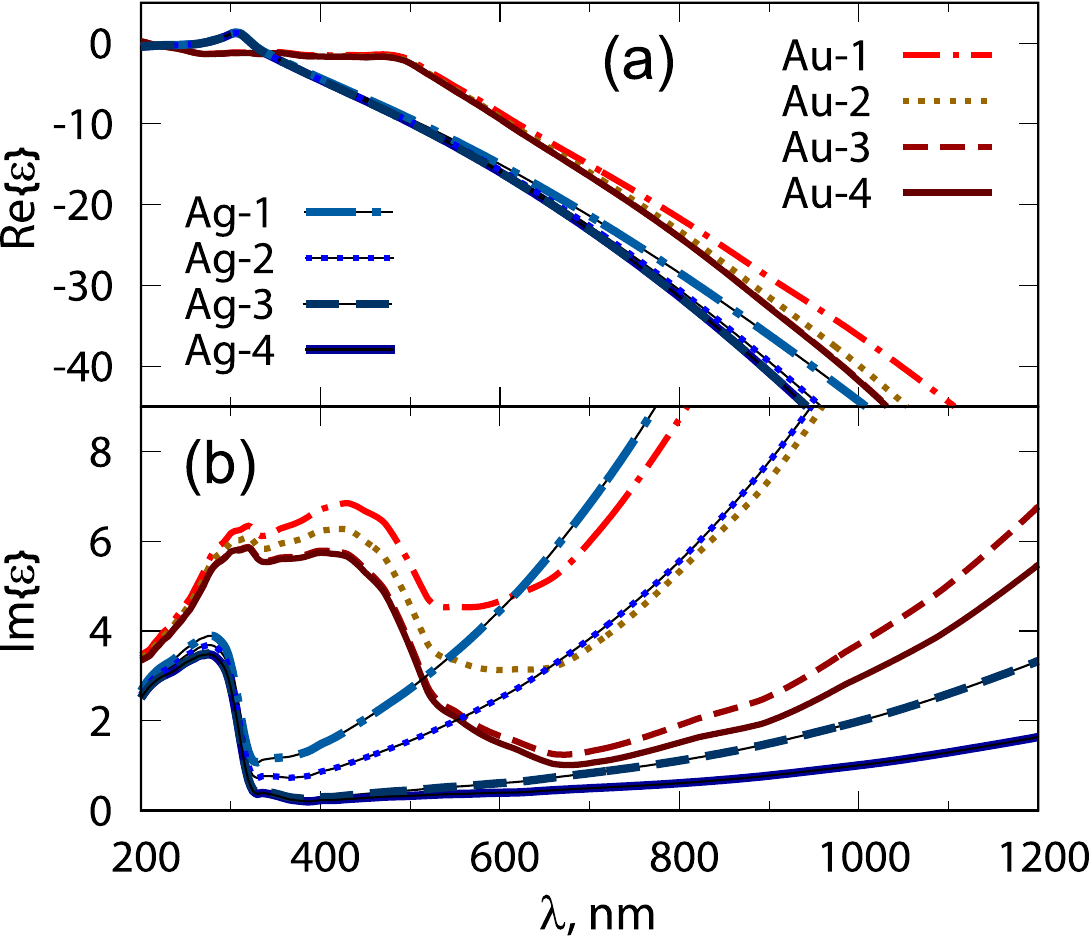}
	\caption{Real (a) and imaginary (b) parts of the dielectric functions of silver and gold spherical nanoparticles calculated using formulas \eqref{gamma-free-size-dep} and \eqref{eps-size-dep} for three values of their diameter: $D=10$ nm (curves Ag-1 and Au-1), $D=20$ nm (Ag-2 and Au-2), and $D=200$ nm (Ag-3 and Au-3). Curves \text{Ag-4} and \text{Au-4} show experimental data for bulk metal \cite{Johnson1972,Babar2015}.}
	\label{fig:sd_me}
\end{figure}

To demonstrate the size-dependent behavior of the dielectric functions of silver and gold nanospheres with sizes smaller than the mean free path of an electron in a bulk metal, we show in Figs.~\ref{fig:sd_me}a and \ref{fig:sd_me}b the dependences of their real and imaginary parts on the  wavelength for three values $r=5$, $10$, and $100$ nm of particle radius. It follows that the real parts $\text{Re}\{\varepsilon_{\text{m}}(\omega,r)\}$ of silver and gold depend weakly on $r$. In contrast, the imaginary parts $\text{Im}\{\varepsilon_{\text{m}}(\omega,r)\}$ depend significantly on the particle size, especially at $\lambda \gtrsim 320$ nm for Ag and at $\lambda \gtrsim 550$ nm for Au, where the contribution of free electrons is predominant.

A comparison of the results of calculations of light absorption spectra carried out for a number of metal and metalorganic nanospheres of different radii $r$ showed that the dielectric functions determined by the equations~(\ref{gamma-free-size-dep}) and (\ref{eps-size-dep}) describe well the spectral manifestations of the size effect for Ag, Au, Cu, and Al \cite{Kreibig1995, Lebedev2008, Lebedev2010}. In~\cite{Kondorskiy2023b, Kondorskiy2023a} the same expressions were applied to nanoparticles of more complex shapes, including nanoprisms and nanorods. It was found that even in the case of non-spherical geometry, expressions~(\ref{gamma-free-size-dep})--(\ref{eps-size-dep}) allow us to accurately describe the available experimental data on the extinction spectra of metal nanoparticles. To this end, it is possible to introduce the effective radius, $r_{\text{eff}}$, of a sphere with the same volume, $\mathcal{V}$, as that of the nanoparticle under consideration, i.e. $r_{\text{eff}}=\left[3 \mathcal{V}/(4\pi)\right]^{1/3}$. For silver nanoprisms, this approach provides a good agreement between theory and experiment upon variation of the prism sizes over a wide range~\cite{Kondorskiy2023a}.

For very small sizes of a metal nanoparticle, $r \lesssim 2 \pi v_{\text{F}} / \omega$, to calculate its dielectric function, one should, strictly speaking, take into account the nonlocal nature of the relationship between the induction vector, $\mathbf{D}$, and the electric field strength, $\mathbf{E}$, leading to spatial dispersion effects. The theory of light absorption by spherical metal particles of small radius, with the effects of spatial dispersion accounted for, has been developed in a number of works (see, e. g., \cite{Fuchs1987, Ruppin1976}). In this case, the permittivity of a nanoparticle is a tensor even in an isotropic medium, and the distinct direction is defined by the wave vector $\mathbf{k}$. Furthermore, when the medium has an inversion center, we can introduce the quantities  $\varepsilon_{\text{l}}\left(\omega,k\right)$ and $\varepsilon_{\text{t}}\left (\omega,k\right)$, referred to as longitudinal and transverse permittivities, respectively. If $\textbf{E}\parallel\textbf{k}$, then $\textbf{D}=\varepsilon_{\text{l}}\textbf{E}$, and if $\textbf{E}\perp \textbf{k}$, then $\textbf{D}=\varepsilon_{\text{t}}\textbf{E}$. Thus, both transverse and longitudinal waves can propagate in the medium. Their dispersion laws are given by
\begin{equation}
	k_{\text{t}}^{2}=\left(\frac{\omega}{c}\right)^{2}\varepsilon_{\text{t}}\left(\omega,k_{\text{t}}\right), \qquad \varepsilon_{\text{l}}\left(\omega,k_{\text{l}}\right)=0. \label{nonloc}
\end{equation}
\noindent Within the framework of the hydrodynamic model of electron gas \cite{Aleksandrov1999}, which includes the spatial dispersion, the permittivity can be written as
\begin{equation}
	\varepsilon\left(\omega,k\right)=1-\frac{\omega _{\text{p}}^{2}}{\omega^{2}+i\omega\gamma-\beta^{2}k^{2}},\qquad \beta^{2}=\frac{3}{5} v_{\text{F}}^{2},
	\label{hydrodyn-eps}
\end{equation}
\noindent so that the longitudinal wave vector turns out to be
\begin{equation}
	k_{\text{l}}^{2}=\frac{5}{3v_{\text{F}}^{2}}\left(\omega^{2}+i\omega\gamma-\omega _{\text{p}}^{2}\right). \label{k-along}
\end{equation}
Calculations performed within the framework of this theory show that the nonlocal effect leads to a slight shift in the position of light absorption maxima by metal nanoparticles towards the short-wavelength spectral region as their radius, $r$, decreases \cite{Ruppin1976}.  The shift value depends on light wavelength $\lambda$ and Fermi velocity $v_{\text{F}}$, and it is stronger the smaller the size of a sphere. Simultaneously the absorption peak amplitude slightly decreases. For silver nanospheres with $r = 1$ nm, the shift, $\Delta \lambda$, of the absorption maximum position turns out to be equal to $11$ nm and quickly decreases with the increasing radius. At $r = 10$ nm this shift becomes equal to only $\Delta \lambda = 1$ nm, and at $r>10$ nm it is negligible. For gold nanoparticles, nonlocal effects are even weaker.

\subsection{Dipole and multipole plasmon resonances}\label{section:plasmon-multipoles}

When the frequency of the external electromagnetic field coincides with one of the eigen-frequencies of the plasma oscillations, one can observe the phenomenon of localized surface plasmon-polariton resonance. Under resonant conditions the induced oscillations of free electrons being coherent with electromagnetic field oscillations lead to a strong enhancement of the local field near the nanoparticle as compared to the external field strength. As a result, the light absorption and scattering cross sections increase drastically. The frequencies and widths of resonant peaks of light absorption and scattering depend on the particle size, shape and permittivities of the metal and the environment \cite{Kelly2003, Klimov2009, Lam_JRLR2018}. In an aqueous solution, silver and gold nanospheres exhibit plasmon-polariton resonances in the visible region. For many other metals they are located in the UV-region \cite{Creighton1991, Leb-Medv2012}.

For a homogeneous metal sphere, the frequency of the dipole plasmon mode can be determined in the quasistatic approximation ($kr\ll 1$) from the relation $\text{Re}\left\{\varepsilon_{\text{m}}(\omega )\right \} =-2\varepsilon_{\text{h}}$, where $\varepsilon_{\text{h}}$ is the permittivity of the host medium. Substituting here  expression \eqref{eps-intra} for the contribution of intraband transitions to the dielectric function, $\varepsilon_{\text{m}}$, and neglecting its imaginary part, we have
\begin{equation}
	\omega^{(\text{dip})}_{\text{pl}} \equiv \omega_{\text{Fr}}=\frac{\omega_{\text{p}}}{\sqrt{2\varepsilon_{\text{h}}+\varepsilon^{\infty}_{\text{m}}}}.
\label{eq:Frohlich}
\end{equation}
This is the expression for the Fr\"{o}hlich frequency \cite{Bohren1998}, which determines the position of the resonant electric dipole plasmon mode, $\omega^{(\text{dip})}_{\text{pl}}$, expressed in terms of the plasma frequency, $\omega_{\text{p}}$, of electrons in the metal (\ref{eps-intra}). Analytical consideration \cite{Bohren1998, Klimov2009} of the condition for the excitation of plasmon resonance of the electric type of the $n$-th order of multipolarity leads to the well-known result: $\text{Re}\left\{{\varepsilon_{\text{m}}(\omega )}/{\varepsilon_{\text{h}}}\right\} =-({n+1})/{n}$.

\begin{figure*}[ht]
	\centering\includegraphics[width=0.98\textwidth]{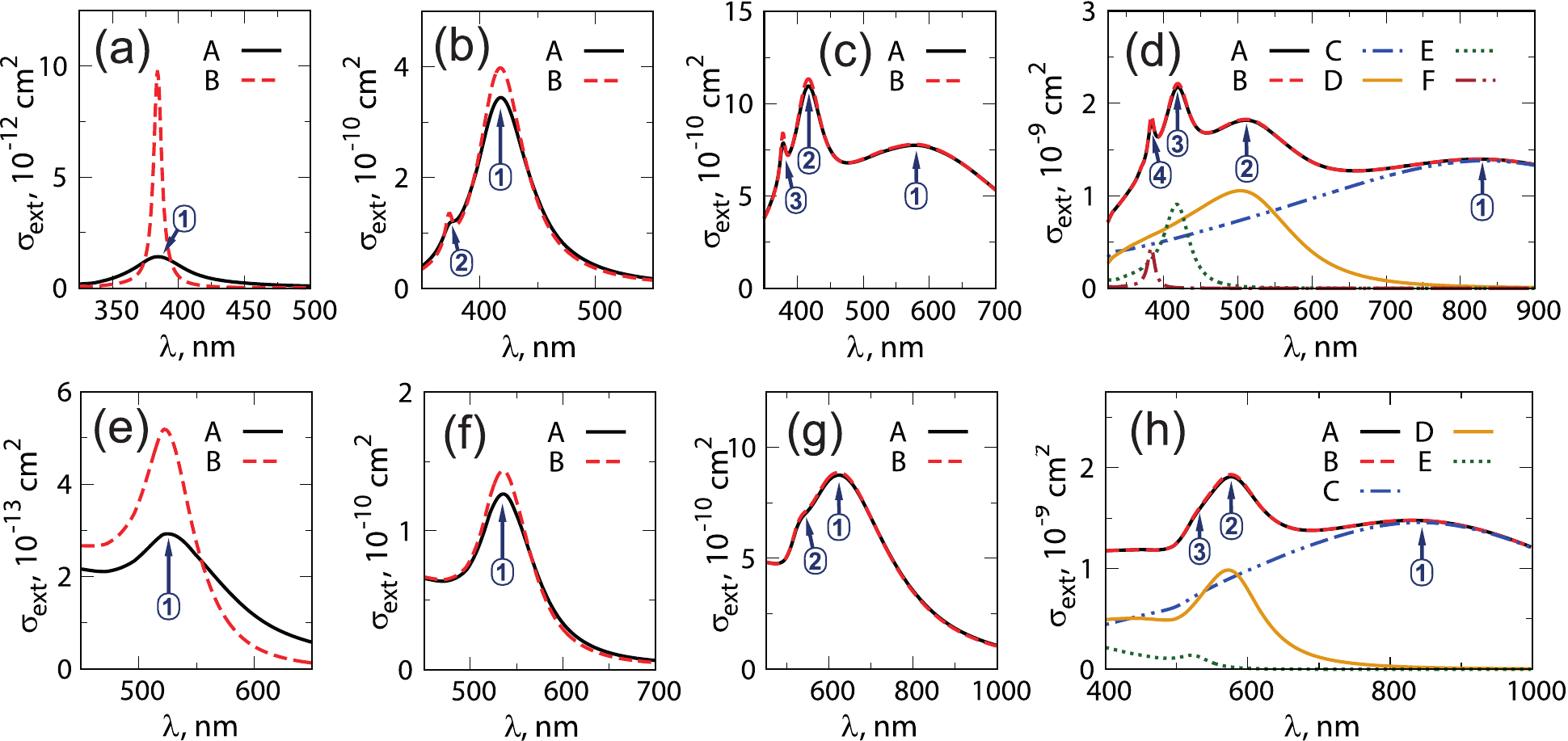}
	\caption{Extinction cross sections of silver (a)--(d) and gold (e)--(h) nanospheres with diameters: $D = 10$ nm (a,e), $60$ nm (b, f), $140$ nm (c, g) and $220$ nm (d, h) in aqueous solution as functions of light wavelength in vacuum, $\lambda$ \cite{Kond-Leb_JRLR2021}. Black solid curves (A) are calculations within the framework of the Mie theory using size-dependent local dielectric functions, $\varepsilon(\omega,D)$, of silver and gold particles (see formulas (\ref{gamma-free-size-dep}) and (\ref{eps-size-dep})). Red dotted curves (B) show similar results obtained using the dielectric functions, $\varepsilon_{\text{bulk}}(\omega)$, of bulk metals. Plasmon resonances are designated by numbers 1, 2, 3, and 4, which indicate the order of a multipole. For particles of large diameter, $D=220$ nm, curves C, D, E, and F in panels (d) and (h) show the contributions of multipoles of different orders ($n=1$, $2$, $3$, and $4$, respectively) into the total extinction cross section, $\sigma_{\text{ext}}$.}
	\label{fig:multipole}
\end{figure*}

Figure \ref{fig:multipole} shows the results of calculations of the extinction cross sections, $\sigma_{\text{ext}}$, of silver and gold nanospheres in an aqueous solution, obtained within the framework of the Mie theory. The dielectric function of the aqueous solution (i.e the permittivity of the host medium surrounding the particle) varies slightly in the visible spectral range and is $1.77 < \varepsilon_{\text{h}} < 1.82$ at $350 \;\text{nm} < \lambda < 700 \;\text{nm}$. Comparison of the wavelength dependences of cross sections, $\sigma_{\text{ext}}$, for different values of the sphere diameter $D$ = 10, 60, 140 and 220 nm allows us to track changes in the behavior of the extinction spectra. In particular, it is possible to determine changes in the positions and maximal intensities of peaks of the localized surface plasmon resonances with an increase in the particle size. 

At $D = 10$ nm, the spectral peaks of silver (Fig. \ref{fig:multipole}a) and gold (Fig. \ref{fig:multipole}d) particles correspond to the electric dipole localized plasmon resonance ($n = 1$). An increase in the particle size (Figs. \ref{fig:multipole}b-g and \ref{fig:multipole}d-h) leads to a qualitative change in the spectral behavior of their extinction cross sections. This is primarily due to the increasing role of multipole plasmon resonances of a higher order, $n > 1$. In particular, Fig. \ref{fig:multipole}b-g clearly shows the sequential formation of new spectral peaks for silver particles: first, a quadrupole at $D = 60$ nm ($n = 2$, Fig. \ref{fig:multipole}b); then an octupole at $D = 140$ nm ($n = 3$, Fig. \ref{fig:multipole}c); and finally, a hexadecapole at $D = 220$ nm ($n = 4$, Fig. \ref{fig:multipole}d). It is evident that the contribution of multipole resonances ($n > 1$) drastically changes the spectral behavior of the extinction cross sections as compared to the case of small diameter particles.

The results of similar calculations for gold particles (Fig. \ref{fig:multipole}e-h) also demonstrate the sequential formation of new spectral peaks of multipole plasmon resonances with increasing diameter $D$. However, their specific effect on the behavior of the $\sigma_{\text{ext}}(\lambda) $ dependences differs significantly from the previous case (Fig. \ref{fig:multipole}a-d). Note that since the widths of plasmon resonances for gold are larger than for silver, some of the peaks partially merge with each other. As a result, the contribution to the extinction cross section of multipole resonances of an order $n > 1$ becomes significant at larger values of $D$ than for resonances of the same order $n$ in the case of silver particles.

The results shown in Fig.~\ref{fig:multipole} suggest how the size effect in the dielectric function (\ref{eps-size-dep}) of silver and gold nanoparticles, $\varepsilon_{\text{m}}(\omega, r)$, may impact their optical properties. We compared the results for $\sigma_{\text{ext}}(\lambda)$ calculated together with the size effect associated with free electron scattering from the metal/water interface (black solid curves) and without it (red dotted curves), i.e., by using the permittivity of the bulk metal, $\varepsilon_{\text{bulk}}(\omega)$ \cite{Johnson1972,Babar2015}. The size effect is especially strong for a silver particle (see Fig.~\ref{fig:multipole}a--~\ref{fig:multipole}d), if its size, $D=10$ nm, is significantly less than the length of  free electron path, $l_{\infty}^{\text{Ag}}=53.3$ nm, in a bulk sample. Then the peak of the dipole plasmon resonance, calculated using the dielectric function $\varepsilon^{\text{Ag}}_{\text{bulk}}(\omega)$ at $D=10$ nm, turns out to be much narrower than that observed in the experiment \cite{Kometani2001,Lebedev2010}. In contrast, the use of the size-dependent dielectric function of silver, $\varepsilon_{\text{Ag}}(\omega, r)$, leads to the peak's broadening by a factor of 6.7 at $D=10$ nm and by a factor of 1.2 at $D=60$ nm. The corresponding decrease in the maximum intensity of the dipole peak also turns out to be most significant at small particle sizes $D\ll l_{\infty}^{\text{Ag}}$. 

When the particle diameter, $D$, becomes larger than $l_{\infty}^{\text{Ag}}=53.3$ nm, the effect of the size-dependent dielectric function of silver on extinction spectrum is slight (see Fig.~\ref{fig:multipole}c, \ref{fig:multipole}d). A similar analysis for Au (see Fig.~\ref{fig:multipole}e--\ref{fig:multipole}h) shows that the influence of the size effect in the dielectric function on the spectral properties of gold nanoparticles is weaker than that of silver \cite{Leb-Medv2012, Leb-Medv2013a}. The reason is that in the frequency range $\hbar \omega\lesssim 3.5$ eV the main contribution to the permittivity of silver is made by the intraband transitions, while in gold the role of interband transitions becomes noticeable even at $\hbar \omega \gtrsim 1.7$ eV. In addition, the mean free path of an electron in gold is lower than that in silver.

\subsection{Effect of nanoparticle shape on optical spectra}

\begin{figure}[ht]
	\centering\includegraphics[width=0.94\linewidth]{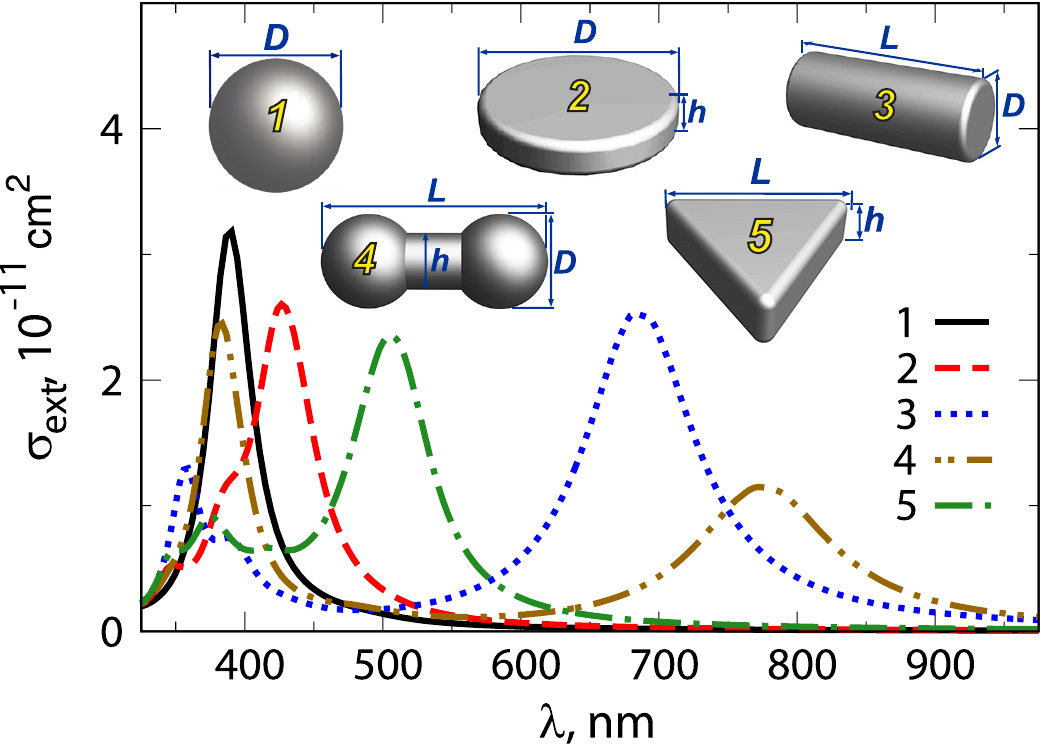}
	\caption{Extinction cross sections of silver nanoparticles in water as functions of light wavelength in vacuum, $\lambda$ \cite{Mekshun2020}. The particles have different shapes, but the same volume ($\mathcal{V} = 8500$ nm$^{3}$): (1) -- the sphere with a diameter $D=25$ nm; (2) -- the disk with a diameter \textit{D} = 27 nm and a height \textit{h} = 15 nm; (3) -- the rod with a diameter \textit{D} = 15 nm and a length \textit{L} = 49 nm; (4) -- the dumbbell with length \textit{L} = 42 nm, a ball diameter \textit{D} = 20 nm, and a waist diameter \textit{h} = 10 nm; (5) -- the triangular prism with an edge length \textit{L} = 33 nm and a height \textit{h} = 15 nm. The cross sections are averaged over all possible particle orientations.}
	\label{fig:plasm_shp_eff}
\end{figure}

For a particle of an elongated (e.g., spheroid and rod) or oblate (disk, prism, and star) shape the plasmonic peak is divided into two, so that the longitudinal and transverse plasmon resonances appear in optical spectra. This holds true even for nanoparticles of sufficiently small sizes for which the quasistatic approximation is definitely justified. The more the longitudinal and transverse dimensions of the nanoparticle differ from each other, the greater the spectral distance between the peak positions is. The peak shifted toward the short-wavelength region corresponds to the electron oscillations perpendicular to the long axis of the elongated particle, and the second peak (shifted toward the long-wavelength region of the spectrum) is due to the electron oscillations along its long axis (see Fig.~\ref{fig:plasm_shp_eff}).

These spectral features of elongated and oblate nanoparticles are clearly illustrated in Figure \ref{fig:plasm_shp_eff}. There we compare results of our calculations for the extinction cross sections of silver nanoparticles of various shapes. All calculations were performed using the FDTD method for nanoparticles having the same volume of $\mathcal{V}=8500$ nm$^3$. We considered the following particle shapes: the sphere with a diameter of $D=25$ nm (black solid curve 1); the disk with a diameter \textit{D} = 27 nm and a height \textit{h} = 15 nm (red dashed curve 2); the rod with a diameter \textit{D} = 15 nm and a length \textit{L} = 49 nm (blue dotted curve 3); the dumbbell with a length \textit{L} = 42 nm, a ball diameter \textit{D} = 20 nm, and a waist diameter \textit{h} = 10 nm (brown dash-double-dotted curve 4); the triangular prism with an edge length \textit{L} = 33 nm and a height \textit{h} = 15 nm (green dash-dotted curve 5). The results shown in Fig. \ref{fig:plasm_shp_eff} represent the cross sections averaged over three equiprobable particle orientations, corresponding to the nanoparticles randomly oriented in the solution. Calculations of the extinction spectra of nanoparticles of various shapes were also performed elsewhere (see, e.g., \cite{Kelly2003, Lam_JRLR2018, Mekshun2020, Kond-Leb_JRLR2021}). 

Figure \ref{fig:plasm_shp_eff} clearly demonstrates the significant influence of the geometrical shape of nanoparticles on the behavior of the extinction cross sections, the positions of spectral peaks maxima, and their total number. It is evident that for a small silver sphere the extinction spectrum contains only one peak corresponding to the electric dipole plasmon resonance (black solid curve 1), whose position of maximum is well described in the quasistatic approximation by the Fr\"{o}hlich formula \eqref{eq:Frohlich}. Whereas, for a nanorod with a given value of aspect ratio, $L/D = 3.3$, there are two spectral peaks (blue dotted curve 3) corresponding to longitudinal (long-wavelength peak) and transverse (short-wavelength peak) plasmon resonances. A similar situation is observed in the extinction spectrum of a dumbbell (brown dashed-dotted curve 4). Here for the aforementioned geometrical parameters the spectral distance between the peaks of the longitudinal and transverse resonances reaches $\Delta\lambda\approx 400$ nm. 

The plasmon resonance splitting is also observed in nanoplatelets, i.e. nanoparticles of oblate geometrical shape. This is illustrated in Figure \ref{fig:plasm_shp_eff} with a nanodisk (red dashed curve 2) and a triangular nanoprism (green dash-dotted curve 5), both having two pronounced spectral peaks in their optical spectra. Thus, by varying the shape of the particle, it is possible to shift the positions of the plasmon resonance peak into a given wavelength range. This specific feature of plasmonic nanoparticles is particularly useful for creating nanostructures and nanomaterials with desired optical properties.

\begin{figure*}[ht]
	\centering\includegraphics[width=0.94\textwidth]{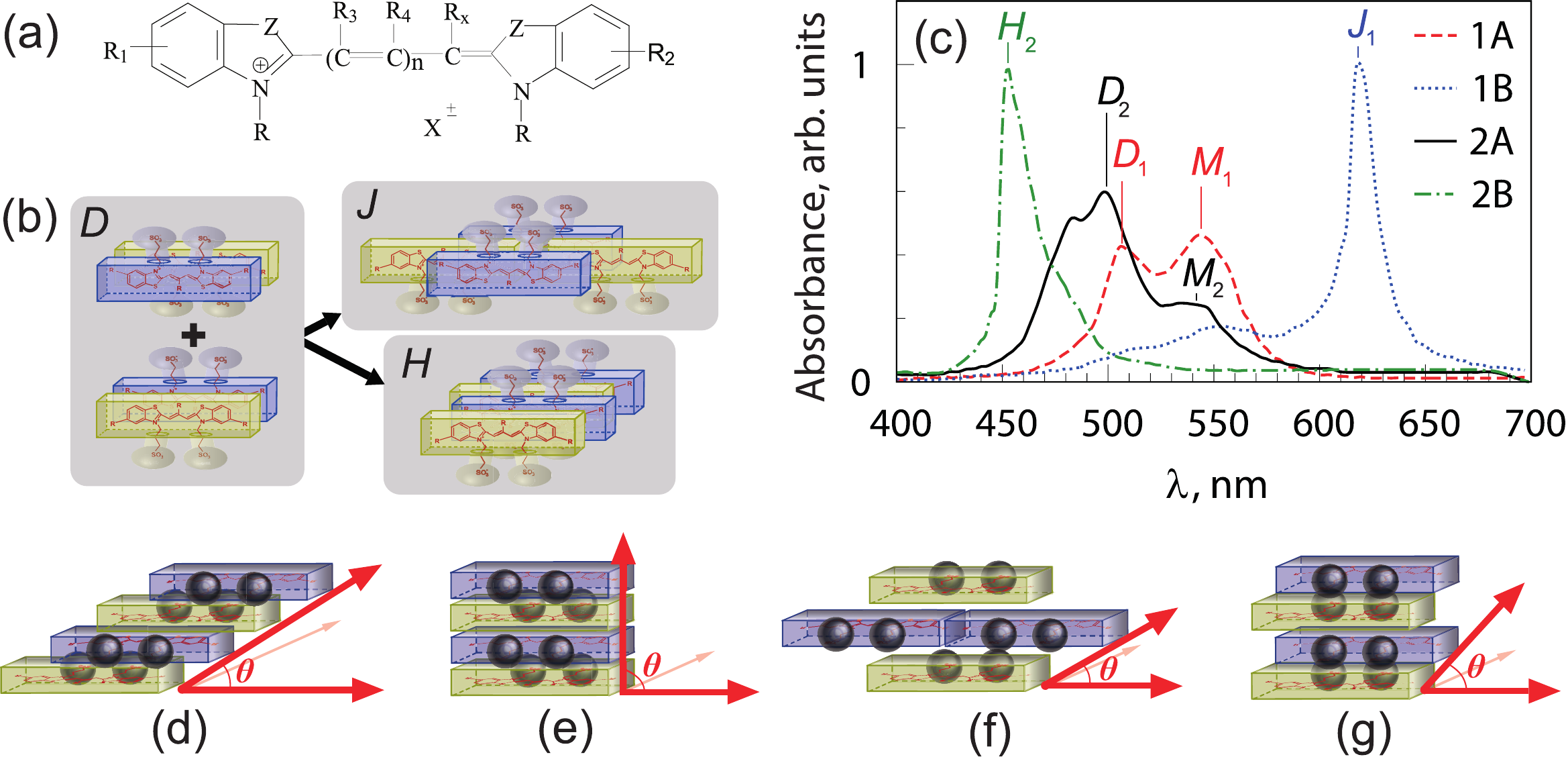}
	\caption{Structure and spectra of dye molecular aggregates: (a) Typical structural formula of cyanine dye~\cite{Shapiro2008}; Z can be O, S, Se, NR, -CH=CH-, etc.; R, R$_1$-R$_\text{x}$ -- various substituents; $n$ can take values from 0 to 7; X$^\pm$ -- counterion. (b) Pathways to form J- and H-aggregate blocks (J and H) from dye dimers (D). (c) Absorption spectra of an aqueous solution of \textit{meso}-ethyl-substituted thiatrimethine cyanine (1A) and (1B), and \textit{meso-}methyl-substituted thiatrimethine cyanine (2A) and (2B)~\cite{Shapiro-MDJH-spectra}. (1A) and (2A) -- results before aggregation, (1B) and (2B) -- results after aggregation. Vertical lines and designations around them in panel (c) indicate the positions of the peak maxima corresponding to various forms of the dye in solution: M -- monomers; D -- dimers; J -- J-aggregates; H -- H-aggregates. Spectra (1A) and (1B) are normalized to the J-band maximum, while spectra (2A) and (2B) are normalized to the H-band maximum. (d)--(g) Types of molecular packing in aggregates: (d) ''staircase''; (e) ''deck''; (f) ''brickwork''; (g) ''ladder''.}
	\label{fig:J-aggr_form}
\end{figure*}

The splitting of the longitudinal and transverse resonances in structures of elongated and oblate shapes discussed above is observed even for small-sized nanoparticles for which the quasistatic approximation is definitely justified, i.e., even when the peaks of multipole resonances ($n>1$) do not yet make a noticeable contribution to the light extinction and scattering cross sections. However, a significant increase in particle size in at least one direction leads to the inapplicability of the quasistatic approximation. Then, for a correct description of the optical spectra one needs also to account for the contributions of high-order multipoles (quadrupole, octupole, etc.) on top of the contribution by the first dipole term.
Therefore, for the rod of a nanometer transverse diameter $D$ and a large longitudinal size $L$ exceeding the de Broglie wavelength of light, $\lambdabar = \lambda/(2\pi)$, the behavior of the extinction spectra becomes more complicated. This was clearly shown in the experimental work \cite{Payne2006} for gold nanorods in heavy water with narrow size distribution. Measurements were performed in the visible and near-IR regions in the wavelength range of $400 \lesssim \lambda \lesssim 2000$ nm. The diameter of the rods in \cite{Payne2006} was $D=85$ nm, and the length reached $L = 1175$ nm. In a number of works, calculations of the optical spectra of micrometer-sized nanoparticles of various elongated and oblate shapes took into account $n$-th order multipole contributions ($n > 1$) \cite{Khlebtsov2007, Zenin2020, Ray2021}.

\section{Ordered molecular aggregates of organic dyes}

\subsection{Formation of molecular aggregates and simple excitonic model}

\subsubsection{Types of molecular packing in aggregates}

Molecular J-aggregates are ordered nanoclusters of non-covalently bound organic molecules in which the intermolecular interactions are significantly weaker than the intramolecular interactions. The monomer molecules in such a complex retain their individuality to a great extent, so that collective electronic excitations of the entire system can be correctly described in terms of small-radius excitons (Frenkel excitons) \cite{Frenkel1931}. It is very important that an excitation localized on a specific molecule is not stationary due to the translational symmetry and intermolecular coupling, so that it would travel through the aggregate. In contrast, the stationary excitation is delocalized, being a superposition of localized excitonic states, with the region of coherence covering the entire ordered aggregate \cite{Frenkel1931, Davydov1971}.

The most studied aggregates are polymethine (cyanine) dyes. The frequently used cyanine dyes are described by a structural formula (Fig.~\ref{fig:J-aggr_form}a), according to which they are cationic chromophores with heterocyclic nuclei (based on benzoxazole, benzthiazole, benzimidazole, quinoline, etc.) linked by a polymethine chain containing an odd number of methine groups~\cite{Shapiro2008}. Their compounds absorb light excellently in the visible and near-IR spectral regions due to the alternation of the signs of $\pi$-charges along the chromophore chain and their change to the opposite when excited by external radiation. The molar extinction coefficients reach the values of $\epsilon = (2-3) \cdot 10^{5}$~l$\cdot$ mol$^{-1}\cdot$cm$^{-1}$, which correspond to the absorption cross sections of $\sigma = (3-5) \cdot 10^{-16}$~cm$^{2}$.

Basically, the molecules of cyanine dyes are almost flat with angles between the planes of heterocycles less than $15^\circ$. For this reason, cyanine dyes are inclined to form  aggregates with a ''plane-to-plane'' structure possessing a shift of molecules relative to each other for optimal $\pi$-$\pi$-charge interaction of oppositely charged methine groups of neighboring molecules. Aggregates of cyanine dyes are formed in solutions \cite{Wurthner2011}, on crystal surfaces, including metal substrates \cite{Bellessa2004, Symonds2008}, on the surfaces of metal nanoparticles \cite{Kometani2001, Shapiro2015, Todisco2015, Takeshima2020}, on anionic platforms of magnesium dye complexes \cite{Shapiro_OE2018, Shapiro_QE2018}, and can also assemble on DNA templates \cite{Asanuma2012}. Besides, dye aggregates can be embedded in polymer matrices and grow along polymer threads \cite{Ciardelli2013}.

Depending on the concentration, dyes in solutions can be monomers M, dimers D, as well as J- and H-aggregates (see Fig. \ref{fig:J-aggr_form}). When the concentration of the solution increases and it is cooled, the dye first dimerizes, and then H- or J-aggregates are formed from the dimers in accordance with the ''block'' formation mechanism \cite{Shapiro2006, Shapiro-MDJH-spectra}. Accordingly, the simplest aggregate consists of two dimers (four monomers). However, at present, it is aggregates of a large number of monomers that are of significant interest for many practical applications, so that the length and width of the supramolecular system ranges from $1$ to $300$ $\mu$m and from $100$ nm to $1$ $\mu$m, respectively \cite{Shapiro_OE2018}. The characteristic thickness of J-aggregate layers can vary greatly. For example, the J-aggregate  thickness usually ranges from $0.5$--$1$ nm \cite{Kometani2001} to $3$--$5$ nm \cite{DeLacy2013} on the surfaces of metal nanoparticles, while the layers on metal films reach thickness of about $10$--$20$ nm \cite{Todisco2015}.

The main physical properties of aggregates are determined by the type of molecular packing~\cite{Shapiro2006}. The most studied is the ''brickwork'' type packing of linear and planar aggregates~\cite{Kobayashi1996}. Other well-known types of packing are ''ladder'' and ''deck of cards''. Recently, the main focus in the studies of molecular aggregates has been shifted towards supramolecular assemblies with a more complex geometrical structure. For example, there have been many works on tubular \cite{Brixner2017} and columnar \cite{Hecht-Wurthner2021} aggregates, cross-stacked aggregates \cite{Ma2021}, as well as helical nanofibers \cite{Lee-Schenning-2009}. In this review, we present only a number of basic results and theoretical treatments for the simplest of linear aggregates with one molecule in a unit cell as this review is aimed at the discussion and analysis of the most prominent results in the optics of plexcitonic systems. Most of the experimental and theoretical works in this area are devoted to the study of hybrid metalorganic nanoparticles and nanostructures containing the conventional J-aggregates of cyanine dyes as their organic component.

\subsubsection{Dispersion relation for simplest aggregates}

Now we come to the basic aspects of the delocalized Frenkel exciton theory \cite{Frenkel1931, Davydov1971, Bardeen2014} using the example of conventional J- and H-aggregates within the framework of the simple McRae-Kasha model \cite{McRae1958,Kasha1964,Kasha1965} based on point dipole-dipole interaction between monomer molecules. In this model, charge transfer and vibronic effects are neglected, and in the potential energy of interaction, only the predominant contribution of the nearest neighbors is taken into account. Thus, the energy of intermolecular interaction, which determines the behavior of the optical spectra of aggregates to a great extent, can be written as
\begin{equation}
U_{\mathrm{dd}}=\frac{\mathbf{d}_{1}\cdot \mathbf{d}_{2}-3\left(\mathbf{d}_{1}\cdot \mathbf{n}\right) \left(\mathbf{d}_{2}\cdot \mathbf{n}\right)}{r^{3}}.
\label{dipole-dipole-inter}
\end{equation}
\noindent Here $\mathbf{d}_{1}$ and $\mathbf{d}_{2}$ are transition dipole moments of ''1'' and ''2'' molecules; $\mathbf{n} = \mathbf{r}/r$, $\mathbf{r}$ is the displacement vector connecting their centers of mass. 

To derive the exciton dispersion law and to calculate the optical spectra of an aggregate, it is necessary to perform a basis transformation from the localized to delocalized states of Frenkel excitons (see, e.g., \cite{Davydov1971}). More specifically, the basic concepts of the molecular exciton theory can be formulated as follows: (i) any state of a molecular aggregate is represented as a product of the eigen-states of its constituent molecules; (ii) the localized excited state of an individual monomer transfers to other molecules (predominantly to nearest neighbors) through the dipole-dipole interaction; (iii) due to the presence of translational symmetry of the aggregate, its eigen-state is represented as a coherent superposition of local excitations. 

For conventional linear aggregates with one molecule per unit cell ($|\mathbf{d}_1|=|\mathbf{d}_2| \equiv d$), containing $N$ molecules in total, it leads to the following dispersion relation \cite{McRae1958, Davydov1971}:
\begin{equation}
\begin{split}
	\mathcal{E}_{q}\!\left(\theta\right) = \mathcal{E}_{\text{M}} + \mathcal{D} + 2\, \mathcal{U}_{\text{dd}} &\left(1 - 3 \cos^{2}\!{\theta}\right) \cos{\left(ql\right)}, \\
	&\mathcal{U}_{\text{dd}} = d^2\!/l^3.
\end{split}
	\label{eq:conventional_aggregate_dispersion_law}
\end{equation}
Here $q$ is the exciton wave number; $\theta$ is the angle between the transition dipole moment of a monomer and the axis of the aggregate; $\mathcal{E}_{\text{M}}$ is the transition energy in the isolated monomer molecule; $\mathcal{D}$ is the gas-to-crystal shift determined by a change in the energy of electrostatic interaction of the monomer with the environment and other aggregated molecules upon the local excitation \cite{Davydov1971}. Parameter $\mathcal{U}_{\text{dd}}$ determines the scale of the dipole-dipole coupling between the nearest neighbors; the value of $l$ corresponds to the distance, $r$, between the centers of neighboring molecules in expression \eqref{dipole-dipole-inter}.

A discrete set of possible values of the exciton wave number, $q \equiv q_j$, appears in formula \eqref{eq:conventional_aggregate_dispersion_law} due to the finite size of linear aggregates and is determined by the expression: $q_j = \pi j /\left[l\left(N + 1\right)\right]$, where $j$ is the excitonic state number ($j = 1,\,\dots,\,N$). According to the terminology accepted in the theory of molecular aggregates, such $q_{j}$ values correspond to the so-called open boundary conditions (see, for example, \cite{Didraga2004}). Thus, the energy width of the excitonic band is given by
\begin{equation}
	W_{\text{ex}} = 4\,\mathcal{U}_{\text{dd}} \left|1 - 3 \cos^{2}\!{\theta} \right| \cos{\left(\frac{\pi}{N+1}\right)},
\end{equation}
and the energy positions of the top and the bottom of the band can be written as
\begin{equation}
	\begin{split}
		&\mathcal{E}_{\text{top}}(\theta) = \mathcal{E}_{\text{M}} + \mathcal{D} + W_{\text{ex}}/2, \\
		&\mathcal{E}_{\text{bottom}}(\theta) = \mathcal{E}_{\text{M}} +  \mathcal{D} - W_{\text{ex}}/2.
	\end{split}
\end{equation}

\begin{table*}[ht]
	\caption{Abbreviations, names, peak positions and widths (Full Width at Half Maximum, FWHM) for a number of dye J-aggregates according to~\cite{Scheblykin2001, Kometani2001,Yoshida2009a,Yoshida2010,Uwada2007,Bellessa2009,Fofang2008,Shapiro2015,Melnikau2016}.}
	\label{table-Jaggr}
	\smallskip 
	\center{\begin{tabular}{|c|l|c|c|c|}
			\hline
			\hspace{7pt} Abbreviation \hspace{7pt} & \hspace{110pt} Name & \hspace{7pt} $\hbar\omega_{\text{ex}}$, eV \hspace{7pt} & 
			\hspace{7pt} $\lambda_{\text{ex}}$, nm \hspace{7pt} & \hspace{7pt} \begin{tabular}{c} $\gamma_{\text{ex}}$, meV \\ (FWHM)\end{tabular} \hspace{7pt} \\ \hline
			OC & {\footnotesize 3,3'-disulfopropyl-5,5'-dichlorooxacyanine triethylammonium salt} & 3.04 & 407 & 39 \\ \hline
			TC & {\footnotesize 3,3'-disulfopropyl-5,5'-dichlorothiacyanine sodium salt} & 2.61 & 475 & 66 \\ \hline
			PIC & {\footnotesize 1,1'-disulfopropyl-2,2'-cyanine triethylammonium salt} & 2.13 & 582 & 33 \\ \hline
			TDBC & {\footnotesize 
				\begin{tabular}{l}
					5,5',6,6'-tetrachloro-1-1'-diethyl-3,3'-di(4-sulfobutyl)-\\ 
					benzimidazolo-carbocyanine
				\end{tabular}
			} & 2.12 & 585 & 48 \\ \hline
			NK2567 & {\footnotesize 2,2'-dimethyl-8-phenyl-5,5',6,6'-dibenzothiacarbocyanine chloride} & 1.79 & 693 & 52 \\ \hline
			D725 & {\footnotesize
				\begin{tabular}{l}
					3,3'-di(g-sulfopropyl)-6,6'-dimethoxy-8,10-dimethylene-\\thiacarbocyanine betaine triethylammonium salt
				\end{tabular}
			} & 1.71 & 725 & 13 \\ \hline
			JC1 & {\footnotesize 5,5',6,6'-tetrachloro-1,1',3,3'-tetraethylimidacarbocyanine iodide} & 2.09 & 592 & 28 \\ \hline
			TPP & {\footnotesize 5,10,15,20-tetraphenyl-21H,23H-porphyrin} & 2.82 & 440 & 90 \\ \hline
			Thia(Et) & {\footnotesize 
				\begin{tabular}{l}
					3,3'-disulfopropyl-5,5'-dichloro-9-ethylthiacarbocyanine\\potassium salt
				\end{tabular}
			} & 2.00 & 621 & 67 \\ \hline
			Thia(Ph) & {\footnotesize 
				\begin{tabular}{l}
					3,3'-disulfopropyl-5,5'-dichloro-9-phenylthiacarbocyanine\\triethylammonium salt
				\end{tabular}
			} & 1.85 & 671 & 85 \\ \hline
	\end{tabular}}
\end{table*}

\begin{figure}[t]
	\centering\includegraphics[width=0.9\linewidth]{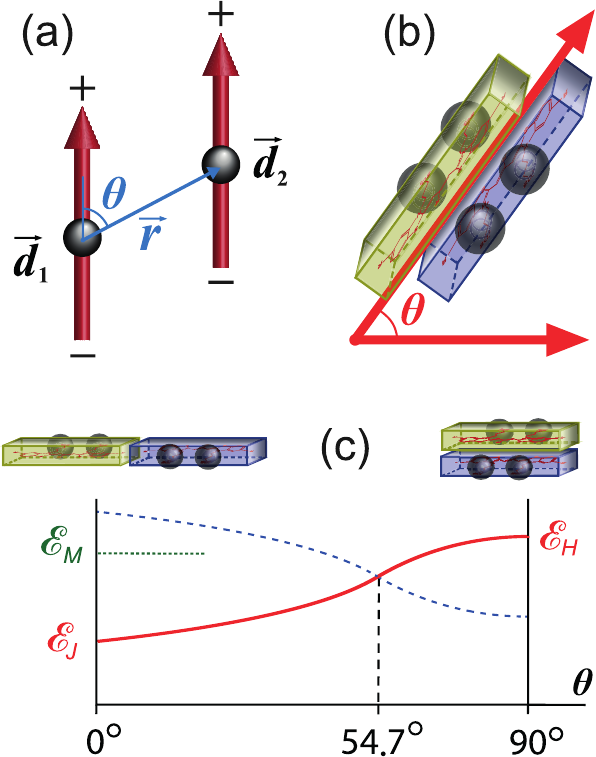}
	\caption{(a) Schematic view of a dye dimer in the form of point-like dipoles, $r$ is the distance between dipoles, $\left|\mathbf{d}_{1}\right|=\left|\mathbf{d}_2\right| \equiv d$ is the magnitude of the transition dipole moment in the monomer molecule, $\theta$ is the angle between the directions of the transition dipole moments, $\mathbf{d}_{1}$ and $\mathbf{d}_{2}$, and the aggregate axis. (b) Relative positions of the monomer molecules in the dimers constituting  the dye aggregates. (c) Dependence of phototransition energies of dye dimers on the $\theta$ angle. The red solid and blue dashed curves are the energies of allowed and forbidden phototransitions to the edges of the excitonic band, respectively. Green dotted horizontal line is the energy, $\mathcal{E}_{\text{M}}$, of the excited state of the monomer. J-aggregates are formed at $\theta <\theta_{\text{m}}$, H-aggregates at $\theta_{\text{m}}<\theta <\pi /2$ ($\theta_{\text{m}}=54.7^{\circ}$ is the magic angle).}
	\label{fig:en-diagram}
\end{figure}

It should be noted that the dispersion relation \eqref{eq:conventional_aggregate_dispersion_law} is applicable to linear aggregates with one molecule per unit cell. Such a structure is seen, for example, in aggregates of cyanine dyes TC, OC, Thia(Et) and TDBC (see Table \ref{table-Jaggr}), which were widely used as an organic component in plexcitonic systems of the ''core-shell'' type. For aggregates with a more complex geometrical shape, the excitonic band has a significantly more complex structure determined as a rule with a numerical simulation. However, for linear aggregates with two molecules in a unit cell, the dispersion relation can be presented in analytical form \cite{MorLeb2023, MorLeb2024}.

The optical properties of molecular aggregates are determined to a great extent by their excitonic band structure. For the linear aggregates considered above, the sign of the excitonic coupling energy, $U_{\mathrm{dd}}$, of the dye molecules plays a key role.
When the coupling energy is negative ($U_{\mathrm{dd}}<0$), a J-aggregate is formed, and the radiative transition to the bottom of the excitonic band is the most efficient, while the transition to the top of the band is forbidden.
In contrast, for positive coupling energy ($U_{\mathrm{dd}}>0$), an H-aggregate is formed, and the transitions to the bottom of the excitonic band is forbidden, while the radiative transitions to the top of the band are most intense. Correspondingly, for J-aggregates the major peak in the absorption spectrum shifts to the red region and undergoes substantial narrowing as compared to the monomer spectrum. In contrast, for H-aggregates the spectral maximum is shifted toward the blue region, and its band width is usually substantially larger compared to the typical width of the J-aggregate spectral band. However, in some cases the H-aggregate band is not much wider than the J-aggregate band. This is shown, for example, in Fig. \ref{fig:J-aggr_form}c, when both types of molecular aggregates are formed from the same dye.

Note that the exciton dispersion law  \eqref{eq:conventional_aggregate_dispersion_law} derived for the simplest case of a linear aggregate with one molecule per unit cell is suitable for the description of both possible types of aggregation. Moreover, expression \eqref{eq:conventional_aggregate_dispersion_law} clearly shows that the aggregation type (J or H), as well as the corresponding qualitative behavior and quantitative characteristics of the optical spectra depend directly on the angle $\theta$, which determines the orientation of the transition dipole moment in a monomer relative to the line connecting centers of neighboring molecules (see Fig. \ref{fig:en-diagram}a). The condition $U_{\mathrm{dd}}=0$ yields the so-called magic angle, $\theta_{\text{m}} = \arccos{\left(1\!/\!\sqrt{3}\right)} = 54.7^ {\circ}$, separating J- and H-types of aggregation (see Fig.~\ref{fig:en-diagram}c). In J-aggregates, the molecules are packed in a ''head-to-tail'' arrangement with the orientation angle, $\theta$, of the transition dipole moment, $0 \leq \theta < \theta_{\text{m}}$. In this case, the dipole-dipole coupling energy (\ref{dipole-dipole-inter}) is negative ($U_{\mathrm{dd}}<0$). In contrast, H-aggregates show a ''side-by-side'' packing with the orientation angles $\theta$ of the transition dipole moments in the range of $\theta_{\text{m}} < \theta \leq \pi /2$, resulting in a positive sign of the coupling energy ($U_{\mathrm{dd}}> 0$).

\subsubsection{\label{section:PL-absorption}Photoluminescence and photoabsorption}

The probabilities of radiative decay per unit time are very different for J- and H-aggregates. For a J-aggregate consisting of $N$ molecules, the rate of spontaneous decay increases $N$ times compared to a monomer (superradiance effect \cite{Dicke1954}). In contrast, in H-aggregates, after light absorption, there is a rapid intraband relaxation leading to the effective population of the state with the lowest energy, with the dipole transition to this state forbidden. Regarding the quantum yield of luminescence, we note that its value in J-aggregates is typically large, while in H-aggregates it is usually small. However, there are fluorescent H-aggregates \cite{Gierschner2013a, Gierschner2013b} in which the slow radiative decay rate dominates over the even slower nonradiative decay rate, resulting in a fairly high quantum yield.

Further we present the analytical expression for the light absorption coefficient, $K_{q}$, of a linear aggregate with one molecule per unit cell for an individual transition from its ground state $|g\rangle$ to a given excitonic state $|e_{q}\rangle$ with a wave number of $q$. It was recently obtained \cite{MorLeb2023,MorLeb2024} within the framework of the McRae-Kasha model \cite{McRae1958} and can be written as 
\begin{equation}
K_{q} = N_{\text{aggr}} \frac{\pi^2 d^2}{\hbar \omega c} \mathcal{E}_{q}^2\!\left(\theta\right)\left(1 + \frac{1}{2}\!\cos^2\!{\theta}\!\right) f_{N}\!\left(ql\right) a_{q}(\omega),
\label{eq:sigma_conventional_aggregate}
\end{equation}
\begin{equation}
{f}_{N}\!\left(ql\right) = \begin{cases} \frac{2}{N + 1}\cot^2\left({\frac{ql}{2}}\right),& j=1,\;3,\;5,\; ... \,, \\ 0,& \mathit{j}=2,\; 4,\;6,\; ... \, \end{cases}
\label{eq:f_M-1}
\end{equation}
\noindent Here $N_{\text{aggr}}$ is the concentration of aggregates in the ground state; $N$ is the number of molecules in the aggregate; $\omega$ is the photon frequency, $c$ is the speed of light. Possible $q$ values are determined from open boundary conditions are $q \equiv q_j = \pi j /\left[l\left(N + 1\right)\right]$, where $j$ is the excitonic state number ($j = 1,\,\dots,\,N$), $l$ is the unit cell length. The function $a_{q}(\omega)$ describes the resonant spectral band contour, normalized by the relation $\int a_q(\omega) d\omega = 1$, with a maximum at $\hbar \omega = \mathcal{E}_{q}(\theta)$. Formula \eqref{eq:sigma_conventional_aggregate} is averaged over the polarization directions of the incident light. 

In \cite{MorLeb2023, MorLeb2024} the result in \eqref{eq:sigma_conventional_aggregate} is extended to linear aggregates with two monomer molecules in a nonplanar unit cell. In addition, expressions are obtained for the absorption coefficient of polarized light by an aggregate, whose axis has a given orientation on a substrate or inside a flow-through cuvette. The resulting coefficient, $K_{\text{tot}}(\omega) = \sum_{q} K_{q}(\omega)$, of light absorption at the frequency $\omega$ by a molecular aggregate is determined by the sum over transitions $|g\rangle \rightarrow |e_q\rangle$ into all possible states of the excitonic band $|e_{q_j}\rangle$ ($j = 1,\,\dots\,,N$). Transitions to excitonic states with the lowest wavenumber, $q$, make the most of the contribution to the resulting absorption coefficient.

The excitonic approach presented above, based on the works of Frenkel \cite{Frenkel1931}, Davydov \cite{Davydov1971}, McRae and Kasha \cite{McRae1958,Kasha1964,Kasha1965}, have turned out to be successful for the classification and reasonable explanation of the photophysical properties of aggregates of a number of dyes. Its development is associated with the generalization of the simplest model of point dipole-dipole interaction as a result of a more accurate consideration of the effects of long-range and short-range interactions between molecules \cite{Hestand2018,Bawendi2019,Deshmukh2019}, as well as with the inclusion of vibronic coupling and intermolecular charge transfer effects in the theory \cite{Hestand2018, Bawendi2019, Brixner2017, Zhu2017}. A variety of molecular packing types in ordered aggregates of various shapes and topologies have been considered in the literature, including two-dimensional herringbone and HJ aggregates, tubular, columnar and helical structures. It is then possible to explore a wide range of new photophysical and photochemical phenomena  (see \cite{Levitz2018, Doria2018, Deshmukh2019, Shapiro_OE2018}, as well as reviews \cite{Shapiro2006,Wurthner2011,J-Aggregates2012,Bricks2018,Hestand2018, Otsuki2018, Ma2021, Hecht-Wurthner2021} and references therein). Their detailed discussion is beyond the scope of this article.

\subsubsection{Nonlinear optical properties}

The optical properties of molecular J-aggregates, which manifest themselves at relatively low electromagnetic excitation energy densities, have been discussed in Section \ref{section:PL-absorption}. At the same time, as has been established in several studies \cite{Bogdanov1991, Wang1991, Zhuravlev1992, Shelkovnikov1993, Gadonas1994, Spano1994, Markov2000, Shelkovnikov2002, Shelkovnikov2012, Lee2018}, molecular aggregates have extremely high resonant nonlinear optical susceptibilities and picosecond-scale relaxation times of the nonlinear response when exposed to optical radiation. For example, for the resonant cubic susceptibility $\chi^{(3)}$ of molecular aggregates PIC-I (pseudoisocyanine iodide) characteristic values of 
$\chi^{(3)}\sim 10^{-9}$ cm$^2\cdot$ V$^{-2}$ were obtained \cite{Gerasimova2000, Markov2001, Markov2004} based on z-scan experimental data using nanosecond laser pulses. Such values of $\chi^{(3)}$ are record-breaking among most organic and inorganic materials, including composite glasses, and are by $5$--$6$ orders of magnitude greater than those of polyconjugated polymers. Therefore, studies of various nonlinear optical effects involving J-aggregates were initially carried out given both their fundamental significance and potential applications in photonics and optoelectronics, for example, for efficient and high-speed nonlinear optical switches. In addition, giant nonlinear optical susceptibilities of molecular aggregates have been used in a number of works on nonlinear effects of their interaction with laser pulses of various time scales: from nanosecond \cite{Gerasimova2000, Markov2001, Markov2004} to femtosecond \cite{Knoester1996, Sasaki2001, Bednarz2001, Kano2002, Rehhagen2020, Belko2022, Jumbo-Nogales2022, Nishimura2004, Dijkstra2008, Ginsberg2009, Milota2009, Abramavicius2009, Bolzonello2016, Quenzel2022, Peruffo2023, Russo2024}. 

Specifically, a pump-probe technique was utilized to study the collective properties of multiexcitonic states in aggregates \cite{Knoester1996, Sasaki2001, Bednarz2001, Kano2002, Rehhagen2020, Belko2022, Jumbo-Nogales2022}, four-wave mixing was employed to investigate the coherence times of delocalized excitons \cite{Fidder1993, Minoshima1994, Kobayashi1996, Gadonas1997}, and hole burning was used to identify features of the hierarchical structure of cyanine dye aggregates \cite{Hirschmann1988, Kobayashi1996}. In recent years, two-dimensional electronic spectroscopy has become an efficient method for the examination of supramolecular structures, including J-aggregates \cite{Nishimura2004, Dijkstra2008, Ginsberg2009, Milota2009, Abramavicius2009, Bolzonello2016, Quenzel2022, Peruffo2023, Russo2024}. It is also appropriate here to note a series of works on the effects of optical bistability of molecular aggregates \cite{Malyshev1996, Malyshev1998, Glaeske2002, Zabolotskiy2006, Klugkist2007, Klugkist2008, Zabolotskiy2008, Nesterov2013} and Metal/J-aggregate composite systems \cite{Zabolotskiy2016}, since these effects are of certain interest in nonlinear optics and spectroscopy \cite{Boyd2008}. At the same time, despite the undoubted significance of these studies of nonlinear optical properties of ordered molecular aggregates, there have not been that many works on plexcitonic nonlinear effects in hybrid metalorganic nanostructures. In other words, plexcitonic nonlinear optics has not yet emerged into a separate extensive field of research, as is the case with linear optics of the systems involving metals and dye aggregates. Thus, we are not looking specifically into any of the above works here.

\subsection{Dielectric properties of dye aggregates}

\subsubsection{Scalar model of excitonic shell}

\begin{figure}[b]
	\centering\includegraphics[width=0.94\linewidth]{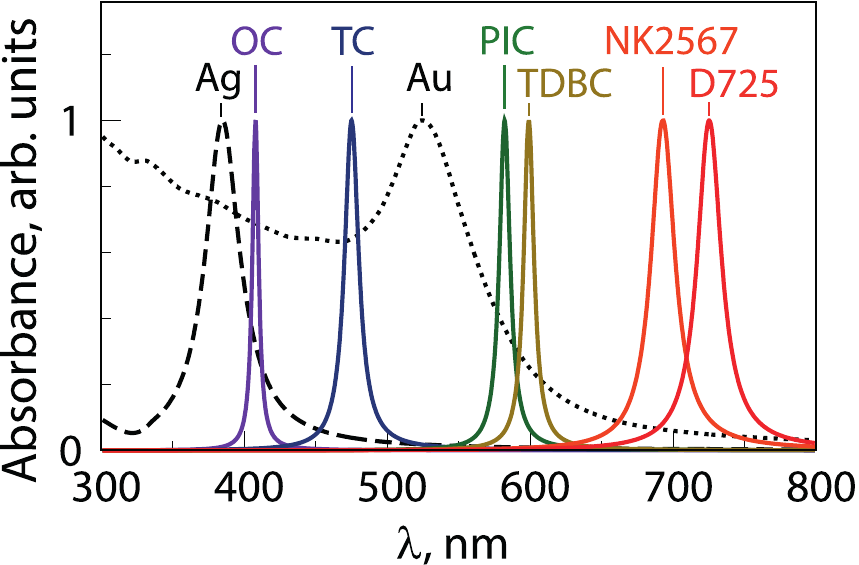}
	\caption{Normalized light absorption J-bands (solid curves) of aggregates of dyes: OC, TC, PIC, TDBC, NK2567 and D725, calculated by using the scalar Lorentzian model (\ref{eps-J}) with parameters taken from the papers~\cite{Kometani2001,Yoshida2010,Uwada2007,Bellessa2009,Fofang2008,Shapiro2015}. The dashed curves show the absorption spectra of silver and gold spherical nanoparticles with a radius of $r=10$ nm in an aqueous solution, calculated within the framework of the Mie theory.}
	\label{fig:j-bands}
\end{figure}

For a reliable analysis of the plexcitonic coupling effects in metalorganic nanostructures using classical electrodynamics it is necessary to correctly describe the dielectric properties not only of their plasmonic component, but of the excitonic one too. Within the framework of the commonly used isotropic model, the local dielectric function of the dye J-aggregate shell is described via the Lorentz oscillator model
\begin{equation}
\varepsilon_{\text{J}}(\omega)=\varepsilon^{\infty}_{\text{J}}+\frac{f \omega_{\text{ex}}^2}{\omega_{\text{ex}}^{2}-\omega ^{2}-i\omega \gamma_{\text{ex}}}.
\label{eps-J}
\end{equation}
\noindent Parameters of the Lorentz contour of the J-band of a dye aggregate are determined experimentally; $\omega_{\text{ex}}$ is the transition frequency associated with the center of the J-band; $\mathit{\gamma}_{\text{ex}}$ is its full width at half maximum (FWHM); $\varepsilon^{\infty}_{\text{J}}$ is the value of the permittivity at frequencies far from the absorption J-band. The dimensionless parameter $f$ in equation \eqref{eps-J} is the effective oscillator strength, which reflects the resonant contribution of the dye J-band to the dielectric properties of the excitonic shell. According to \cite{Wooten1972}, $f$ is proportional to the concentration of dye molecules. In a series of papers within the excitonic and plexcitonic research field, this quantity is also referred to as the reduced oscillator strength or strength of resonance. For a number of dyes the center positions and widths of the J-band included in formula \eqref{eps-J} are given in Table \ref{table-Jaggr}. Note that the values of the effective oscillator strength $f$ and the constant $\varepsilon^{\infty}_{\text{J}}$, reported by different authors for J-aggregates of the same dyes, sometimes differ greatly from each other.

Individual dye aggregates usually manifest optical anisotropy, and their polarizability is of a tensor nature. However, when a material consists of many aggregates with random orientation, its macroscopic permittivity no longer exhibits orientational effects. In this case, the isotropic model \eqref{eps-J} can be appropriate for describing the spectra of dye J-aggregates. To obtain more accurate results in a wide spectral range and to correctly reproduce the asymmetry of the left and right wings of the J-absorption band (see Fig. \ref{fig:J-aggr_form}), the imaginary part of the dielectric function of J-aggregates can be restored from experimental data on light extinction spectra, and its real part can be calculated using the Kramers-Kronig relation.

The coupling regime of the Frenkel exciton in the J-aggregate shell and the surface plasmon-polariton in the metal core of the Metal/J-aggregate nanoparticle essentially depends on the optical constants of its constituent materials as well as on the relationship between the incident light frequency and resonant frequencies of the studied system. Therefore, one of the most important physical parameter determining the coupling regime is the spectral detuning between the center of the absorption J-band of the excitonic shell, $\omega_{\text{ex}}$, and the peak maximum, $\omega_{\text{pl}}$, of the plasmon resonance of the metal core. To demonstrate the considerably different relative positions of excitonic and plasmonic peaks, Figure~\ref{fig:j-bands} displays the normalized absorption peaks for J-aggregates of several common dyes together with peaks of plasmon resonances of silver and gold nanospheres with a radius of $r=10$ nm. Such dye J-aggregates are listed in Table \ref{table-Jaggr}, complete with their names, abbreviations, maxima positions and full widths of spectral peaks. Note that for the dyes (e.g., PIC and TPP) whose J-aggregates have several peaks in the visible and near-IR ranges, Table \ref{table-Jaggr} shows parameters for the most intense of them.

\subsubsection{Tensor model of excitonic shell}\label{subsubsect:tensor}

\begin{figure}[ht]
	\centering\includegraphics[width=0.94\linewidth]{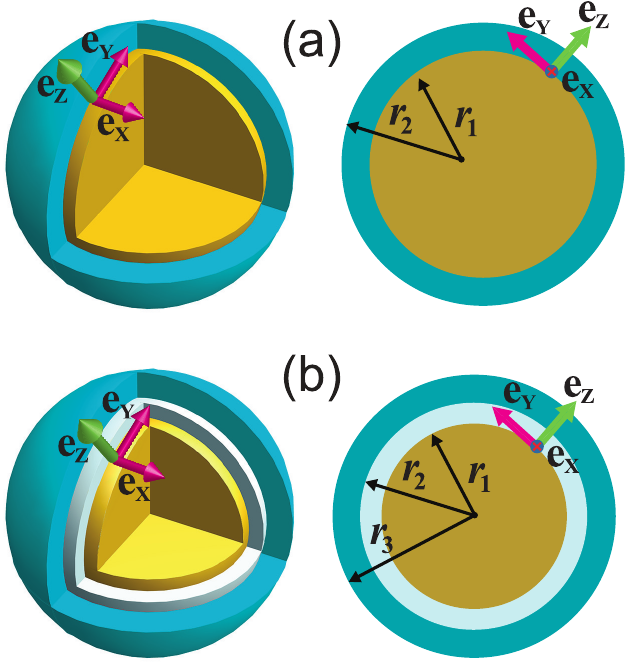}
	\caption{Schematic view of two-layer and three-layer nanospheres: (a) a particle with a metal core covered with a J-aggregate dye shell, $r_{1}$ is the core radius, $r_{2}$ is the outer shell radius; (b) a particle with a metal core, an intermediate passive dielectric spacer, and an outer J-aggregate shell. $r_{2}$ is the intermediate shell radius ($\ell_{\text{s}} = r_{2} - r_{1}$ is the spacer thickness), $r_{3}$ is the outer shell radius ($l_{\text{J}} = r_{3} - r_{2}$ is the J-aggregate shell thickness).}
	\label{fig:2-3_layer-particles}
\end{figure}

In many situations, it is fundamentally imperative to take into account anisotropic and orientational effects in molecular aggregates to explain and correctly describe the behavior of their optical spectra. As was recently shown in \cite{KML_OE2022, Kondorskiy2024}, the pronounced tensor nature of the dielectric function of dye aggregates also becomes crucial in exploring plexcitonic effects in hybrid metalorganic nanosystems. In other words, the incorporation of the anisotropic and orientational effects of the outer excitonic shell into the theoretical description of the optical properties of Metal/J-aggregate and Metal/Spacer/J-aggregate nanoparticles requires a significant modification of its local permittivity compared to expression \eqref{eps-J}. First, the dielectric function of the aggregate should be represented in a tensor form.  Secondly, the more complex nature of the J-band should be taken into consideration, in particular the presence of several resonant peaks with a specific polarization. If the molecular aggregates in the outer shell have favored direction of their orientation on the nanoparticle surface (see Fig. \ref{fig:2-3_layer-particles}), the components of the dielectric function tensor of the organic shell should be described as a following sum of Lorentzians \cite{KML_OE2022}
\begin{equation}
\varepsilon_{(\parallel,\perp)}\left( \omega \right) =\varepsilon^{\infty}_{(\parallel,\perp)} +\sum\limits_{n}\frac{f^{(\parallel,\perp)}_{n}(\omega^{(\parallel,\perp)}_{n})^{2}}{(\omega^{(\parallel, \perp)}_{n})^{2}-\omega^{2}-i\omega \gamma^{(\parallel, \perp)}_{n}}. \label{eps-J_tensor}
\end{equation}
\noindent Here the tensor components $\varepsilon_{\parallel}$ and $\varepsilon_{\perp}$ correspond to the directions parallel (\textit{longitudinal} component) and perpendicular (\textit{transverse} component) to the orientation axis of the aggregate. The parameters in (\ref{eps-J_tensor}) are chosen based on the measured absorption spectra of the J-aggregate, as described in \cite{KML_OE2022}. 

According to \cite{KML_OE2022}, the most adequate explanation to the available experimental data can be gleaned by considering the following possible arrangements of molecules in the J-aggregate shell: (i)~\textit{normal} orientation of aggregates, when their axes are directed along the normal to the nanoparticle surface (i.e. along the unit vector $\mathbf{e}_z$ in Fig. \ref{fig:2-3_layer-particles}); (ii)~\textit{tangential} orientation of aggregates, when their axes lie in the tangent plane to the nanoparticle (i.e. in the plane formed by the unit vectors $\mathbf{e}_x$ and $\mathbf{e}_y$) but do not have any particular direction in this plane; (iii)~\textit{equiprobable} orientation of the aggregate axis in space. In these three cases, the dielectric functions are described by tensors whose form is obtained by proper averaging over the possible directions of the J-aggregate axis. Provided that the tensor components are oriented along the $\mathbf{e}_{x}$, $\mathbf{e}_{y}$ and $\mathbf{e}_{z}$ axes (see Fig.~\ref{fig:2-3_layer-particles}), the corresponding expressions for the dielectric tensors take the form
\begin{equation}
\begin{split}
	&\widehat{\varepsilon }_{\text{norm}}=\left(
	\begin{array}{ccc}
		\varepsilon _{\bot } & 0 & 0 \\ 
		0 & \varepsilon _{\bot } & 0 \\ 
		0 & 0 & \varepsilon _{\Vert } \end{array}
	\right),
	\\
	&\widehat{\varepsilon }_{\text{tang}}=\left( 
	\begin{array}{ccc}
		\frac{1}{2}\varepsilon _{\Vert }+\frac{1}{2}\varepsilon _{\bot } & 0 & 0 \\ 
		0 & \frac{1}{2}\varepsilon _{\Vert }+\frac{1}{2}\varepsilon _{\bot } & 0 \\ 
		0 & 0 & \varepsilon _{\bot }
	\end{array}
	\right),
	\\
	&\widehat{\varepsilon }_{\text{epo}}=\left( 
	\begin{array}{ccc}
		\frac{1}{3}\varepsilon _{\Vert }+\frac{2}{3}\varepsilon _{\bot } & 0 & 0 \\ 
		0 & \frac{1}{3}\varepsilon _{\Vert }+\frac{2}{3}\varepsilon _{\bot } & 0 \\ 
		0 & 0 & \frac{1}{3}\varepsilon _{\Vert }+\frac{2}{3}\varepsilon _{\bot }
	\end{array}
	\right).
\end{split}
\label{eq:3tensors}
\end{equation} 
\noindent To compare our calculations \cite{KML_OE2022} with those obtained with conventional isotropic theoretical models, one more case should be considered. It corresponds to a fully (iv)~\textit{isotropic} J-aggregate shell, described by the scalar expression~(\ref{eps-J}) for a single excitonic peak or by its direct generalization
\begin{equation}
\varepsilon_\text{iso}\left( \omega \right) =\varepsilon^{\infty}_{\text{J}} +\sum\limits_{n}\frac{f_{n}\omega_{n}^{2}}{\omega_{n}^{2}-\omega^{2}-i\omega \gamma_{n}},
\label{eps-iso}
\end{equation}
\noindent which accounts for multiple absorption bands in a number of aggregates.

Importantly, the choice of scalar or tensor models corresponding to one or another orientation of the molecular aggregate in the outer shell is primarily determined by the physicochemical properties of the surface on which the aggregation occurs, i.e. metal or organic spacer layer. Most calculations and theoretical analysis of the optical properties of Metal/J-aggregate and Metal/Spacer/J-aggregate hybrid plexcitonic nanosystems were performed using the scalar isotropic model for the description of the permittivity of the outer dye shell. However, recent works \cite{Kondorskiy2024, KML_OE2022} clearly point to the importance of using the tensor models of aggregate permittivity to explain and quantitatively describe the available experimental data.

\section{Summary of theory of light absorption and scattering by hybrid nanostructures}

\subsection{Physical approaches to description of plexcitonic coupling}

Physical interpretation of the plexcitonic coupling effects in hybrid nanostructures that include metal and excitonic subsystems is usually given within the framework of classical or semiclassical approaches. In classical electrodynamics, these subsystems are treated using the dielectric functions of the corresponding materials, written, for example, in the Drude and Lorentz models. The theoretical basis for calculating optical spectra and explaining various phenomena caused by the plexcitonic coupling in organic or inorganic components of hybrid nanostructures is the solution to Maxwell's equations. A notable example here is the analytical solution of the problem of scattering and absorption of a plane electromagnetic wave by a multilayer concentric sphere (extended Mie theory). Additionally, numerical solutions obtained using the FDTD method or other computer simulation techniques are available for composite nanostructures of arbitrary shapes.

In composite nanosystems containing metals as the plasmonic subsystem and organic or inorganic materials (e.g., J-aggregates) as the excitonic subsystem, the classical approach provides a reliable self-consistent explanation for the majority of phenomena observed in experiments to date. These include, in particular, the phenomena of resonances and antiresonances during the interaction of an exciton with dipole and multipole plasmons, the effects of a deep dip in the light absorption spectra and the replication of the spectral bands of plexcitonic nanoparticles and their dimers. Although these phenomena are traditionally interpreted in terms of ''plasmons'' and ''excitons'', it is not formally required to describe quasiparticles within the framework of the classical electrodynamics. However, since the terms ''plasmon'' and ''exciton'' are well justified from a physical point of view, their use remains valid even within the purely classical approach. Recently, the term ''plexciton'' has come into use, referring to a hybrid quasiparticle.

In classical electrodynamics, the plasmon--exciton coupling phenomenon can be interpreted as follows. The electric and magnetic fields in each component of the plexcitonic system obey the known boundary conditions for Maxwell's equations. Consequently, the spatial field distributions in both the plasmonic and excitonic subsystems are determined by the geometrical structure and material permittivity of both of them. In this case, plexcitonic coupling arises due to the dependence of the field distribution in the plasmonic subsystem on that in the excitonic subsystem and vice versa. In other words, new hybrid normal modes of field oscillations arise in the system with eigen-frequencies distinct from those of the individual plasmonic and excitonic resonances. In the simplest case of a two-layer sphere of small radius, $r \ll \lambdabar$, the interaction between a dipole plasmon resonance with a frequency of $\omega^{(\text{dip})}_{\text{pl}}$ and an excitonic resonance with a frequency of $\omega_{\text{ex}}$ results in two new normal modes with $\omega_{+}$ and $\omega_{-}$ frequencies emerging in the system. The difference between $\omega_{+}$ and $\omega_{-}$ frequencies of these new normal modes determines the effective plasmon--exciton coupling constant, $g$. The value of $g$ drastically affects the optical spectra behavior of hybrid systems containing plasmons and excitons. This corresponds to qualitatively different regimes of near-field plexcitonic coupling: weak, strong and ultrastrong. The plexcitonic coupling has a near-field nature, because its efficiency decreases rapidly with the increasing distance, $\ell$, between the metal (plasmonic) and organic molecular or inorganic semiconductor (excitonic) components.

In the theory of plexcitonic coupling, the semiclassical approach is also quite common. In this approach, the excitonic subsystem is described as a quantum-mechanical two-level or multilevel system \cite{Meystre2007, Grynberg2010}, while the electromagnetic field and the plasmonic subsystem are treated in a conventional classical way. In the plexcitonic context, the essence of the semiclassical approach is most clearly described in the review article \cite{Barnes2015} using the example of the propagating surface plasmon-polariton interacting with a set of two-level quantum emitters. In the semiclassical approach, the polarization of the excitonic material is expressed through the mean value of the dipole moment operator of its constituent particles (quantum dots or organic molecules). As a result, it becomes possible to describe the dynamics of a quantum emitter interacting not only with fairly weak fields, but also with a strong local field. This allows one to include nonlinear optical effects in the theory \cite{Boyd2008}. Thus, with the semiclassical theory it is possible to go beyond the scope of linear electrodynamics and describe the dependence of the optical response of a plexcitonic system on the incident light intensity.

The authors of \cite{Barnes2015} also discussed the next step toward a quantal description of plexcitonic systems; there the electromagnetic field is treated on the basis of secondary quantization formalism using photon creation and annihilation operators. This approach, in which the field and emitters are quantized, remains essentially semiclassical, since the plasmonic subsystem is still considered at the classical level. Herewith the entire ensemble of quantum emitters constituting the excitonic subsystem is considered as a unified many-particle quantum system interacting with the quantized electromagnetic field. Within the framework of such a description of the exciton-polariton interaction, splitting of the hybrid modes occurs in the plexcitonic system even when photons are absent, in line with the vacuum Rabi splitting in quantum optics (for more details, see \cite{Barnes2015}).

Within the framework of a consistent quantum approach, it becomes necessary to quantize all three subsystems of the hybrid system (plasmonic, excitonic, and electromagnetic field). A self-consistent description of plexcitonic effects can be given using general methods developed in the fields of quantum statistical physics and physical kinetics of open systems. There are several fundamental approaches for solving many-body problems, including those based on quantum field theory methods \cite{Abrikosov1963} and on the generalizations of the Bogolyubov hierarchy (Bogolyubov-Born-Green-Kirkwood-Yvon hierarchy, BBGKY) \cite{BBGKY2014}. However, applying such general quantum approaches to describe plexcitonic phenomena requires solving extremely complicated systems of equations, which significantly limits their applicability for solving specific problems in the physics of plexcitonic systems. Some attempts to apply quantum methods to study the coupling effects between plasmonic and excitonic subsystems (such as molecules or quantum dots) have been made in \cite{Nordlander2011, White2012}. The theoretical consideration in these works was performed within the formalism of Zubarev's Green's functions \cite{Zubarev1960}, but some essential parameters of the system were chosen empirically rather than obtained through direct calculations. In \cite{Nordlander2011} the light absorption spectra were calculated for two different plexcitonic systems: (a) a quantum emitter near a metal nanoparticle; (b) a quantum emitter in the gap of a metal dimer. Similarly, in \cite{White2012} the photoabsorption spectra of a system comprising a quantum emitter and a metal dimer were calculated at various temperatures.

Note here that most of the available experimental data on the optics and spectroscopy of hybrid metalorganic nanostructures pertains to a range of problems for which the use of a purely classical electrodynamical approach is quite sufficient. Therefore, we briefly outline the Mie theory below, extended to the case of two-layer, three-layer and multilayer concentric spheres. We also list efficient numerical methods used in specific calculations of the optical absorption, scattering and extinction spectra in the Metal/J-aggregate and Metal/Spacer/J-aggregate systems. Simple formulas of the quasistatic approximation are also given, applicable for the explanation of the corresponding experiments for the ''core-shell'' spherical particles of small radius.

\subsection{Extended Mie theory for multilayer spherical particles}

For hybrid nanoparticles with a metal core whose size exceeds the Fermi wavelength of an electron in a metal $\lambdabar_{\text{F}}\sim 1$ nm, a reliable quantitative description of light absorption and scattering spectra can be given within the framework of classical electrodynamics of continuous media, when the materials of a particle are described by using the local dielectric functions. For the simplest spherical geometry, the exact solution to the problem of absorption and scattering of light is a generalization of the standard Mie theory for a homogeneous sphere to the case of multilayer concentric spheres. This approach was developed in a series of works \cite{Aden1951,Guttler1952,Ruppin1968,Irimajiri1979,Bhandari1985,Wu1991,Sinzig1994,Fuller1993}. First, the standard Mie theory was extended to the case of particles with one additional outer layer. A generalization to an arbitrary number of layers was performed in \cite{Bhandari1985} using the matrix formalism and in \cite{Wu1991, Sinzig1994} based on recurrence relations for the light scattering coefficients of a multilayer spherical particle.

Figure \ref{fig:2-3_layer-particles}a shows a schematic view of a hybrid particle consisting of a core with radius $r_1$ and a shell with thickness $l=r_2-r_1$, surrounded by a passive medium with permittivity, $\varepsilon_{\text{h}}\left(\omega\right)$, and magnetic permeability, $\mu_h=1$. The materials composing the concentric spherical layers are assumed to be homogeneous and isotropic with complex frequency-dependent dielectric functions $\varepsilon_1\left(\omega\right)$ and $\varepsilon_2\left(\omega\right)$ and magnetic permeabilities $\mu_1=\mu_2=1$. A linearly polarized plane monochromatic wave $\propto \exp\left(-i\omega t + ik_{\text{h}} z\right)$ is incident on the particle. The incident wave is partially scattered and absorbed by the particle.

The resulting exact expressions for the cross sections of absorption, $\sigma_{\text{abs}}$, scattering, $\sigma_{\text{scat}}$, and extinction, $\sigma_{\text{ext}}$, of light from a multilayer spherical particle, applicable for an arbitrary relationship between the wavelength and its overall radius, can be written as series expansions in multipoles 
\begin{equation}
	\sigma_{\text{abs}} =\frac{\pi }{2k_{\text{h}}^{2}}\sum_{n=1}^{\infty} \left( 2n+1 \right) 
	\left( 2- \left\vert 2a_{n}-1\right\vert^{2}-\left\vert 2b_{n}-1\right\vert ^{2}\right), 
	\label{sigma-abs-sum}
\end{equation}
\begin{equation}
	\sigma_{\text{scat}} =\frac{2\pi }{k_{\text{h}}^{2}}\sum_{n=1}^{\infty } \left( 2n+1\right) \left( \left\vert a_{n}\right\vert ^{2}+\left\vert b_{n}\right\vert ^{2}\right)
	\label{sigma-scat-sum}
\end{equation}
\begin{equation}
	\sigma_{\text{ext}} = \frac{2\pi }{k_{\text{h}}^{2}}\sum_{n=1}^{\infty}\left(2n+1\right)\text{Re} \left\{a_{n}+b_{n}\right\}. 
	\label{sigma-ext-sum}
\end{equation}
\noindent similar to the case of a homogeneous sphere \cite{Stratton1948}. Here $a_n$ and $b_n$ are the expansion coefficients of the transverse electric (TE) and transverse magnetic (TM) modes of the scattered wave, respectively; $n$ is the  multipole order; $k_{\text{h}}=\omega\sqrt{\varepsilon_{\text{h}}}/c$ is the modulus of the wave vector of light in the medium surrounding the particle. 

The expansion coefficients $a_n$ and $b_n$ in formulas \eqref{sigma-abs-sum}-- \eqref{sigma-ext-sum} are determined by the specifics of the problem and depend on the geometrical parameters of the hybrid particle and on the permittivities of its constituent materials. For bilayer ''core-shell'' nanoparticles, the general expressions for the complex coefficients $a_n$ and $b_n$ can be written as 
\begin{equation}
	a_{n}=-\frac{X_{n}^{(a)}}{Y_{n}^{(a)}},\qquad b_{n}=-\frac{X_{n}^{(b)}}{%
		Y_{n}^{(b)}},  \label{coeff-sol}
\end{equation}
where the functions $X_{n}^{(a)}$, $Y_{n}^{(a)}$ and $X_{n}^{(b)}$, $Y_{n}^{(b)}$ can be expressed
in compact form through determinants \cite{Leb-Medv2012}. For the contributions of the TE modes of order $n$, expressions for $X_{n}^{(a)}$ and $Y_{n}^{(a)}$ take the form
\begin{equation}
	X_{n}^{(a)}=\left|%
	\begin{array}{cccc}
		j_{n}\left(k_{1}r_{1}\right) & j_{n}\left(k_{2}r_{1}\right) &
		y_{n}\left(k_{2}r_{1}\right) & 0 \\
		u_{n}^{\prime}\left(k_{1}r_{1}\right) & u_{n}^{\prime}\left(k_{2}r_{1}\right)
		& v_{n}^{\prime}\left(k_{2}r_{1}\right) & 0 \\
		0 & j_{n}\left(k_{2}r_{2}\right) & y_{n}\left(k_{2}r_{2}\right) &
		j_{n}\left(k_{\text{h}}r_{2}\right) \\
		0 & u_{n}^{\prime}\left(k_{2}r_{2}\right) & v_{n}^{\prime}\left(k_{2}r_{2}%
		\right) & u_{n}^{\prime}\left(k_{\text{h}}r_{2}\right)%
	\end{array}%
	\right|,  \label{Xna}
\end{equation}
\begin{equation}
	Y_{n}^{(a)}=\left|%
	\begin{array}{cccc}
		j_{n}\left(k_{1}r_{1}\right) & j_{n}\left(k_{2}r_{1}\right) &
		y_{n}\left(k_{2}r_{1}\right) & 0 \\
		u_{n}^{\prime}\left(k_{1}r_{1}\right) & u_{n}^{\prime}\left(k_{2}r_{1}\right)
		& v_{n}^{\prime}\left(k_{2}r_{1}\right) & 0 \\
		0 & j_{n}\left(k_{2}r_{2}\right) & y_{n}\left(k_{2}r_{2}\right) &
		h_{n}^{\left(1\right)}\left(k_{\text{h}}r_{2}\right) \\
		0 & u_{n}^{\prime}\left(k_{2}r_{2}\right) & v_{n}^{\prime}\left(k_{2}r_{2}%
		\right) & w_{n}^{\prime}\left(k_{\text{h}}r_{2}\right)%
	\end{array}%
	\right|.  \label{Yna}
\end{equation}
Here, $k_{1}=\omega\sqrt{\varepsilon_{1}}/c$ and $k_{2}=\omega\sqrt{\varepsilon_{2}}/c$ are the magnitudes of the wave vectors of the light in the core and shell, respectively; $\varepsilon_{1}\equiv
\varepsilon_{\text{m}}\left(\omega\right)$ and $\varepsilon_{2} \equiv
\varepsilon_{\text{J}}\left(\omega\right)$ are the complex permittivities of the metal core and organic shell at the frequency of the incident light, $\omega$; $r_{1}$ is the core radius; $r_{2}$  is the outer radius of the particle; $j_n(z) $,
$y_n(z)$, and $h_n^{\left(1\right)}\left( z\right)$ are spherical Bessel, Neumann and Hankel functions; $u_{n}\left(z\right)=z
j_{n}\left(z\right)$, $v_{n}\left(z\right)=zy_{n}\left(z\right)$, and $w_{n}\left(z\right)=z h_{n}^{\left(1\right)}\left(z\right)$ are spherical Riccati--Bessel, Riccati--Neumann and Riccati--Hankel functions; and primes denote differentiation of a function with respect to its argument.

Similarly, the final expressions for the functions $X_{n}^{(b)}$ and $Y_{n}^{(b)}$, which determine the contributions of the TM modes, can be represented as in \cite{Leb-Medv2012}
\begin{equation}
\resizebox{.9\hsize}{!}{$
	X_{n}^{(b)}= \left|%
	\begin{array}{cccc}
		j_{n}\left(k_{1}r_{1}\right) & \sqrt{\frac{\varepsilon_{2}}{\varepsilon_{1}}}%
		\;j_{n}\left(k_{2}r_{1}\right) & \sqrt{\frac{\varepsilon_{2}}{\varepsilon_{1}%
		}}\;y_{n}\left(k_{2}r_{1}\right) & 0 \\
		u_{n}^{\prime}\left(k_{1}r_{1}\right) & \sqrt{\frac{\varepsilon_{1}}{%
				\varepsilon_{2}}}\;u_{n}^{\prime}\left(k_{2}r_{1}\right) & \sqrt{\frac{%
				\varepsilon_{1}}{\varepsilon_{2}}}\;v_{n}^{\prime}\left(k_{2}r_{1}\right) & 0
		\\
		0 & \sqrt{\frac{\varepsilon_{2}}{\varepsilon_{\text{h}}}}\;j_{n}\left(k_{2}r_{2}%
		\right) & \sqrt{\frac{\varepsilon_{2}}{\varepsilon_{\text{h}}}}\;y_{n}%
		\left(k_{2}r_{2}\right) & j_{n}\left(k_{\text{h}}r_{2}\right) \\
		0 & \sqrt{\frac{\varepsilon_{\text{h}}}{\varepsilon_{2}}}\;u_{n}^{\prime}%
		\left(k_{2}r_{2}\right) & \sqrt{\frac{\varepsilon_{\text{h}}}{\varepsilon_{2}}}%
		\;v_{n}^{\prime}\left(k_{2}r_{2}\right) & u_{n}^{\prime}\left(k_{\text{h}}r_{2}%
		\right)%
	\end{array}%
	\right|,  
$}
\label{Xnb}
\end{equation}
\begin{equation}
	\resizebox{.9\hsize}{!}{$
	Y_{n}^{(b)}=
	\left|%
	\begin{array}{cccc}
		j_{n}\left(k_{1}r_{1}\right) & \sqrt{\frac{\varepsilon_{2}}{\varepsilon_{1}}}%
		\;j_{n}\left(k_{2}r_{1}\right) & \sqrt{\frac{\varepsilon_{2}}{\varepsilon_{1}%
		}}\;y_{n}\left(k_{2}r_{1}\right) & 0 \\
		u_{n}^{\prime}\left(k_{1}r_{1}\right) & \sqrt{\frac{\varepsilon_{1}}{%
				\varepsilon_{2}}}\;u_{n}^{\prime}\left(k_{2}r_{1}\right) & \sqrt{\frac{%
				\varepsilon_{1}}{\varepsilon_{2}}}\;v_{n}^{\prime}\left(k_{2}r_{1}\right) & 0
		\\
		0 & \sqrt{\frac{\varepsilon_{2}}{\varepsilon_{\text{h}}}}\;j_{n}\left(k_{2}r_{2}%
		\right) & \sqrt{\frac{\varepsilon_{2}}{\varepsilon_{\text{h}}}}\;y_{n}%
		\left(k_{2}r_{2}\right) & h_{n}^{\left(1\right)}\left(k_{\text{h}}r_{2}\right) \\
		0 & \sqrt{\frac{\varepsilon_{\text{h}}}{\varepsilon_{2}}}\;u_{n}^{\prime}%
		\left(k_{2}r_{2}\right) & \sqrt{\frac{\varepsilon_{\text{h}}}{\varepsilon_{2}}}%
		\;v_{n}^{\prime}\left(k_{2}r_{2}\right) & w_{n}^{\prime}\left(k_{\text{h}}r_{2}%
		\right)%
	\end{array}%
	\right|.
$}
\label{Ynb}
\end{equation}
Combined with (\ref{sigma-abs-sum})--(\ref{sigma-ext-sum}), equations (\ref{coeff-sol})--(\ref{Ynb}) allow one to calculate not only the total absorption, scattering and extinction cross sections but also the contributions of individual terms
in the multipole series that correspond to TM and TE modes of various orders. In practical calculations, the maximum number of terms of the series that must be taken into account in formulas (\ref{sigma-abs-sum})--(\ref{sigma-ext-sum}) can be estimated using the relation \cite{Barber1990}
\begin{equation}
n_{\max }=\left( kr\right) +4.05(kr)^{1/3}+2,  
\label{n_max}
\end{equation}
\noindent where $r=r_2$ is the outer radius of the concentric spheres.

For three-layer (see Fig. \ref{fig:2-3_layer-particles}) and, particularly, for multilayer particles, the expressions for the expansion coefficients $a_{n}$ and $b_{n}$ turn out to be quite cumbersome. Therefore, it is convenient to present them not in the form of determinants, but to use recurrence relations for their calculation. This approach has been utilized in \cite{Wu1991, Sinzig1994, Leb-Medv2013b}.

\subsection{Formulas of quasistatic approximation}

For a particle with a radius much smaller than the light wavelength, one can use the quasistatic approximation and limit ourselves to the contribution of the electric dipole term ($n=1$). Then the extinction and scattering cross sections takes the form \cite{Bohren1998}
\begin{equation}
\sigma _{\text{ext}}\left( \omega \right) = 4\pi k_{\text{h}}\;\text{Im}\left\{\alpha \left( \omega \right) \right\} ,
\quad \sigma _{\text{scat}}\left( \omega \right) =\frac{8\pi }{3}
k_{\text{h}}^{4}\left\vert \alpha \left( \omega \right)
\,\right\vert ^{2}.
\label{sigma-abs-static}
\end{equation}
\noindent Here $k_{\text{h}}$ is the modulus of the wave vector of incident radiation in the passive environment with the dielectric constant of $\varepsilon_{\text{h}}$; $\alpha$ is the particle polarizability. The simplest form is the dipole polarizability of a bare metal particle with a radius $r$ and permittivity $\varepsilon \equiv \varepsilon_{\text{m}}$ \cite{Bohren1998}:
\begin{equation}
\alpha =r^3 \frac{\varepsilon_{\text{m}}-\varepsilon_{\text{h}}}{\varepsilon_{\text{m}}+2\varepsilon_{\text{h}}}\,.
\label{alpha-metal}
\end{equation}
According to \eqref{alpha-metal} the polarizability of a sphere is proportional to its volume, $\alpha \propto r^3$, therefore, the value of the scattering cross section for spheres of small radius in the quasistatic approximation ($k r \ll 1$) turns out to be much smaller than the absorption cross section: $\sigma_{\text{scat}} \ll \sigma_{\text{abs}}$. In such a situation the extinction cross section, $\sigma_{\text{ext}} = \sigma_{\text{abs}} + \sigma_{\text{scat}}$, in equation \eqref{sigma-abs-static} is almost equal to the absorption cross section, $\sigma_{\text{ext}} \approx \sigma_{\text{abs}}$ (see \cite{Bohren1998}). For a two-layer particle, instead of $\varepsilon_{\text{m}}$ in (\ref{alpha-metal}), one should use the effective permittivity, ${\varepsilon}_{2}^{\text{eff}}$ , of the ''core-shell'' system, which is equivalent to the permittivity of a homogeneous sphere and is calculated as below \cite{Irimajiri1979}
\begin{equation}
{\varepsilon}_{2}^{\text{eff}} =\varepsilon_{2}
\frac{2\left(1-\left(r_{1}/r_{2}\right)^{3}\right)+
\left(1+2\left(r_{1}/r_{2}\right)^{3}\right)\left(\varepsilon_{1}/
\varepsilon_{2}\right)} {\left(2+\left(r_{1}/r_{2}\right)^{3}\right)+
\left(1-\left(r_{1}/r_{2}\right)^{3}\right)
\left(\varepsilon_{1}/\varepsilon_{2}\right)}.  
\label{twolayer}
\end{equation}
Here $r_1$ and $r_2$ are the inner and outer radii of the concentric spheres; $\varepsilon_1=\varepsilon_{\text{m}}$ and $\varepsilon_2=\varepsilon_{\text{J}}$ are the dielectric functions of the metal core and J-aggregate shell, respectively. Thus, the dipole polarizability $\alpha$ of a two-layer particle with volume $\mathcal{V}=4\pi r_2^3/3$ is represented as \cite{Guttler1952}
\begin{equation}
\alpha =\frac{
	\left(\varepsilon_{1}\!-\!\varepsilon _{2}\right) 
	\left(2\varepsilon _{2}+\varepsilon _{\text{h}}\right) r_{1}^{3}+
	\left(\varepsilon _{2}\!-\!\varepsilon_{\text{h}}\right) 
	\left(2\varepsilon_{2}+\varepsilon_{1}\right) r_{2}^{3}
}{
	2\left( \varepsilon _{1}\!-\!\varepsilon_{2}\right) \left(\varepsilon _{2}\!-\!\varepsilon _{\text{h}}\right) \left(r_{1}/r_{2}\right) ^{3}+\left( \varepsilon _{2}+2\varepsilon_{\text{h}}\right) \left( 2\varepsilon _{2}+\varepsilon _{1}\right) 
}.
\label{alpha}
\end{equation}
\noindent Together with (\ref{sigma-abs-static}), this expression makes it possible to estimate the electric dipole contribution to the cross sections of extinction and  scattering of light by a hybrid two-layer particle.

Whenever the radius of a three-layer particle is much smaller than the light wavelength, the extinction, $\sigma _{\text{ext}}$, and scattering, $\sigma _{\text{scat}}$, cross sections can be written as (\ref{sigma-abs-static}) and are expressed through its dipole polarizability $\alpha$ and a wave number $k_{\text{h}}$ of the incident radiation in the environment with dielectric constant $\varepsilon_{ \text{h}}$. According to \cite{Scaife1998} the effective dipole polarizability of a three-layer particle, $\alpha \left( \omega \right)$, has the form
\begin{equation}
{\alpha }=r_3^3 \,\frac{A\left( \varepsilon
_{3}-\varepsilon_{\text{h}}\right) -B\left( 2\varepsilon _{3}+\varepsilon_{\text{h}}\right) \left( r_{2}/r_{3}\right) ^{3}}{A\left( 2\varepsilon
_{3}+\varepsilon_{\text{h}}\right) -B\left( \varepsilon _{3}-\varepsilon_{\text{h}}\right) \left( r_{2}/r_{3}\right) ^{3}}\,.
\label{alpha-3rd}
\end{equation}
\noindent The coefficients $A$ and $B$ in (\ref{alpha-3rd}) are calculated using the formulas
\begin{equation}
A =\left( 2\varepsilon _{3}+\varepsilon _{2}\right) \left( 2\varepsilon
_{2}+\varepsilon _{1}\right) +2\left( \varepsilon _{3}-\varepsilon _{2}\right) \left( \varepsilon
_{2}-\varepsilon _{1}\right) \left( r_{1}/r_{2}\right) ^{3},
\label{alpha-3rd-A-1} 
\end{equation}
\begin{equation}
B =\left( \varepsilon _{3}-\varepsilon _{2}\right) \left( 2\varepsilon
_{2}+\varepsilon _{1}\right)  
 +\left( \varepsilon _{3}+2\varepsilon _{2}\right) \left( \varepsilon
_{2}-\varepsilon _{1}\right) \left( r_{1}/r_{2}\right) ^{3},
\label{alpha-3rd-B-1}
\end{equation}
\noindent and $\mathcal{V}={4\pi r_3^3}/3$ is the total volume of a three-layer particle.

\subsection{Simulation of optical properties of non-spherical hybrid nanoparticles}

As already noted in Section \ref{section:plasmon-multipoles}, unlike spherical particles, spheroidal particles have a distinct direction along their symmetry axis. Therefore, contrary to a metal sphere, light absorption and scattering spectra of a homogeneous metal spheroid are split into two peaks of plasmon resonances -- longitudinal and transverse. The problem of scattering and absorption of electromagnetic waves by spheroidal particles is divided into two cases: prolate spheroid $a>b$ (formed by rotating an ellipse around its long axis $a$) and oblate spheroid $a <b$ (formed by rotating the ellipse around minor axis $b$). For a spheroid consisting of a metal core and a J-aggregate shell, general formulas for the cross sections for light scattering and absorption turn out to be very cumbersome for both prolate and oblate spheroids \cite{Farafonov1993}. However, when the size of the composite particle is sufficiently small compared to the radiation wavelength ($a, b \ll \lambdabar$), extinction cross sections can be calculated using simple analytical formulas of the quasistatic approximation for a two-layer spheroid \cite{Wang1982} and ellipsoid \cite{Ambjornsson2006}. 

For hybrid nanostructures with shapes more complex than spherical, spheroidal, cubic or cylindrical, solutions in the form of analytical or special functions are generally not available. Light absorption and scattering spectra of both single-layer and multilayer metalorganic nanostructures can be calculated based on a variety of numerical methods for solving Maxwell's equations for electromagnetic fields inside and outside nanostructures. The most widely used methods are: 1) the Finite-Difference Time-Domain (FDTD) method \cite{Taflove2005}; 2) $T$-matrix method \cite{Waterman1971, Mishchenko2002}; 3) Multiple Multipole method (MMP) \cite{Hafner1990, Moreno2002}; 4) Discrete Dipole Approximation (DDA) \cite{Yurkin2007}; 5) Volume Integral Equations Method (VIEM) \cite{Lee-Mal1995}. Depending on the type of problem being solved, one or another method has an advantage. In contrast to the quasistatic approximation, approaches based on the use of these numerical methods make it possible to determine the cross section of absorption and scattering of light for an arbitrary relationship between the nanoparticle size and the light wavelength, taking into account the contribution of all electric and magnetic multipoles.

\section{\label{Sect5}Analytical models of plasmon--exciton coupling}

\subsection{\label{Sect5-analytical-hybrid-freqs}Simple formulas for frequencies of hybrid plexcitonic modes}

The coupling between a surface plasmon in the metal core and a Frenkel exciton in the dye shell leads to the formation of hybrid plexcitonic states of the two-layer particle. The eigen-frequencies of hybrid modes of two-layer Metal/J-aggregate nanosphere and, hence, the spectral positions of corresponding peaks in its absorption cross section can be determined approximately using an analytical approach \cite{Leb-Medv2012}. It is a generalization of a simple means for finding the spectral position of the dipole plasmon resonance peak in a homogeneous metal sphere. In this approach the eigen-frequencies of hybrid modes can be found in the quasistatic approximation using the relation, $\text{Re}\left\{ \varepsilon_2^{\text{eff}}(\omega)\right\} =-2\varepsilon _{\text{h}}$, containing the effective permittivity $\varepsilon _{2}^{\text{eff}}$ of a two-layer sphere (\ref{twolayer}). For the permittivity of the J-aggregate shell, the conventional isotropic model (\ref{eps-J}) is used. The consideration \cite{Leb-Medv2012} is valid in the frequency range, where the dominant contribution to the  permittivity of the metal core is determined by free electrons: $\varepsilon_{\text{m}} = \varepsilon_{\text{intra}}$. Then, neglecting the imaginary parts of dielectric functions \eqref{eps-intra} and \eqref{eps-J} (i.e., putting $\gamma_{\text{intra}}= 0$ and $\gamma_{\text{ex}}= 0$), we  obtain the cubic equation
\begin{equation}
C_3 x^3 + C_2 x^2 + C_1 x + C_0 =0, \quad x=\omega^2   
\label{cubic-eq}
\end{equation}
\noindent for the square of the eigen-frequencies of hybrid particle modes. The $C_{j}$ coefficients ($j=0,\,1,\,2,\,3$) are determined by the parameters of the dielectric functions of the plasmonic and excitonic subsystems of the hybrid particle and by the ratio $\zeta =\left( r_{1} /r_{2}\right)^{3}$ of concentric sphere volumes with radii $r_{1}$ and $r_{2}$. Their specific form is presented in \cite{Leb-Medv2012}.

Using substitution of $y=x+C_{2}/3C_{3}$, equation \eqref{cubic-eq} can be rewritten in the canonical form
\begin{multline}
	y^{3}+py+q=0, \\
	p = -a^{2}/3+b, \quad
	q=2\left( a/3\right) ^{3}-ab/3+c,
	\label{eq:cubic-canon}
\end{multline}
where $a=C_{2}/C_{3}$, $b=C_{1}/C_{3}$ and $c=C_{0}/C_{3}$. Since all the coefficients $C_i$ and $p$, $q$ are real numbers, the number of real roots of equation \eqref{cubic-eq} depends on the sign of $Q=\left( p/3\right)^{3}+\left( q/2\right)^{2}$. It follows that $Q$ is negative in the cases under consideration and, hence, $p < 0$. Therefore, within the framework of the isotropic model \eqref{eps-J} equations \eqref{cubic-eq} and \eqref{eq:cubic-canon} have three different real roots except for the limiting cases $r_1 =0$ and $r_1 = r_2$. It is convenient to represent these roots ($x_{i}\equiv \omega_{i}^{2}$) in the trigonometric form
\begin{equation}
\begin{split}
	&\omega_{1}^{2} =-\frac{a}{3}+2\sqrt{-\frac{p}{3}}\;\cos \left( \frac{
		\beta }{3}\right), \\
	&\omega_{2,3}^{2} =-\frac{a}{3}-2\sqrt{-\frac{p}{3}}\;\cos \left( \frac{
		\beta }{3}\pm \frac{\pi }{3}\right) ,  
\end{split}
\label{omega_123}
\end{equation}
here the $\beta$ parameter is defined by the equation
\begin{equation}
	\cos \beta =-\frac{q}{2\sqrt{-\left( p/3\right) ^{3}}}. \label{abc}
\end{equation}
The eigen-frequencies of hybrid modes depend on the plasma frequency, $\omega_{\text{p}}=(4\pi n_{\text{e}} e^2/m_{\text{e}})^{1/2}$, from the Drude-like formula (\ref{eps-intra}); on the exciton frequency $\omega_{\text{ex}}$ and on the effective oscillator strength $f$ of the dye J-band \eqref{eps-J}; as well as on the ratio $\zeta =\left(r_{1}/r_{2}\right)^{3}$. We recall here that the eigen-frequency, $\omega^{(\text{dip})}_{\text{pl}}$, of electric dipole surface plasmon-polariton is expressed through plasma frequency, $\omega_{\text{p}}$, by using the Fr\"{o}hlich formula \eqref{eq:Frohlich}. Formulas \eqref{omega_123} and \eqref{abc} determine the positions of three maxima in the absorption spectrum of a Metal/J-aggregate particle. They are valid in the quasistatic approximation for the case of the outer shell being isotropic and can be used when interband transitions in the metal core (for example, Ag) do not make a noticeable contribution to the dielectric function $\varepsilon_{\text{m}}\left( \omega,r\right)$.

Below we present the explicit expressions for the coefficients $a$, $b$ and $c$ in formulas (\ref{eq:cubic-canon}) -- (\ref{abc}) in dependence on the frequencies $\omega_{\text{p}}$ and $\omega_{\text{ex}}$, the effective oscillator strength $f$, and on the $\zeta$ ratio in the particular case when $\varepsilon_{\text{h}}=\varepsilon^{\infty}_{\text{J}}=\varepsilon^{\infty }_{\text{m}}=1$:
\begin{equation}
\resizebox{.95\hsize}{!}{$
a =-\frac{1}{3}\left[ 3\left( f+2\right) \omega _{\text{ex}}^{2}+\omega _{\text{p}
}^{2}\right], \quad c =-\frac{\omega _{\text{ex}}^{4}\omega _{\text{p}}^{2}}{9}\left[ 3+f\left(
1+2\zeta \right) \right]
$}
\label{ac}
\end{equation}
\begin{equation}
\resizebox{.95\hsize}{!}{$
	b =\frac{\omega _{\text{ex}}^{2}}{9}\left\{\omega_{\text{ex}}^{2}\left[ 2f^{2}\left(
	1-\zeta \right) +9\left( f+1\right) \right] \right.
	+\left. \omega _{\text{p}}^{2}\left[ f\left( 1+2\zeta \right) +6\right]
	\right\}. 
$}
\label{b}
\end{equation} 
In general, coefficients $a$, $b$ and $c$ and the frequencies of the hybrid modes depend, among other factors, on the permittivity $\varepsilon_{\text{h}}$ of the host medium and on the constants of $\varepsilon^{\infty}_{\text{m}}$ and $\varepsilon^{\infty}_{\text{J}}$.

\begin{figure}[b]
	\centering\includegraphics[width=0.94\linewidth]{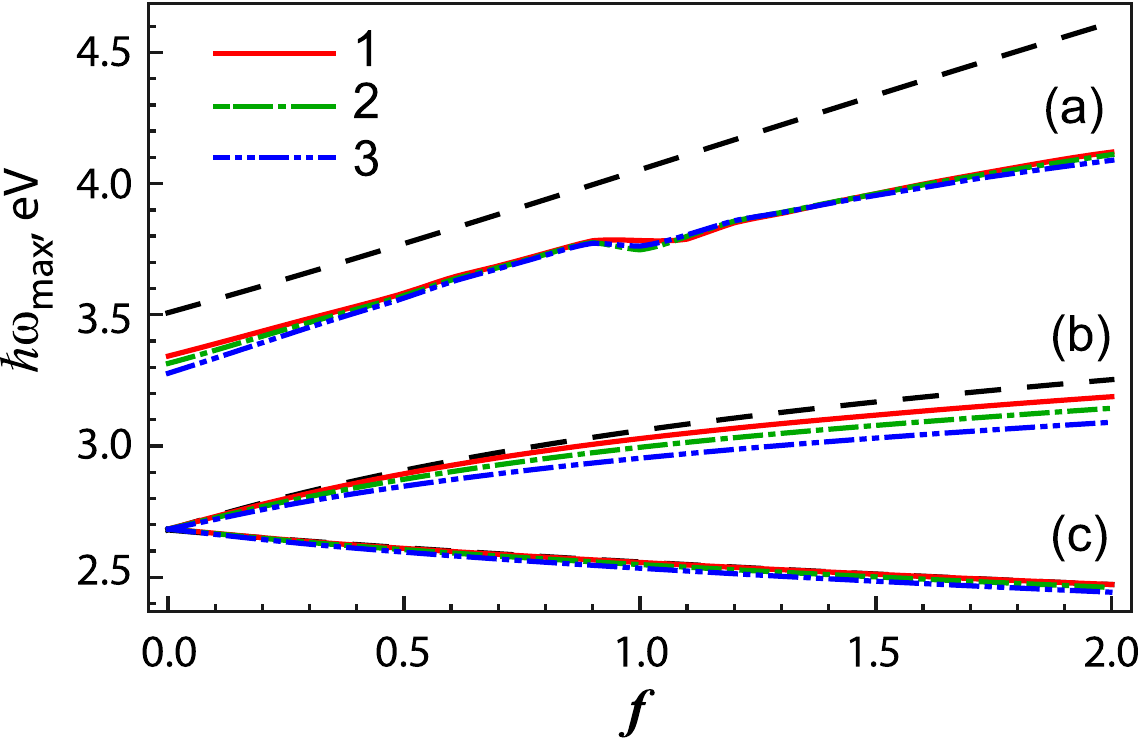}
	\caption{Spectral positions of the maxima, $\hbar \omega_{\max}$, of (a) ''high-energy'', (b) ''medium-energy'' and (c) ''low-energy'' absorption peaks for Ag/J-aggregate particles versus the effective oscillator strength, $f$, of an excitonic transition \cite{Leb-Medv2012}. Calculations have been performed for $r_{1}=10$ nm, $r_{2}=12$ nm (solid red curve 1); $r_{1}=15$ nm, $r_{2}=18$ nm (green dash-dotted curve 2); $r_{1}=20$ nm, $r_{2}=24$ nm (blue dash-double-dotted curve 3) with a fixed ratio, $r_{1}/r_{2}=5/6$, of the core radius to an overall radius of the particle. Black dashed curves are the analytical results (\ref{abc}) obtained for the same value of $r_{1}/r_{2}$.}
	\label{f-position(Ag/TC)}
\end{figure}

A comparison of the results of the model above with exact calculations within the framework of the extended Mie theory has shown their good agreement in the range of applicability of the quasistatic approximation and justified the use of the isotropic model for the J-aggregate shell permittivity of a plexcitonic particle. We demonstrate this in Fig. \ref{f-position(Ag/TC)} which shows the results \cite{Leb-Medv2012} for the spectral positions of the maxima of ''high-energy'' (a), ''medium-energy'' (b), and ''low-energy'' (c) peaks of light absorption by spherical Ag/J-aggregate particles depending on the effective oscillator strength $f$ of the dye J-band. The corresponding numerical calculations were performed for several values of the inner, $r_1$, and outer, $r_2$, particle radii: $r_{1}=10$ nm, $r_{2}=12$ nm (curves 1); $r_{1}=15$ nm, $r_{2}=18 $ nm (curves 2); and $r_{1}=20$ nm, $r_{2}=24$ nm (curves 3). The values of $r_1$ and $r_2$ are chosen so that their ratio remains the same: $r_1/r_2= 5/6$. Dashed curves in Fig. \ref{f-position(Ag/TC)} show the results of calculations of the eigen-frequencies of hybrid modes of the particle, performed within the framework of the analytical model \cite{Leb-Medv2012}. All optical constants related to the J-aggregate shell (except for the effective oscillator strength, $f$) are taken here to be equal to the corresponding values for the TC dye (see Table \ref{table-Jaggr}).

Quantitative discrepancies between the results of the analytical model and the numerical calculation are present only for the high-energy peak appearing near the energies of electron transitions between the valence d-band and the conduction sp-band of silver ($\hbar \omega_{\text{g}}=3.7$ eV). In this spectral range, the dielectric function of silver cannot be reliably described by the Drude formula alone. Therefore, to correctly describe the behavior of this hybrid mode for photon energies close to the boundary of interband transitions, the analytical approach \cite{Leb-Medv2012} should be supplemented with a correct evaluation of the contribution to the dielectric function $\varepsilon_{\text{m}}(\omega,r)$ of the interband transitions of electrons in the silver core. In addition, it is necessary to correctly account for the spectral width ($\gamma_{\text{ex}}$) of the shell J-band and damping coefficients of free ($\gamma_{\text{intra}}$) and bound ($\gamma_{\text{inter}}$) electrons. Note that since the core radius in calculations shown in Fig. \ref{f-position(Ag/TC)} does not exceed 20 nm, no higher-order multipole resonances ($n>1$) make any noticeable contribution to the absorption cross section. Therefore, hybrid modes of the composite system arise here only as a result of the interaction of the Frenkel exciton of the J-aggregate shell with a dipole plasmon in the core. 

Figure \ref{f-position(Ag/TC)} clearly demonstrates that the positions of the maxima in the light absorption spectra strongly depend on the effective oscillator strength in the dye J-band \cite{Leb-Medv2012}. An analysis of the dependencies of the frequencies of hybrid modes of metalorganic nanoparticles on the values of $f$ and $\varepsilon^{\infty }_{\text{J}}$ was also performed in \cite{Antosiewicz2014, Thomas2018} and recently in \cite{Kondorskiy2024} in relation to the isotropic and anisotropic excitonic shell.

\subsection{\label{Sect-Coupl-Ocs-Model}Classical model of coupled oscillators}

\subsubsection{General formulas}

The effects of plexcitonic coupling are often treated with the model of coupled damped oscillators. It describes a system of dipoles in an external monochromatic field coupled through local electromagnetic fields. Such models have been extensively used in various applications of the vibration theory. Starting with \cite{Nitzan1981,Gersten1981a,Gersten1981b}, they have been applied to analyze the effects of plasmon interaction in metal nanostructures with various organic and inorganic systems, including excitons in quantum dots and in molecular dye J-aggregates. According to \cite{Shah2013, Barnes2015, Thomas2018} the system of differential equations for two coupled dipoles can be written as
\begin{gather}
{\ddot {p}}_{\text{pl}}+\gamma _{\text{pl}}{\dot {p}}_{\text{pl}
}+\omega _{\text{pl}}^{2}p_{\text{pl}} = A_{\text{pl}}\left[{E}_{0}\cos
\left( \omega t\right) +Mp_{\text{ex}}\right], \label{Eq-1} \\ 
 {\ddot {p}}_{\text{ex}}+\gamma _{\text{ex}}{\dot {p}}_{\text{ex}
}+\omega _{\text{ex}}^{2}p_{\text{ex}} = A_{\text{ex}}\left[{E}_0\cos
\left( \omega t\right) +Mp_{\text{pl}}\right].  
\label{Eq-2}
\end{gather}
\noindent Here $p_{\text{pl}}$ and $p_{\text{ex}}$ are the dipole moments of the plasmonic and excitonic subsystems, ${E}_0$ and $\omega$ are the amplitude and frequency of the external electric field, respectively. The parameters $A_{\text{pl}}$ and $A_{\text{ex}}$ are related to the classical oscillator strengths, $f_{\text{pl}}$ and $f_{\text{ex}}$, and $M$ is the dipole coupling constant \cite{Shah2013}. The system of equations (\ref{Eq-1}) and (\ref{Eq-2}) allows for an exact solution (see, for example, \cite{Thomas2018}):
\begin{eqnarray}
p_{\text{pl}}(t) = \text{Re}\{p_{\text{pl}}^{(0)}e^{-i\omega t}\},\; \; 
p_{\text{ex}}(t) = \text{Re}\{p_{\text{ex}}^{(0)}e^{-i\omega t}\}, \label{Eq-3}
\end{eqnarray}

\begin{equation}
\begin{split}
	&p_{\text{pl}}^{(0)} = \frac{{E}_{0}A_{\text{pl}}\left[1+\left(A_{\text{ex}}M\right)/B_{\text{ex}}(\omega )\right] }{B_{\text{pl}}(\omega )\left[ 1-\left(A_{\text{pl}}A_{\text{ex}}M^{2}\right)/B_{\text{pl}}(\omega )B_{\text{ex}}(\omega)\right] 
	}, \\
	&p_{\text{ex}}^{(0)} = 
	\frac{{E}_{0}A_{\text{ex}}\left[ 1+\left(A_{\text{pl}}M\right)/B_{\text{pl}}(\omega)\right] }{B_{\text{ex}}(\omega )\left[ 1-\left(A_{\text{pl}}A_{\text{ex}}M^{2}\right)/\left(B_{\text{pl}}(\omega )B_{\text{ex}}(\omega)\right)\right] },
\end{split}
\label{Eq-4_Eq-5}
\end{equation}
\noindent where $B_{j}(\omega)=\omega_{j}^2-\omega^2-i\gamma_{j}\omega$, $j=\text{pl},\text{ex}$. In this case, the cross sections of extinction and scattering  of light by the plexcitonic system have the form \cite{Shah2013}:
\begin{equation}
\begin{split}
	&\sigma_{\text{ext}}=\frac{4\pi n_{\text{h}} \omega }{c}
	\text{Im}\left\{ 
	\frac{p_{\text{pl}}^{(0)}+p_{\text{ex}}^{(0)}}{{E}_{0}}\right\},  \\
	&\sigma_{\text{scat}}=\frac{8\pi }{3}\left( \frac{n_{\text{h}}\omega }{c}\right)
	^{4}\left\vert \frac{p_{\text{pl}}^{(0)}+p_{\text{ex}}^{(0)}}{{E}_{0}}
	\right\vert^{2},  
\end{split}
\label{Eq-6_Eq-7}
\end{equation}
where $n_{\text{h}}$ is refractive index of the host medium.

The model of coupled dipoles has seven parameters: $\omega_{j}$, $\gamma_{j}$, $A_{j}$ ($j$ = pl, ex) and $M$ included in the system of equations (\ref {Eq-1}), (\ref{Eq-2}) and in the expressions (\ref{Eq-6_Eq-7}) for light extinction and scattering cross sections. The classical expression for the coupling energy $V$ of oscillators, similar to the expression for the coupling energy $V = \hbar g$ of two states in the quantum model, can be easily obtained in the resonance case: $\omega _{\text{pl}}=\omega_{\text{ex}}\equiv \omega_{0}$. Here classical and quantum considerations equally reproduce optical spectra with two symmetrical resonant peaks, and their energy splitting corresponds to double coupling energy of the oscillators. Within the framework of the classical model, these resonant frequencies are associated with the zeros of the real parts of the denominators in formulas (\ref{Eq-4_Eq-5}). Assuming that the damping constants, $\gamma_{i}$, are small compared to $\omega$ and $\omega_{0}$, we can find two zeros separated by energy by the value: 
\begin{equation}
	V = ({\hbar M}/{2\omega _{0}})\sqrt{A_{\text{pl}}A_{\text{ex}}}.
	\label{eq:V_through_A}
\end{equation}
Formula \eqref{eq:V_through_A} relates the effective plexcitonic coupling energy $V$ to the coupling constant $M$, appearing in the system of equations \eqref{Eq-1} and \eqref{Eq-2} for coupled dipoles.

For a single excitonic oscillator (a single molecule, a molecular aggregate or a quantum dot) interacting with a plasmonic oscillator, coefficients $A_{\text{pl}}$ and $A_{\text{ex}}$ can be represented in a standard form \cite{ Thomas2018} through the classical oscillator strengths, $f^{\text{(cl)}}_{\text{pl}}$ and $f^{\text{(cl)}}_{\text{ex}}$, that determine their polarizabilities:
\begin{align*} 
	&A_{\text{pl}}=(e^{2}/m_{e})f^{\text{(cl)}}_{\text{pl}},\\
	&A_{\text{ex}}=(e^{2}/m_{e})f^{\text{(cl)}}_{\text{ex}}.
\end{align*} 
Thus, formula \eqref{eq:V_through_A} can be written as  \begin{equation*}
	V=\frac{\hbar M}{2\omega _{0}}\frac{{e^{2}}}{m_{e}}\sqrt{f^{\text{(cl)}}_{\text{pl}}f^{\text{(cl)}}_{\text{ex}}}. 
\end{equation*} 
When considering $N_{\text{ex}}$ excitonic oscillators interacting with a plasmon, coefficient $A_{\text{ex}}$ can be written as follows \cite{Thomas2018}:
\begin{equation*}
	A_{\text{ex}}= N_{\text{ex}}(e^{2}/m_{e})f^{\text{(cl)}}_{\text{ex}}.
\end{equation*} 
So, the expression for the plexcitonic coupling energy of dipoles is given by:
\begin{equation}
V = \frac{\hbar M}{2\omega _{0}}\frac{e^{2}}{m_{e}}\sqrt{N_{\text{ex}}}\sqrt{f_{\text{pl}}f_{\text{ex}}}, 
\label{V_CL-3}
\end{equation}
Expression \eqref{V_CL-3} demonstrates a general important feature of proportionality of the plexcitonic coupling energy to the square root of a number of excitonic oscillators, $V \propto \sqrt{N_{\text{ex}}}$, which can be monomer molecules, molecular aggregates or quantum dots. In a review article \cite{Barnes2015} on electromagnetic coupling of surface plasmons with quantum emitters, this dependence was demonstrated in several other ways. It should be noted that under the usual experimental conditions the density of excitons in ordered molecular organic materials is very low \cite{Barnes2015, Davydov1971}. For the plexcitonic systems consisting of a metal core with an outer J-aggregate shell of the dye, this means that the average number of aggregates in the excited state, $N^{\ast}_{\text{ex}}$, is significantly less than their total number, $N_{\text{ex}}$, which is included in formula \eqref{V_CL-3}: $N^{\ast}_{\text{ex}} \ll N_{\text{ex }}$. The average number of excitons depends on several factors, including the type of plexcitonic system under consideration and the power density of the radiation incident on this system. A detailed discussion of this topic is beyond the scope of this article.

\subsubsection{Resonance and antiresonance}\label{Sect-resonance-antiresonance}

\begin{figure}[ht]
	\centering\includegraphics[width=0.94\linewidth]{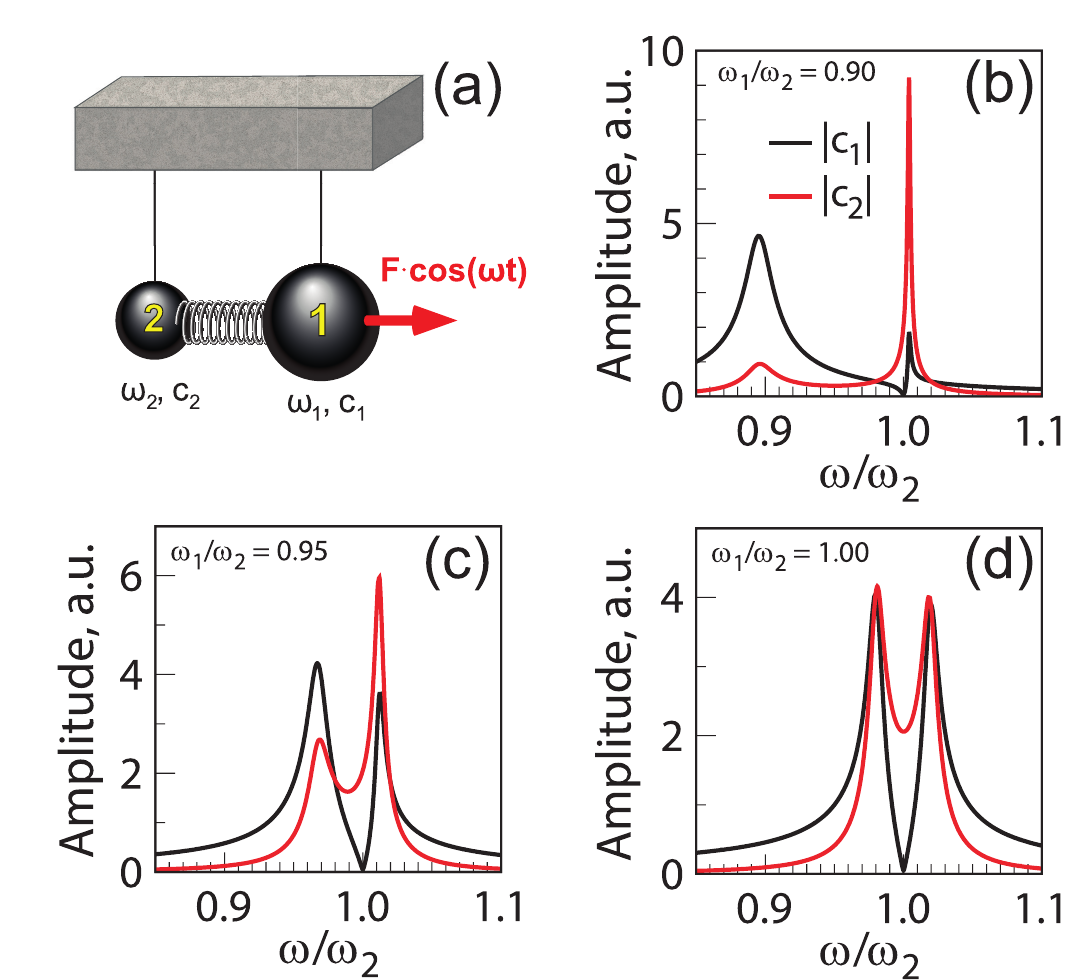}
	\caption{(a) Schematic view of coupled damped oscillators ''1'' (driven by periodic force) and ''2'' with eigen-frequencies of $\omega_{1}$ and $\omega_2$, respectively. (b)--(d) Amplitudes $|c_1|$ and $|c_2|$ of the oscillators ''1'' and ''2'', respectively, as functions of a dimensionless frequency, $\omega/\omega_2$ (see formulas \eqref{c1} and \eqref{c2}). Calculations were performed for three values of the frequency ratio: $\omega_{1}/\omega_{2} = 0.90$ (b), $0.95$ (c), $1.00$ (d); factor $\left|F\right|/\left(m\omega_{2}\right)$ that determines the oscillation amplitude is equal to $0.1$ cm$\cdot$s$^{-1}$. Other parameters of the model are as follows: $\gamma_1/\omega_{2} = 2 \cdot 10^{-2}$, $\gamma_{2}/\omega_{2} = 1 \cdot 10^{-3}$, $\Omega/\omega_{2} = 0.2$.}
	\label{Antiresonance}
\end{figure}

The phenomenon of resonance, well-known in various areas of physics, consists of an enhanced response of a system to an external excitation at a frequency close to the eigen-frequency of the system. In some cases, the opposite phenomenon of antiresonance may occur, when the system's reaction is suppressed under certain conditions. The phenomena of resonance and antiresonance can be illustrated using a model of coupled oscillators with eigen-frequencies $\omega_{1}$ and $\omega_{2}$. Not infrequently, only one of them can be considered to be driven by a periodic force (see Fig. \ref{Antiresonance}a). This situation is realized, for example, in the metalorganic nanoparticles of the ''core-shell'' type studied here (see Sect. \ref{Sect6} below). For such particles, the interaction of light with their plasmonic subsystem turns out to be dominant, since the volume $\mathcal{V}_{\text{pl}}$ of the plasmonic core substantially exceeds the volume $\mathcal{V}_{\text{ex}}$ of the external excitonic shell. Therefore, the polarizability of the metal core, $\alpha_{\text{pl}}$, is significantly higher than that of the outer excitonic shell of the particle, $\alpha_{\text{pl}} \gg \alpha_{\text{ex}}$. Below, for the sake of generality, we will use index $1$ for the oscillator, driven by a periodic force $F \equiv F_{1}$ with a frequency $\omega$, without specifying whether this oscillator is plasmonic or excitonic. The system is described by coupled differential equations
\begin{eqnarray}
\overset{..}{x}_{1}+\gamma _{1}\overset{.}{x}_{1}+\omega
_{1}^{2}x_{1}-\Omega^{2} x_{2} &=& \frac{F}{m} \exp \left( i\omega t\right) ,
\label{syst_coupl-eq1} \\
\overset{..}{x}_{2}+\gamma _{2}\overset{.}{x}_{2}+\omega
_{2}^{2}x_{2}-\Omega^{2} x_{1} &=& 0,  
\label{syst_coupl-eq2}
\end{eqnarray}
\noindent where $\gamma_{1}$ and $\gamma_{2}$ are the damping constants of the first and second oscillators, $\Omega$ is their coupling constant, and $m \equiv m_{1}$ is the mass of the force-driven oscillator.

The free motion of oscillators ($F=0$) in the absence of coupling ($\Omega=0$) represents their oscillations with given eigen-frequencies $\omega_1$ and $\omega_2$. Solving equation (\ref{syst_coupl-eq2}) for two coupled oscillators in the form of forced oscillations, $x_1=c_{1}\exp {(i\omega t)}$ and $x_2=c_{2}\exp {(i\omega t)}$, we get the following expressions for the oscillation amplitudes \cite{Joe2006}:
\begin{align}
c_{1}\left( \omega \right) &=\frac{\left( \omega _{2}^{2}-\omega
^{2}+i\gamma _{2}\omega \right) \left(F/m\right)}{\left( \omega _{1}^{2}-\omega
^{2}+i\gamma _{1}\omega \right) \left( \omega _{2}^{2}-\omega ^{2}+i\gamma
_{2}\omega \right) -\Omega^{4}},\ \   \label{c1} \\
c_{2}\left( \omega \right) &= -\frac{\Omega^{2} \left(F/m\right)}{\left( \omega
_{1}^{2}-\omega ^{2}+i\gamma _{1}\omega \right) \left( \omega
_{2}^{2}-\omega ^{2}+i\gamma _{2}\omega \right) -\Omega^{4}}.  
\label{c2}
\end{align}
The phase difference between two coupled oscillators is given by
\begin{equation}
\varphi_{2}-\varphi_{1}=\pi -\arctan \left(\frac{\gamma_{2}\omega}{\omega_{2}^{2}-\omega ^{2}}\right) .  
\label{phase-diff}
\end{equation}

Figure \ref{Antiresonance} shows the results of our calculations performed using formulas \eqref{c1} and \eqref{c2} for amplitudes $|c_{1}|$ and $|c_{2}|$ of two coupled oscillators at different eigen-frequencies ratios, $\omega_{1}/\omega_{2}$. Cases of large, intermediate and zero frequency detuning are illustrated in Figs. \ref{Antiresonance}b, \ref{Antiresonance}c and \ref{Antiresonance}d, respectively. It follows that in such a system there are two resonances at frequencies $\omega_{-}$ and $\omega_{+}$, which, neglecting damping ($\gamma_{1} = 0$, $\gamma_{2} = 0 $), are expressed through eigen-frequencies of the oscillators $\omega_{1}$ and $\omega_{2}$ (see, e.g., \cite{Thomas2018}):
\begin{equation}
\omega^{2}_{\pm} = \frac{\omega_{1}^{2}+\omega_{2}^{2}}{2} \pm \sqrt{\left(\frac{\omega_{1}^{2}-\omega_{2}^{2}}{2}\right)^2 + \Omega^{4}}.
\label{eq:omega_pm}
\end{equation}

The behavior of the spectral profile of the first oscillator significantly depends on the detuning of its eigen-frequency $\omega_{1}$ from the eigen-frequency of the second oscillator $\omega_{2}$. In the case of large detuning (b), the first oscillator exhibits resonances with symmetrical and asymmetrical profiles near frequencies $\omega_1$ and $\omega_2$, respectively. The first resonance in the spectrum of oscillator ''1'' excited by an external force is described by a symmetrical Lorentz-type contour and demonstrates a standard increase in the oscillation amplitude near the eigen-frequency $\omega_{1}$. The second resonance is characterized by an asymmetrical profile (see Fig. \ref{Antiresonance}b). At the eigen-frequency of the second coupled oscillator $\omega_{2}$, the amplitude of the first oscillator vanishes, suggesting complete suppression of the amplitude of the driven oscillator ''1'' at the eigen-frequency of the coupled oscillator ''2''. This is an antiresonance phenomenon caused by the destructive interference of oscillations resulting from the action of a periodic external force and interaction with the second oscillator. Antiresonance can be observed in coupled mechanical, acoustic, electromagnetic and quantum-mechanical systems \cite{Joe2006, Miroshnichenko2010}. 

As the frequency detuning between $\omega_{1}$ and $\omega_{2}$ decreases, the difference in amplitudes $|c_{1}|$ and $|c_{2}|$ at the peak maxima becomes less pronounced (see Fig. \ref{Antiresonance}c). The shape of the curve describing two resonant peaks and the dip between them, corresponding to antiresonance, also becomes somewhat more symmetrical. If the eigen-frequencies of the oscillators coincide, $\omega_{1}=\omega_{2}$, the shape of the dip becomes almost symmetrical (see Fig. \ref{Antiresonance}d), and the antiresonance manifests itself in the splitting of the spectral peak of oscillator ''1''. Note that in all three cases, both resonant peaks of oscillator ''2'' are symmetrical (red curves in Fig. \ref{Antiresonance}b--\ref{Antiresonance}d), although its amplitude falls to a minimum between the peaks.

Based on the model of coupled oscillators, a reasonable theoretical interpretation can be offered to a number of experimentally observed phenomena in the optics of plexcitonic nanosystems. Moreover, it is possible to provide a simple explanation for some limiting cases of the model that differ from each other by a qualitatively diverse behavior of the light absorption and scattering spectra. 

\subsubsection{Effect of induced transparency}\label{Sect-Induced-Transparency}

One of the most discussed phenomena in the optics of plexcitonic nanosystems is a pronounced dip in the light absorption, scattering and extinction spectra, observed in many experimental works. This phenomenon occurs when the frequencies of the plasmon and exciton coincide exactly (or are sufficiently close to each other), $\omega_{0} \equiv \omega_{\text{pl}} = \omega_{\text{ex}}$, and it leads to a sharp increase in the light transmission at the resonant frequency of $\omega_{0}$. As a result, this effect manifests itself as an intense peak in the transmission spectra. Correspondingly, it is often referred to as an induced transparency (IT) effect \cite{DeLacy2015, Antosiewicz2014, Krivenkov2019}.

A theoretical explanation for this phenomenon can be given within the framework of the classical coupled oscillators model outlined above. To this end, we need to derive a general expression for the averaged power, $\overline{P}\left(\omega\right)$, absorbed by the plexcitonic system over the period of the incident light wave.
Using expressions \eqref{c1} and \eqref{c2} for the oscillation amplitudes of $c_{1}$ and $c_{2}$ we arrive at the following result:
\begin{equation}
\resizebox{1.\hsize}{!}{$
\begin{split}
& \overline{P}\left( \omega \right) = \frac{\omega}{2} \operatorname{Im}\left\{ c_{1}\left(\omega\right) F^{\ast}\right\} = \frac{\omega^2 \left\vert
F\right\vert ^{2}}{2m} \\&\times 
\frac{\gamma_{2}\left(\Omega^{4}+\gamma _{1}\gamma_{2}\omega^{2}\right) + \gamma_{1} \left(\omega^{2}-\omega_{2}^{2}\right)^{2}
}{
\left[ \left(\omega^{2}-\omega _{1}^{2}\right) \left(\omega^{2}-\omega_{2}^{2}\right)-\Omega^{4}\right]^{2}+\omega^{2}\left[\gamma_{1}^{2} \left(\omega^{2}-\omega_{2}^{2}\right)^{2} + \gamma_{2}^{2} \left(\omega^{2}-\omega_{1}^{2}\right)^{2} + 2 \gamma_{1}\gamma_{2}\Omega^{4}\right] + \gamma_{1}^{2}\gamma	_{2}^{2} \omega^{4}
}.
\end{split}
$}
\label{Power-averaged-result}
\end{equation}
This expression is applicable to arbitrary frequency detuning of oscillators $\omega_{1}$ and $\omega_{2}$. It takes into account the fact that in a simplified model of coupled oscillators, the driving force $F$ acts on one of them only (see Fig. \ref{Antiresonance}a). In order for this effect to manifest itself most clearly, the decay rate of one of the oscillators should be much less than that of the other, $\gamma _{2}\ll \gamma _{1}$, since there should be no significant increase in energy dissipation as a result of adding a second oscillator. This means that the energy dissipation occurs predominantly in the plasmonic subsystem ''1'', rather than in the excitonic subsystem ''2''. Correspondingly, the width of the plasmon resonance, $\gamma_{\text{pl}} \equiv \gamma_{1}$, is significantly larger than that of the excitonic band, $\gamma_{\text{ex}} \equiv \gamma_{2}$. This situation is realized in the core-shell plexcitonic systems containing a metal nanoparticle as a core and a dye J-aggregate as a shell. It is these hybrid metalorganic nanoparticles that have been most extensively studied both experimentally and theoretically. In particular, a series of detailed numerical calculations has been performed using the FDTD method in order to give a self-consistent explanation of this effect.

\begin{figure}[t]
	\centering\includegraphics[width=0.94\linewidth]{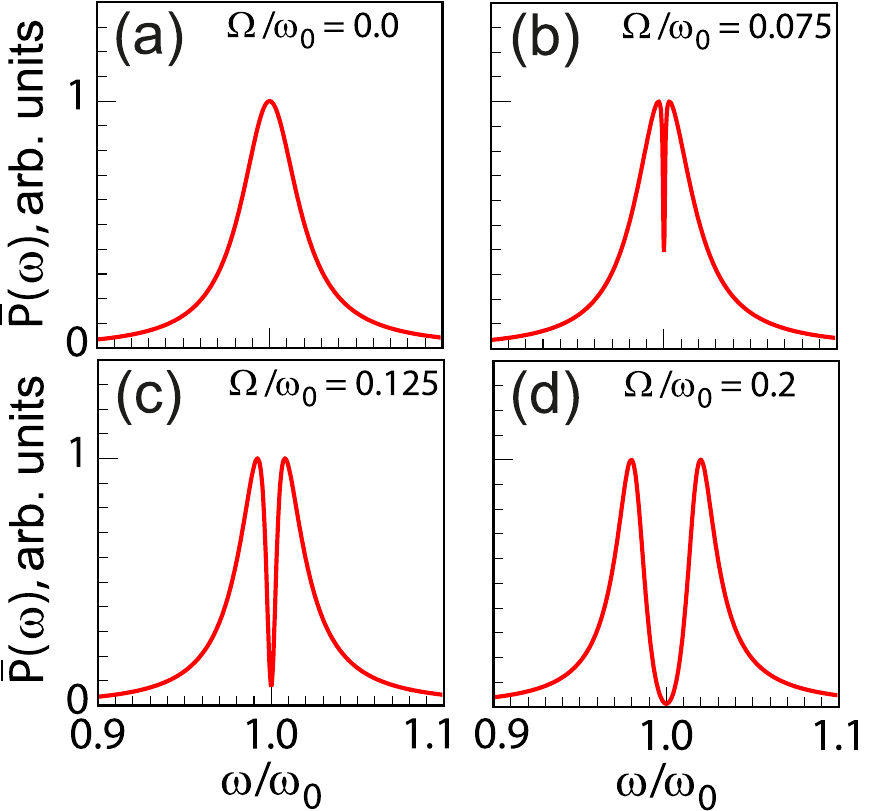}
	\caption{ Averaged power $\overline{P}(\omega)$, absorbed by a system of two coupled oscillators (see Fig. \ref{Antiresonance}a) over the oscillations period of driving force, $T=2\pi/\omega$, as a function of the normalized frequency $\omega/\omega_{0}$ at four different values of the dimensionless coupling parameter: (a) $\Omega/\omega_0=0$; (b) $\Omega/\omega_0=0.075$; (c) $\Omega/\omega_0=0.125$; (d) $\Omega/\omega_0=0.2$. Eigen-frequencies of ''1'' and ''2''  oscillators  coincide, $\hbar\omega_{1} = \hbar\omega_{2} = \hbar\omega_{0} = 2.0$ eV; oscillator damping constants are $\hbar\gamma_{1} = 8 \cdot 10^{-2}$ eV, $\hbar\gamma_{2} = 1 \cdot 10^{-3}$ eV; the efficiency of the driving force is $\left|F\right|^2/\left(m\omega^{2}_{0}\right) = 0.5$ eV.}
	\label{Fig_Power-vs-Detuning}
\end{figure}

Here we offer a simple explanation to the induced transparency effect by applying analytical formula \eqref{Power-averaged-result} obtained within the framework of the coupled oscillator model. It takes a particularly simple form when the resonant frequencies of the coupled plasmonic and excitonic subsystems coincide, $\omega_{0} \equiv \omega_{1}=\omega_{2}$:
\begin{equation}
	\resizebox{1.\hsize}{!}{$
\begin{split}
\overline{P}\left( \omega \right)  &= \frac{\omega^2 \left\vert F\right\vert ^{2}}{2m} \\&\times \frac{\gamma_{2}\left(\Omega^{4}+\gamma _{1}\gamma_{2}\omega^{2}\right) + \gamma_{1} \left(\omega^{2}-\omega_{0}^{2}\right)^{2}
}{
\left[ \left(\omega^{2}-\omega _{0}^{2}\right)^2-\Omega^{4}\right]^{2}+\omega^{2}\left[\left(\gamma_{1}^{2} + \gamma_{2}^{2} \right) \left(\omega^{2}-\omega_{0}^{2}\right)^{2} + 2 \gamma_{1}\gamma_{2}\Omega^{4}\right] +\gamma_{1}^{2}\gamma	_{2}^{2} \omega^{4}
}.
\end{split}
	$}
\label{Power-averaged-equal-eigenfreqs}
\end{equation}
The results of calculations of the average power absorbed by the plexcitonic system, performed by using formula \eqref{Power-averaged-equal-eigenfreqs}, are presented in Fig. \ref{Fig_Power-vs-Detuning}. The figure clearly demonstrates the phenomenon of induced transparency, i.e., it shows a pronounced dip in the light absorption spectrum at the resonant frequency. As $\Omega$ coupling constant increases, the dip deepens and the distance between the positions of the spectral maxima increases.

Importantly, the induced transparency (IT) effect in plexcitonic nanoparticles, i.e. a deep dip in their light absorption spectrum, has a different physical nature than the electromagnetically induced transparency (EIT), which has been studied extensively for many years due to its applications in the science of quantum information and nonlinear optics \cite{Fleischhauer2005, Peng2014}. This effect was discovered when exploring the laser radiation interaction with atomic gases. It consists in the elimination of absorption and refraction at the transition frequency as a result of quantum interference between the system's photoexcitation pathways. According to \cite{Fleischhauer2005}, the cause of the modified optical response of the medium here is the coherence of atomic states induced by a laser field. The EIT effect is usually described in terms of quantum dynamics of a three-level atom interacting with two external laser fields. One of them (probe) propagates through the medium at a frequency close to that of the transition between the atomic ground and excited states, and the other (control) field has its frequency in resonance with the transition between the two excited states of the atom. Electromagnetic coupling in the system results from the interaction of the control field with the transition dipole moment between the two excited states of the atom. A number of optical analogues of the electromagnetically induced transparency are considered in the literature including systems of two directly coupled microtoroidal silica resonators \cite{Peng2014}. A discussion of these topics is given, for example, in \cite{Fleischhauer2005, Peng2014, Krivenkov2019, Kivshar2017}.

\subsubsection{Fano formula and its limiting cases}

As shown above in Sect. \ref{Sect-Induced-Transparency}, the coupled oscillator model describes successfully the resonance and antiresonance phenomena, characterized by symmetrical and asymmetrical band profiles of a plexcitonic system, respectively, as well as a pronounced dip in the absorption spectrum, associated with the effect of induced transparency. In a series of well-known works (see \cite{Kivshar2017, Miroshnichenko2010, Fano-Springer} and references wherein) these phenomena were interpreted in terms of quantum-mechanical Fano model. Below we discuss whether it is justified to use this kind of analogy. When comparing results obtained in the optics of plexcitonic nanostructures with the corresponding results of the Fano theory and its special cases, note that this theory was originally developed in atomic physics for the description of electron scattering on an atom in the presence of an autoionizing state. The result for the dimensionless quantity of $\sigma(\mathcal{E})$ describing the shape of the resonance cross section has the form \cite{Fano1961}
\begin{equation} 
\sigma(\mathcal{E})= \frac{(q+\epsilon)^2}{1+\epsilon^2}, \quad \epsilon = \frac{2(\mathcal{E}-\mathcal{E}_{\text{r}})}{\mathit{\Gamma}_{\text{r}}}.
\label{Fano-eq}
\end{equation}
\noindent Here $\mathcal{E}$ is the energy, $\mathcal{E}_{\text{r}}$ and $\mathit{\Gamma}_{\text{r}}$ are the energy and resonance width (FWHM) in the continuous spectrum band in the Fano model. The cross section profile in this model is primarily determined by the Fano asymmetry parameter, $q$.

\begin{figure}[b]
	\centering\includegraphics[width=0.94\linewidth]{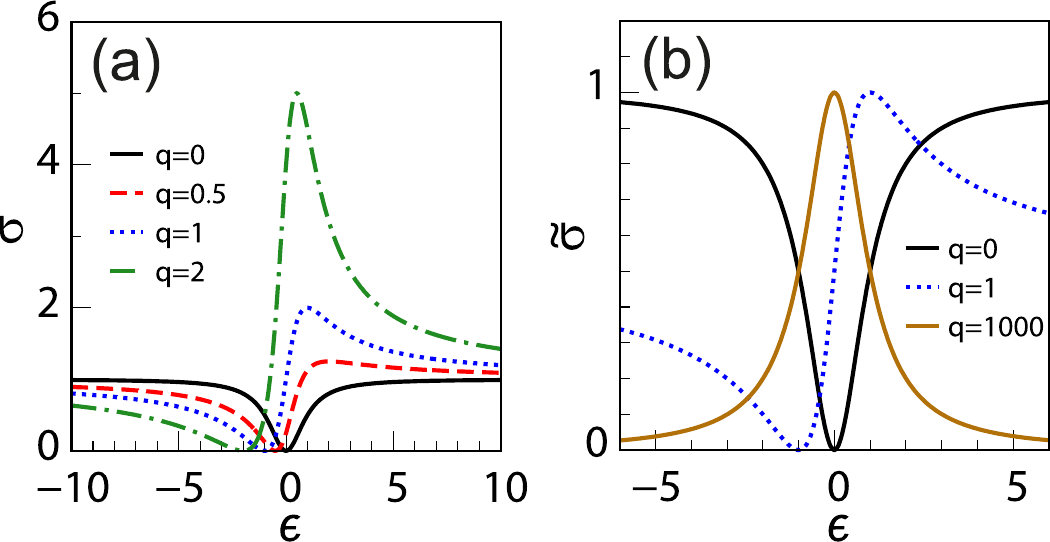}
	\caption{(a) Fano cross section profile \eqref{Fano-eq},  $\sigma(\epsilon)=\left(q+\epsilon\right)^2/\left(1+\epsilon^2\right)$, as a function of the dimensionless energy $\epsilon$ for different values of the $q$ parameter, which determines the asymmetry of the curve shape $\sigma(\epsilon)$. (b) Normalized Fano profile, $\tilde{\sigma}(\epsilon) = \sigma(\epsilon)/(1 + q^2)$. At $q=0$ there is no asymmetry of the cross section profile. The Fano resonance has a maximum at $\epsilon=1/q$ and a minimum at $\epsilon=-q$. The case of $q\rightarrow \infty$ corresponds to the Breit-Wigner resonance when the shape of the $\sigma(\epsilon)$ curve becomes symmetrical and is described by a Lorentz contour.}
	\label{Fano_resonance}
\end{figure}

Now we discuss the limiting cases of the Fano formula (see Fig. \ref{Fano_resonance}) for the profile of the cross section $\sigma(\mathcal{E})$. For the discussion of the spectral behavior of the cross section for large values of the parameter $q$, it is also convenient to introduce the normalized value of the cross section, $\Tilde{\sigma}(\mathcal{E})$, by formula $\Tilde{\sigma}(\mathcal{E}) = \sigma(\mathcal{E})/\left(1 + q^2\right)$. It is evident that the degree of contour asymmetry changes with the varying asymmetry parameter of $q$. When $q=0$, formula \eqref{Fano-eq} describes a symmetrical dip in the light absorption cross section, sometimes called a quasi-Lorentzian profile (see Fig. \ref{Fano_resonance}a and \ref{Fano_resonance}b). A typical asymmetrical spectral contour appeared at $q>0$ as seen in Fig. \ref{Fano_resonance}a. This is the so-called Fano profile calculated for three values of $q=0.5$, $1$, and $2$ (red, blue and green curves, respectively). For $\left|q\right|\gg 1$ the asymmetry of the spectral contour manifests itself only in the tail of the spectral wing far from the resonant energy, $\mathcal{E}=\mathcal{E}_{\text{r}}$ ($\epsilon=0$). Meanwhile, in the vicinity of resonance the contour becomes almost symmetrical, see Fig. \ref{Fano_resonance}b at $q=1000$. Thus, for $\left|q\right| \rightarrow \infty$ Fano formula (\ref{Fano-eq}) transforms into the well-known formula \cite{LL-QM} of the Breit-Wigner theory for resonant scattering,
\begin{equation*}
	\sigma(\mathcal{E}) \propto \frac{\mathit{\Gamma}_{\text{r}}/2}{\left(\mathcal{E} - \mathcal{E}_{\text{r}}\right)^2 + \mathit{\Gamma}_{\text{r}}^{2 }/4},
\end{equation*}
with a Lorentz contour, which is symmetrical relative to the resonant energy of $\mathcal{E}_{\text{r}}$ and has a full width at half maximum of $\mathit{\Gamma}_{\text{r}}$.

Although available experimental and theoretical results on the optics of hybrid plexcitonic systems are often discussed in terms of the Fano model, its applicability for describing the behavior of spectra in each individual case requires, strictly speaking, a special analysis. The reason is that the original Fano theory for autoionization deals with a purely discrete level against the continuum background \cite{Fano1961}. In the optics of plexcitonic nanostructures, an accurate description of similar interference effects within the framework of the Fano model can be achieved only within the limits of an infinitely narrow excitonic band, $\gamma_{\text{ex}}/\gamma_{\text{pl}} \rightarrow 0 $, corresponding to a purely discrete state in the theory \cite{Fano1961}. When extending the Fano theory to plexcitonic systems with two resonances of finite width (plasmonic and excitonic), it should be borne in mind that although the characteristic width of the excitonic resonance is significantly smaller than the plasmonic one ($\gamma_{\text{ex}} \ll \gamma_{\text {pl}}$), they both must be taken into account when calculating the optical spectra of hybrid nanoparticles. Both widths are directly included into a simple classical model of coupled oscillators and into accurate calculations within the framework of the extended Mie theory or FDTD method.

Note still that with the classical coupled oscillator model (see expression \eqref{c1} in Section \ref{Sect-resonance-antiresonance}), an approximate analytical formula can be derived for plexcitonic systems. To some extent it might be considered as an analogue of the simple expression \eqref{Fano-eq} of the Fano theory. To derive this formula we proceed from the coupled oscillators model and assume that the driving force with a frequency of $\omega$ and an amplitude of $F$ acts only on one plasmonic oscillator. In addition, we take a rather small width of the excitonic resonance equal to zero, $\gamma_{\text{ex}} = 0$. A plasmon resonance peak with an eigen-frequency of $\omega_{\text{pl}}$ and a sufficiently large width of $\gamma_{\text{pl}}$ has been considered as a band of a continuous spectrum. Then, the square modulus of the corresponding amplitude of forced oscillations, $\left\vert c_{\text{pl}}\left(\omega\right)\right\vert^2$, can be written as
\begin{equation}
\left\vert c_{\text{pl}}\left(\omega\right)\right\vert^2 = 
\frac{\left\vert F/m \right\vert^{2}}{\left(\omega^2_{\text{pl}} - \omega^2\right)^2 + \gamma^{2}_{\text{pl}} \omega^2} \frac{\left(q + \epsilon(\omega)\right)^2}{1 + \epsilon^2(\omega)},
\label{eq:fano-like-oscillator-1}
\end{equation}
\noindent where the effective parameter $\epsilon(\omega)$ has the form
\begin{equation}
\begin{split}
	\epsilon(\omega) 
	&= 
	\left[\omega^2 - \omega^2_{\text{ex}} + \frac{\Omega^{4} \left(\omega^2_{\text{pl}}-\omega^2_{\text{ex}}\right)}{\left(\omega^{2}_{\text{pl}} - \omega^{2}_{\text{ex}}\right)^2 + \gamma^2_{\text{pl}} \omega^2_{\text{ex}}}\right] \times\\
	&\times
	\frac{\left(\omega^{2}_{\text{pl}} - \omega^{2}_{\text{ex}}\right)^2 + \gamma^2_{\text{pl}} \omega^2_{\text{ex}}}
	{\gamma_{\text{pl}} \omega_{\text{ex}}\Omega^4},
\end{split}
\label{eq:fano-like-oscillator-2}
\end{equation}
\noindent and the Fano asymmetry parameter is given by
\begin{equation}
	q = \frac{\omega^{2}_{\text{ex}} - \omega^2_{\text{pl}}}{\gamma_{\text{pl}} \omega_{\text{ex}}}.
\label{eq:fano-like-oscillator-3}
\end{equation}

It follows that in a general case there is no direct analogy between the above approximate expressions of the model of coupled oscillators and the simple Fano formula \eqref{Fano-eq}. In particular, formulas \eqref{eq:fano-like-oscillator-2} and \eqref{eq:fano-like-oscillator-3} for parameters $\epsilon(\omega)$ and $q$ explicitly include frequencies of both exciton, $\omega_{\text{ex}}$, and plasmon, $\omega_{\text{pl}}$, resonances. Although the characteristic width of the excitonic resonance is significantly smaller than that of the plasmon resonance ($\gamma_{\text{ex}}\ll \gamma_{\text{pl}}$), it should be taken into account for correct reproduction of experimental data. However, in some cases, approximate expressions \eqref{eq:fano-like-oscillator-1}--\eqref{eq:fano-like-oscillator-3} give results that, to a certain extent, can be interpreted with the simple Fano formula. In particular, this is true when a narrow excitonic peak is located far from the plasmon resonance peak on its wing. This is confirmed by comparing the results for the asymmetrical profile shown in Fig. \ref{Antiresonance}b (black curve in the vicinity of the right peak calculated within the coupled oscillator model), with the corresponding asymmetrical Fano profile (see Fig. \ref{Fano_resonance}a for $q=1$ and $q=2$). Another example is the dip in the light absorption spectrum shown in Fig. \ref{Antiresonance}d. It occurs under the resonant condition, $\omega_{\text{ex}}\approx\omega_{\text{pl}}$, and corresponds to the dip in the Fano profile at $q = 0 $ (see Fig. \ref{Fano_resonance}a and \ref{Fano_resonance}b).

\subsection{Effective Hamiltonian models}

\subsubsection{\label{Sect-Simplest-2Level-model}Simplest two-level model}

To clarify the mechanism behind the formation of optical spectra caused by the coupling of Frenkel excitons with localized surface plasmons, one can also rely on an approach based on an appropriate choice of the effective Hamiltonian of the Metal/J-aggregate system. In the simplest case, when one plasmonic mode interacts with one excitonic mode, the effective Hamiltonian describing the effects of plexcitonic coupling in this system has the form:
\begin{equation}
H=\left( 
\begin{tabular}{ll}
$\mathcal{E}_{\text{pl}}$ & $V$ \\ 
$V^\ast$ & $\mathcal{E}_{\text{ex}}$%
\end{tabular}%
\right)  
\label{Hamiltonian_1}
\end{equation}
\noindent Here $\mathcal{E}_{\text{pl}}=\hbar \omega_{\text{pl}}$ is the plasmon resonance energy, i.e., the position of the peak maximum in the photoabsorption spectrum of the metal core of the particle; $\mathcal{E}_{\text{ex}}=\hbar \omega_{\text{ex}}$ is the exciton energy, i.e. position of the absorption peak maximum of the J-aggregate of the dye in the expression \eqref{eps-J}; $V$ is the plexcitonic coupling energy. The eigenvalues of the Hamiltonian (\ref{Hamiltonian_1}), which determine the energy levels of the hybrid plexcitonic system, are given by the formula:
\begin{equation}
\mathcal{E}_{\pm }=\frac{1}{2}\left(\mathcal{E}_{\text{pl}}+\mathcal{E}_{\text{ex}}\right) \pm \frac{1}{2}%
\sqrt{\left(\mathcal{E}_{\text{pl}}-\mathcal{E}_{\text{ex}}\right) ^{2}+4|V|^{2}}.
\label{Energy_plus-minus}
\end{equation}
\noindent  Formula \eqref{Energy_plus-minus} is similar to the well-known formula (90.9) in \cite{LL-QM}, which arises when calculating the electronic terms of a diatomic molecule in the vicinity of their quasi-crossing point in the presence of the coupling between the diabatic potential energy curves. If the excitonic absorption peak is in resonance with the plasmonic one, $\mathcal{E}_{\text{pl}} = \mathcal{E}_{\text{ex}}$, then formula \eqref{Energy_plus-minus} yields a simple expression for the splitting energy $\Delta \mathcal{E}$ of the hybrid modes, 
\begin{equation}
	\Delta \mathcal{E} = \mathcal{E}_{+} - \mathcal{E}_{-} = 2|V|.
	\label{eq:Rabi-simplest}
\end{equation}
In other words, this directly yields the constant, $V\equiv \hbar g$, of the near-field plexcitonic coupling in composite nanoparticles in the absence of losses. The value of this constant, $g$, can thus be found experimentally by measuring the splitting energy $\Delta \mathcal{E}$ of the spectral peaks of plexcitonic resonances in hybrid nanoparticles.

Following the well-known works \cite{Vasa2013Rabi, Jiang2019, Tserkezis2020, Deng2023} on exciton-plasmon-polariton coupling in cavities, the energy splitting of $\Delta \mathcal{E}$ in the absence of energy dissipation in a plexcitonic nanosystem (see \eqref{eq:Rabi-simplest}) is often interpreted in terms of the Rabi splitting, $\Delta \mathcal{E} = \hbar \Omega_{\text{R}} = 2 \hbar |g|$. However, the value of $\Omega_{\text{R}}$ defined through the constant $g$ reflects the effective parameter of the plexcitonic coupling in the composite system and does not depend on the external field strength. It is associated with vacuum field fluctuations and corresponds to the so-called Rabi vacuum frequency \cite{Barnes2015}. Thus the physical meaning of $\Omega_{\text{R}}$ for plexcitonic systems differs from the standard definition of the Rabi frequency, $\Omega_{\text{R}} = 2 \left\vert \mathbf{d}_{fi} \mathbf{{E}}\right\vert/\hbar$, written for a two-level atom interacting with external resonant radiation field ($\mathbf{d}_{fi}$ -- dipole matrix element of the $\left\vert i\right\rangle \rightarrow \left\vert f\right\rangle$ transition).

\subsubsection{Damping in a plexcitonic system}

One can include losses associated with plasmon, $\gamma_{\text{pl}}$, and exciton, $\gamma_{\text{ex}}$, damping into the theoretical consideration presented in Sect. \ref{Sect-Simplest-2Level-model}. To this end, we represent the energies of the plasmonic and excitonic subsystems by complex quantities defined by the following: $\mathcal{E}_{\text{pl}}\rightarrow \mathcal{E}_{\text{pl}}- i\hbar \gamma _{\text{pl}}/2$ and $\mathcal{E}_{\text{ex}}\rightarrow \mathcal{E}_{\text{ex}}- i\hbar \gamma _{\text{ex}}/2$, respectively. The use of the coupled oscillators model yields the well-known formula
\begin{equation}
\begin{split}
E_{\pm }
&=\hbar \omega_{\pm } =\frac{\hbar }{2}\left[ \omega _{\text{pl}}+\omega _{\text{ex}%
}-i\left(\frac{\gamma_{\text{pl}}}{2}+\frac{\gamma _{\text{ex}}}{2}\right) \right] 
\pm \\
&\pm \frac{\hbar }{2}\sqrt{4\left\vert g\right\vert ^{2}+\left[ \delta
+i\left(\frac{\gamma_{\text{pl}}}{2}-\frac{\gamma_{\text{ex}}}{2}%
\right) \right]^{2}},
\end{split}
\label{Epm-complex}
\end{equation}
\noindent used in many works on plexcitonic coupling (see \cite{Dong2010, Melnikau2016} and references therein). Here $g=V/\hbar$ is the plasmon--exciton coupling constant introduced above; $\omega _{\text{pl}}$ and $\omega _{\text{ex}}$ are the angular frequencies of plasmonic and excitonic modes; $\delta =\omega _{\text{pl}}-\omega _{\text{ex}}$ is their detuning. The same expression can be obtained for the complex eigenvalues of the Hamiltonian
\begin{equation}
\mathcal{H}=\hbar \left( 
\begin{tabular}{cc}
$\omega _{\text{pl}}-i\gamma _{\text{pl}}/2$ & $g$ \\ 
$g^\ast$ & $\omega _{\text{ex}}-i\gamma _{\text{ex}}/2$%
\end{tabular}%
\right)   \label{H-complex}
\end{equation}
\noindent Formula (\ref{Epm-complex}) is in accordance with the Jaynes-Cummings quantum optical model \cite{Jaynes-Cummings1963}. The real parts of the eigen-frequencies of hybrid modes \eqref{Epm-complex} of a plexcitonic system are determined by the expressions \cite{MorLeb-QE2024}
\begin{align}
&\text{Re}\!\left\{ \omega _{\pm }\right\} =\frac{\omega _{\text{pl}}+\omega _{%
\text{ex}}}{2} \pm \nonumber\\& 
\pm\frac{1}{2}\left[\left(\gamma_{\text{pl}}\!-\!\gamma _{\text{ex}}\right) ^{2}\delta^{2}+\left(
4|g|^{2}+\delta^{2}-\frac{\left( \gamma _{%
\text{pl}}\!-\!\gamma _{\text{ex}}\right) ^{2}}{4}\right) ^{2}\right] ^{1/4}\nonumber \!\!\!\!\times\\
&\times\cos{\left[\frac{1}{2}\operatorname{\arctan}\left(\frac{4\left(\gamma_{\text{pl}} - \gamma_{\text{ex}} \right) \delta}{16 g^2 + 4 \delta^2 + \left(\gamma_{\text{pl}}-\gamma_{\text{ex}}\right)^2}\right)\right]}
\label{omega_pm}
\end{align}
\noindent In the special case of zero detuning, $\delta = 0$ and  $\omega _{\text{pl}}=\omega _{\text{ex}}$, this expression takes particularly simple form \cite{Gomez2014}
\begin{equation}
	\text{Re}\!\left\{ \Delta \omega \right\} =\text{Re}\left\{ \omega _{+}-\omega
	_{-}\right\} =2\sqrt{|g|^{2}-\frac{\left( \gamma _{\text{pl}}-\gamma _{\text{ex}}\right) ^{2}}{16}}  
\label{ReDeltaomega}
\end{equation}
Consequently, the hybrid modes of the plexcitonic system (the upper, $\mathcal{E}_{+}$, and lower, $\mathcal{E}_{-}$, branches) have the following complex energies
\begin{equation}
\begin{split}
	&\mathcal{E}_{\pm}= \mathcal{E}_{0} \pm \frac{1}{2}\sqrt{4|V|^{2}-\frac{\hbar^{2}}{4}\left( \gamma_{\text{pl}}-\gamma _{\text{ex}}\right) ^{2}}-i\frac{\hbar }{4}\left(\gamma_{\text{pl}}+\gamma _{\text{ex}}\right). \\
	&\mathcal{E}_{0} \equiv \mathcal{E}_{\text{pl}} = \mathcal{E}_{\text{ex}}
\end{split}
\label{E_pm_dissipation}
\end{equation}
The presence of strong coupling regime can be identified by the positive value of energy splitting between hybrid modes ($+$) and ($-$). According to \eqref{ReDeltaomega} and \eqref{E_pm_dissipation}, this corresponds to the fulfillment of the condition
\begin{equation}
2|g|>\left\vert \gamma _{\text{pl}}-\gamma _{\text{ex}}\right\vert /2.
\label{eq:strong-coupling-condition}
\end{equation} 
Here short-wavelength ($+$) and long-wavelength ($-$) resonance peaks appear in the optical spectra. The scale of the coupling constant is often discussed in terms of this frequency, which, with minor losses, reaches the value $\Delta \omega =2|g|$. By contrast, when the coupling constant $g$ is sufficiently small, there is no  splitting of hybrid modes ($+$) and ($-$) in accordance with \eqref{ReDeltaomega} and there is no point discussing the value of $g$ in terms of the frequency splitting, $\Delta \omega$.

\subsubsection{Multilevel models of plexcitonic coupling}\label{Sect-multilevel-hamiltonian}

A number of problems in the optics of plexcitonic nanosystems require the use of models with multiple coupled oscillators, equivalent to models of a multilevel effective Hamiltonian. These, in particular, include problems of a Frenkel exciton interacting with several multipole (dipole, quadrupole, octupole, etc.) plasmons in the Ag/J-aggregate  and Au/J-aggregate systems \cite{Leb-Medv2012, Moritaka2020, Moritaka2023}. Another example is strong plexcitonic coupling in the Ag/J-aggregate/WS$_{2}$ nanostructure between excitons in the WS$_{2}$ monolayer, excitons in the J-aggregate layer and localized surface plasmons of the silver nanoprism. A mathematical description of this problem was done in \cite{Jiang2019} using a model of three coupled oscillators. Yet another example is the problem of the strong coupling of quasidegenerate Frenkel exciton modes in the outer organic dye shell with a variety of plasmonic modes in plexcitonic dimers consisting of a pair of nanoparticles with a silver spherical or disk-shaped core coated with a layer of dye J-aggregate \cite{Kond-Leb_QE2018, Kond-Leb_OE2019,KondorMek2022}. The strong plexcitonic coupling in such systems \cite{Kond-Leb_OE2019} leads to the replication of spectral bands. The phenomenon results in the emergence of twice the number of plasmon--exciton spectral bands (compared to a purely plasmonic dimer) and narrow peaks associated with resonances in the J-aggregate shell. This phenomenon was described in \cite{Kond-Leb_OE2019} using the FDTD method and interpreted within the framework of the multiple coupled oscillator model and the equivalent multilevel effective Hamiltonian method.

\subsection{\label{Sect-coupling-regimes}Plasmon--exciton coupling regimes}

When plasmons interact with excitons, different regimes of plexcitonic coupling are possible, resulting in the qualitatively different behavior of the optical spectra of composite nanosystems containing metal and excitonic components. The efficiency of plexcitonic coupling strongly affects the absorption and emission properties of metalorganic nanosystems \cite{Wiederrecht2007, Dong2010}. The realization of one or another coupling regime depends on a number of factors, the most important of which are: (i) the difference in unperturbed frequencies $\delta =\omega_{\text{pl}} -\omega_{\text{ex}}$ of the plasmon and exciton; (ii) the effective oscillator strength, $f$, of the J-band transition; (iii) the widths of the plasmonic, $\gamma_{\text{pl}}$, and excitonic, $\gamma_{\text{ex}}$, resonances; (iv) sizes of plasmonic and excitonic subsystems. The classification of plexcitonic coupling regimes has been presented in a series of works (see \cite{Antosiewicz2014,Faucheaux2014,Chen2012,Petoukhoff2020} and references therein). In most of them, weak, strong and ultrastrong coupling modes are usually distinguished.

\subsubsection{Weak plexcitonic coupling}

Obviously, the condition determining the weak plexcitonic coupling regime should be opposite in its physical meaning to the strong coupling condition. Using equation \eqref{eq:strong-coupling-condition} where resonant frequencies, $\omega_{\text{pl}} = \omega_{\text{ex}}$, coincide we obtain
\begin{equation}
|g| \ll \left\vert \gamma_{\text{pl}}-\gamma_{\text{ex}}\right\vert /4 .  
\label{weak-coupl-cond}
\end{equation}

Sometimes the weak coupling condition is formulated so that the $g$ constant would be less than one quarter of the width of one of the damping constants, $\left\vert g \right\vert \ll \gamma_{\text{pl}}/4$ or $\left \vert g \right\vert \ll \gamma_{\text{ex}}/4$. Then the interaction, $V = \hbar g$, between the subsystems leads to a relatively weak perturbation of the considered plasmonic mode and excitonic state. In this case, small shifts occur in the positions of the corresponding resonant spectral peaks relative to the unperturbed frequencies of the plasmon, $\omega_{\text{pl}}$, and exciton, $\omega_{\text{ex}}$. Then, the spectrum of a composite plexcitonic nanoparticle is close to a superposition of the spectra of its metal and J-aggregate subsystems. The weak coupling of plasmons with the organic or inorganic (molecular or excitonic) component of the nanostructure leads to such well-known physical phenomena as surface-enhanced Raman scattering (SERS \cite{Fleischmann1974, Campion1998, Cui2015, Walters2018}), plasmon-induced photoabsorption \cite{ Pockrand1979} and fluorescence \cite{Pompa2006,Yeh2008}, as well as fluorescence quenching \cite{Dulkeith2002}.

\subsubsection{Strong plexcitonic coupling}

The strong plexcitonic coupling regime is quite sought after applications. It is realized when the energy of the near-field plasmon--exciton coupling exceeds the average dissipation energy in the system and new hybrid plexcitonic states of the composite system have physical properties that neither its plasmonic nor excitonic components possess \cite{Liu2015,Liu2016}. In the frequency domain this can be described as a splitting of dispersion curves of the unperturbed plasmonic and excitonic modes resulting in new hybrid modes of the composite system. The quantum analogue of the formation of hybrid plasmon--exciton modes is the splitting of the adiabatic potential energy curves of a diatomic molecule in the vicinity of the crossing point of the diabatic curves \cite{LL-QM}. In the time domain, the hybrid modes in the strong coupling regime arise as a coherent energy exchange between the excitonic and plasmonic subsystems \cite{Vasa2013Rabi, Pelton2019}. Such effects were analyzed, in particular, for Metal/J-aggregate composite nanogratings \cite{Vasa2013Rabi} and for quantum dots in the metal dimers gap \cite{Pelton2019}. However, most of the experimental works on the optics of plexcitonic nanoparticles have not studied the temporal dynamics, having considered only various spectral manifestations of plexcitonic effects in the strong coupling regime.

The rigorous criterion of a strong coupling depends somewhat on the context and the particular area of physics. In various works \cite{Khitrova2006, Schlather2013, Zengin2015, Zengin2013, Petoukhoff2020} the condition has been written in different ways. The following two conditions seem to be the most significant from a physical point of view:
\begin{equation}
|g|>\left|\gamma_{\mathrm{pl}}-\gamma_{\mathrm{ex}}\right|/4, \quad 
|g|^2 > (\gamma^2_{\mathrm{pl}}+\gamma^2_{\mathrm{ex}})/8.
\label{E_pm_dissipation_3eqs}
\end{equation}
\noindent The first condition means that the strong coupling regime is realized when the value of $\sqrt{4 g^2-(\gamma_{\text{pl}}-\gamma_{ \text{ex}})^2/4}$ is positive and so is the frequency splitting, $\Delta\omega = \omega_{+} - \omega_{-}$, given by \eqref{ReDeltaomega} for hybrid modes under resonant condition, $\omega_{\text{ex}} = \omega_{\text{pl}}$. However, the term ''strong coupling'' is often formulated in a empirical way: the system is in the strong coupling regime whenever the Rabi splitting is observed experimentally \cite{Barnes2015}. Therefore, as noted in \cite{Wu2021}, an experimental observation of the Rabi splitting requires fulfillment of a more rigorous condition: the minimum difference between the upper and lower energy branches must be greater than the average width of the plasmonic and excitonic resonances:
\begin{equation}
\Delta \mathcal{E} = \operatorname{Re}\{\mathcal{E}_{+} - \mathcal{E}_{-}\} > \frac{\gamma_{\text{pl}}+\gamma_{\text{ex}}}{2}.\nonumber
\end{equation}

\begin{figure}[b]
	\centering\includegraphics[width=0.94\linewidth]{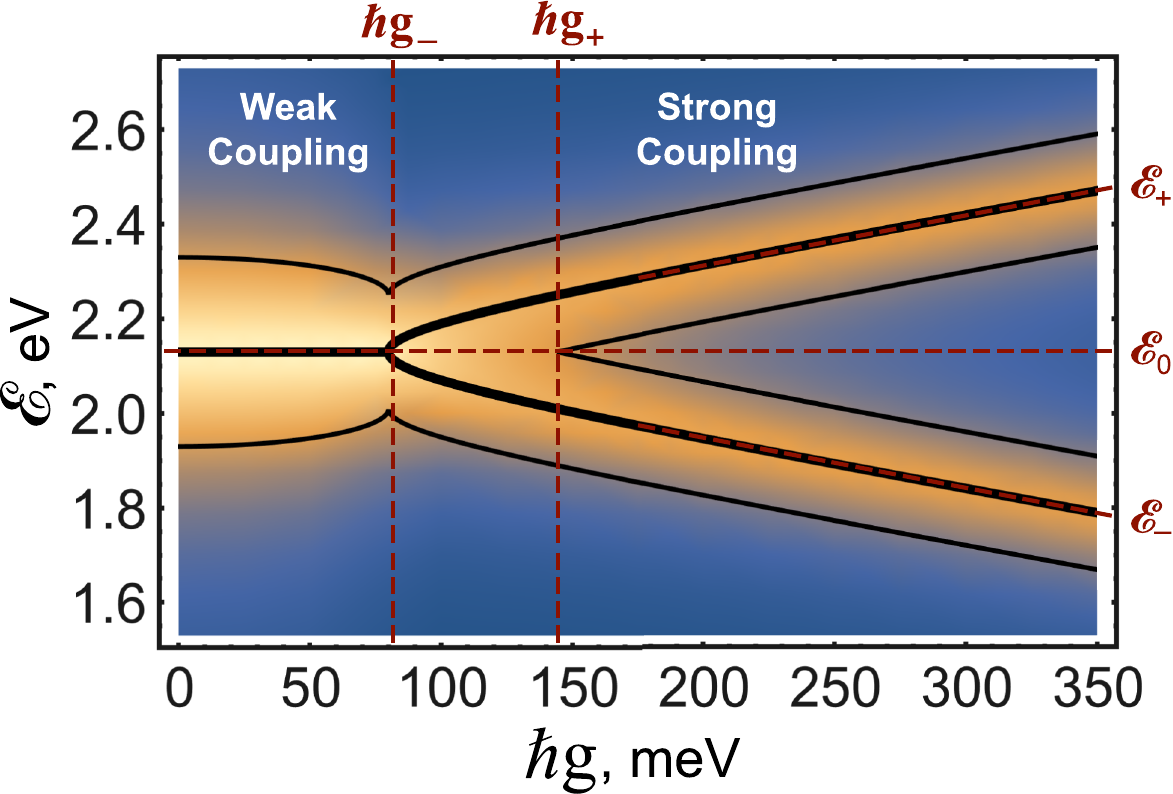}
	\caption{Schematic representation of hybrid modes, demonstrating the transition from weak to strong coupling in a plexcitonic system with a change in coupling energy, $V = \hbar g$. Calculations were performed using formula \eqref{E_pm_dissipation} for the complex energy $\mathcal{E}$ at $\mathcal{E}_{0} = \mathcal{E}_{\text{pl}} = \mathcal{E}_{\text{ex}}$ by taking into account the full widths (FWHM) of the plasmonic, $\hbar\gamma_{\text{pl}}$, and excitonic, $\hbar\gamma_{\text{ex}}$, resonances. Thick black curves show the positions of the maxima in the energy spectrum of the system. The split upper and lower plexcitonic branches correspond to the values of $\mathcal{E}_{+}$ and $\mathcal{E}_{-}$ in formula \eqref{E_pm_dissipation}. Vertical dashed lines indicate coupling energy values corresponding to the rigorous mathematical, $g > g_{-} = \left(\gamma_{\text{pl}}-\gamma_{\text{ex}}\right)/4$, and semiempirical, $g > g_{+} = \sqrt{\left(\gamma^2_{\text{pl}}+\gamma^2_{\text{ex}}\right)/8}$, criteria for the occurrence of a strong coupling regime in the system. The following parameters have been chosen in the calculations: $\mathcal{E}_{0} = 2.13$ eV, $\gamma_{\text{pl}} = 400$ meV, $\gamma_{\text{ex}} = 80$ meV.}
	\label{Fig_regime-criteria}
\end{figure}

Both of these strong coupling criteria are illustrated in Fig. \ref{Fig_regime-criteria} showing the calculations by formula \eqref{E_pm_dissipation} of hybrid modes energies, $\mathcal{E}_{+}$ and $\mathcal{E}_{-}$, of a plexcitonic system depending on the coupling energy, $V = \hbar g$. It follows that under $\hbar g < \hbar g_{-} < \hbar \left|\gamma_{\text{pl}}-\gamma_{\text{ex}}\right|/4$ there is no mode splitting and the influence of the plexcitonic coupling is manifested only in a small change in the width of the single spectral peak. As the interaction energy increases, the plexcitonic branches (${+}$) and (${-}$) split starting with $\hbar g = \hbar g_{-}$. However, as shown in Fig. \ref{Fig_regime-criteria}, these branches become clearly distinguishable at higher values of $\hbar g$ which obviously satisfy the semiempirical criterion formulated above, $\hbar g \gtrsim \hbar g_{+} = \hbar\sqrt{(\gamma^2_{\text{pl}}+\gamma^2_{\text{ex}})/8}$.

\begin{table*}[t]
	\caption{
		Splitting energies, $\Delta \mathcal{E} = \operatorname{Re}\{\mathcal{E}_{+} - \mathcal{E}_{-}\}$, of hybrid modes of a number of composite nanostructures which determine, in accordance with \eqref{ReDeltaomega}, the characteristic values of the plexcitonic coupling constant, $g \approx \Delta \mathcal{E} / (2\hbar)$. Many papers on plexcitonics, when discussing the strong and ultrastrong coupling regimes, refer to $\Delta \mathcal{E}$ quantity as the Rabi splitting energy, $\hbar \Omega_{\text{R}}$.
	}
	\label{tab:RabiSplitting}
	\smallskip
	\center{\footnotesize{\begin{tabular}{|c|c|c|c|}
				\hline
				\hspace{10pt} $\Delta \mathcal{E}$, meV \hspace{10pt} & Form & Composition & Reference \\ \hline
				155 & Nanorod array & \hspace{10pt} Au/Dye(CAS RN 27268-50-4) \hspace{10pt} & Wurtz et al., 2007 \cite{Wurtz2007} \\ \hline
				180 & Dye aggregate on silver film & Ag/TDBC & Bellessa et al., 2004 \cite{Bellessa2004} \\ \hline
				187 & Right-handed nanorod dimer & Ag/TDBC & Zhu et al., 2021 \cite{ ZhuWu2021} \\ \hline
				199 & Right-handed nanorod & Alloy-AgAu/TDBC & Cheng et al., 2023 \cite{ Cheng-Yang2023} \\ \hline
				202 & Left-handed nanorod dimer & Ag/TDBC & Zhu et al., 2021 \cite{ ZhuWu2021} \\ \hline
				205 & Left-handed nanorod & Alloy-AgAu/TDBC & Cheng et al., 2023 \cite{ Cheng-Yang2023} \\ \hline
				207 & Nanoprism & Ag/PIC & DeLacy et al., 2015 \cite{DeLacy2015} \\ \hline
				214 & Nanorod & Au/Ag/TDBC & Wu et al., 2021 \cite{ Wu2021} \\ \hline
				260 & Nanostars & Au/JC1(CAS RN 3520-43-2) & Melnikau et al., 2013 \cite{Melnikau2013} \\ \hline
				265 & Nanorods & Au/JC1(CAS RN 3520-43-2) & Melnikau et al., 2016 \cite{Melnikau2016} \\ \hline
				350 & Nanodisk & Ag/TDBC & Balci et al., 2019 \cite{ Balci2019} \\ \hline
				377 & Nanocubes & Au/PIC & Song et al., 2019 \cite{Song2019} \\ \hline
				400 & Nanoprism & Ag/TDBC & Balci et al., 2013 \cite{Balci2013} \\ \hline
				412 & Silver film with nanoholes & Ag/H$_{4}$TPPS & Salomon et al., 2013 \cite{Salomon2013} \\ \hline
				450 & Nanodisk array & Ag/TDBC & Bellessa et al., 2009 \cite{Bellessa2009} \\ \hline
				550 & Nanodisk & Ag/Heptamethine & Todisco et al., 2018 \cite{Todisco2018} \\ \hline
				700 & \hspace{10pt} Dye molecules in a low-Q resonator \hspace{10pt} & Ag/PMMA/Spiropyran & \hspace{10pt} Schwartz et al., 2011 \cite{Schwartz2011} \hspace{10pt} \\ \hline
	\end{tabular}}}
\end{table*}

The dip in the spectra of plexcitonic particles directly related to the splitting of the hybrid modes ($+$) and ($-$) with an increase in $g$ constant, occurs in the strong coupling regime according to the mathematical criterion, $|g| > \left|\gamma_{\text{pl}}-\gamma_{\text{ex}}\right|/4$. Moreover, several experimental studies \cite{DeLacy2015, Balci2013, Balci2019} have demonstrated that a decrease in the plexcitonic coupling efficiency first leads to a reduction in the dip depth, and eventually to its complete disappearance. This is also confirmed by numerical calculations within the framework of the extended Mie theory. 
Thus, in \cite{Leb-Medv2013b, Moritaka2023} it was shown that the efficiency of the near-field plexcitonic coupling  decreases rapidly with an increase in the thickness, $\ell_{\text{s}}$, of the intermediate dielectric layer between the metal core and the outer J-aggregate shell of three-layer metalorganic nanoparticles. Note in this context that the induced transparency effect of plexcitonic nanoparticles fully corresponds to the dip in the absorption spectra, and hence, certainly occurs in the strong coupling regime. To avoid any misunderstanding, we note that the phenomenon of electromagnetically induced transparency (EIT), as already mentioned in Section \ref{Sect-Induced-Transparency}, has a different physical nature. Therefore, the fact that the EIT phenomenon is attributed to the weak coupling regime (see, e.g., \cite{Kivshar2017}) does not contradict the interpretation of the induced transparency (IT) of plexcitonic particles as a manifestation of the strong coupling regime.

\subsubsection{Ultrastrong plexcitonic coupling}

The case of strong plexcitonic coupling is often distinguished by the ultrastrong coupling regime characterized by an even larger value of the interaction constant, $g$, between the plasmon and the exciton under conditions of coincidence of the frequencies of plasmonic and excitonic resonances, $\omega_{0} = \omega_{\text{ex}} = \omega_{\text{pl}}$. In the ultrastrong coupling regime, the coupling constant $g$ reaches values greater than or of the order of one-tenth of the resonant frequency, $\omega_{0}$. Therefore, the frequencies of the hybrid modes of the plexcitonic system, $\omega_{+}$ and $\omega_{-}$, differ significantly from the eigen-frequencies of its subsystems, $\omega_{0}$. Here, describing the interaction of a hybrid system with incident light at a frequency $\omega$ from the same spectral range as the original eigen-frequencies of $\omega_{0}$, one should take into account the nonresonant contribution to the composite system polarizability, 
$\alpha_{\text{NR}} \propto \left({\mathcal{E}_{\pm} + \hbar \omega}\right)^{-1}$, which is usually assumed to be negligibly small compared to the standard resonant one, 
$\alpha_{\text{R}} \propto \left({\mathcal{E}_{\pm} - \hbar \omega}\right)^{-1}$. A consistent approach to accounting for both contributions to polarizability is presented in \cite{Boyd2008} (see formula 3.2.23 and its derivation).

The realization of such a regime in hybrid metalorganic plexcitonic nanosystems was noted in a number of papers \cite{Balci2013, Todisco2018, Balci2019, Petoukhoff2019, Xiong2019, Schwartz2011}. No strict analytical criterion that determines the occurrence of an ultrastrong plexcitonic coupling regime has yet been established, to the authors knowledge. However, as shown in the papers above this regime is achieved when the coupling constant, $g$, reaches $7$--$10$ \% of the frequency $\omega_{0}$. Then, the frequency splitting of the spectral peaks of hybrid modes, called the Rabi frequency of $\Omega_{\text{R}}$, is higher than approximately $15$--$20$ \% of the frequency $\omega_{0}$. Insights into the efficiency of plasmon--exciton interaction in hybrid metalorganic nanosystems of different compositions and shapes can be obtained from Table \ref{tab:RabiSplitting}, which lists values of the energy splitting, $\Delta \mathcal{E} = \hbar\Omega_{R} = 2 \hbar |g|$, of spectral peaks in the strong and ultrastrong coupling regimes.

\section{\label{Sect6}Absorption, scattering and extinction spectra of two-layer metalorganic ''core-shell'' nanospheres}

\subsection{Near-field coupling of Frenkel exciton with dipole and multipole plasmons}

\begin{figure}[b]
	\centering\includegraphics[width=0.94\linewidth]{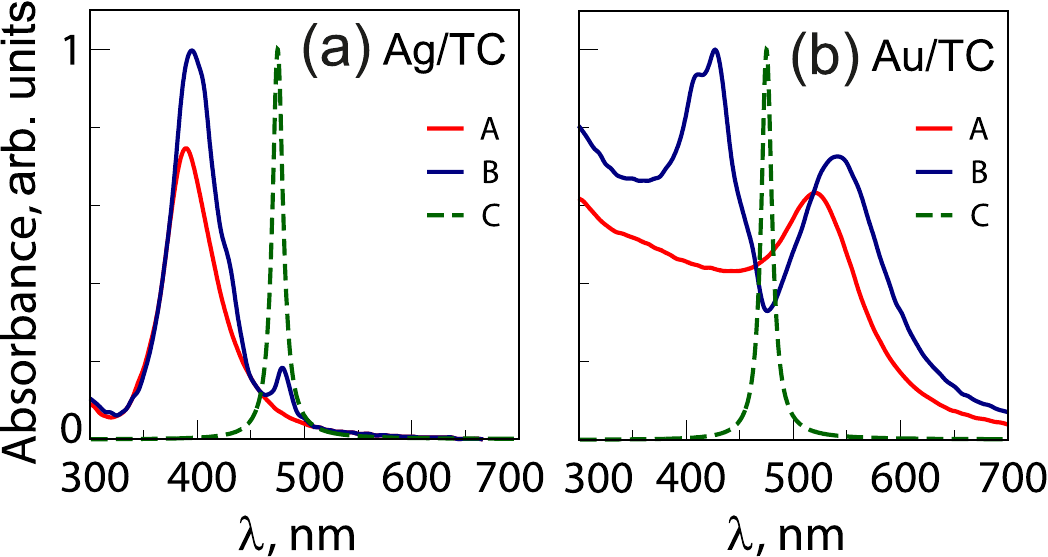}
	\caption{Experimental data ~\cite{Kometani2001} for light absorption spectra of hybrid nanospheres Ag/TC (a) and Au/TC (b) consisting of a metal core and thin J-aggregate layer of TC dye, as a function of light wavelength in vacuum, $\lambda$ (blue curves, B). Red curves (A) in panels (a) and (b) demonstrate corresponding results for light absorption spectra of bare silver and gold cores of these particles. The green dashed curves (C) show the absorption J-band of the TC dye aggregate. The core radius is $r = 5$ nm; the particles are in an aqueous solutions.}
	\label{fig:Au-TC_Ag-TC}
\end{figure}

There are a lot of works on the formation, structure and spectral characteristics of hybrid nanoparticles of of various size and shape, consisting of a metal core and an outer shell of ordered molecular J-aggregates of organic dyes. Among them are works on two-layer metalorganic nanoparticles of spherical shape \cite{Kometani2001, Sato2001, Hranisavljevic2002, Wurtz2003,  Wiederrecht2004, Uwada2007, Wiederrecht2008, Lebedev2008, Yoshida2009a, Lebedev2010, Lekeufack2010, Leb-Medv2012, Vujacic2012, Laban2014, Laban2015, Leb-Medv2013a, Antosiewicz2014, Moritaka2020, Moritaka2023}. There is also a number of experimental and theoretical papers on composite nanoparticles with a metal core and a J-aggregate shell of more complex geometrical shapes: spheroidal \cite{Gulen2010, Zengin2013}, rod-shaped \cite{Wurtz2007, Ni2010, Shapiro2015, Melnikau2016, Wu2021, Guo-Wu2021, Dey2023} and dumbbell-shaped \cite{Kondorskiy2015}, as well as composite nanoparticles in the shape of disks \cite{Todisco2015, Balci2019, Fain2019}, triangular prisms \cite{Balci2013, DeLacy2015, Zengin2015, Balci2016, Lam-Kond-Leb2019}, and stars \cite{Melnikau2013, Vasa2013, Lam-Kond-Leb2019, Melnikau2022}. Most experiments were performed for particles with a gold or silver core coated with one of the cyanine dyes that can aggregate on the surface of a metal core. These works draw a fairly complete picture of the nature and basic patterns in the spectra of light absorption and scattering by Ag/J-aggregate and Au/J-aggregate nanoparticles under various regimes of the near-field coupling of Frenkel excitons with localized surface plasmon-polaritons at different values of the optical constants of the metal core and J-aggregate shell, or the size and shape of hybrid nanoparticles.

Importantly, although the shape of hybrid particles has a significant impact on the behavior of their absorption, scattering, and luminescence spectra, many important consistent patterns can still be demonstrated using nanospheres as an example. We will first discuss the behavior of the extinction spectra of Metal/J-aggregate small-radius nanospheres in an aqueous solution. Corresponding experimental data \cite{Kometani2001} for silver and gold particles coated with a J-aggregate cyanine dye TC, with a small core radius $r_{1}=5$ nm and shell thickness $l_{\text{J}}=1$ nm are shown in Fig. \ref{fig:Au-TC_Ag-TC}.

In both cases, the absorption spectra are affected by the interaction of Frenkel excitons in the outer shell with plasmons localized in the metal core. However, the behavior of these spectra turns out to be qualitatively different for Ag/TC and Au/TC particles, reflecting two different regimes: weak and strong plexcitonic coupling. For Ag/TC particles, the absorption spectrum of the plexcitonic system is close to the superposition of the spectra of the plasmon resonance and the excitonic J-band of the molecular aggregate. This corresponds to a weak plexcitonic coupling regime, which is caused by quite large spectral distance between the positions of the maxima of the unperturbed peaks of the plasmonic ($\lambda^{(\text{Ag})}_{\text{pl}} = 384$ nm) and excitonic ($\lambda_{\text{ex} }=475$ nm) resonances on the wavelength scale (see Fig. \ref{fig:j-bands}). It is a qualitatively different situation for Au/TC particles. Here a pronounced dip appears in the vicinity of the resonant frequency of the J-band of the dye aggregate. The dip can be considered as a vivid example of strong plexcitonic coupling regime. This coupling regime arises because the resonant frequencies of the unperturbed plasmonic ($\lambda^{(\text{Au})}_{\text{pl}} = 525$ nm) and excitonic subsystems are significantly closer to each other ($\left\vert\Delta \lambda\right\vert = 50$ nm) in the case of Au/TC than in the case of Ag/TC ($\left\vert\Delta \lambda\right\vert = 91$ nm). The presence of a peak or a dip near the absorption J-band of the TC aggregate is interpreted as a result of constructive or destructive interference of the contributions of plasmonic and excitonic subsystems to the total spectrum of the hybrid particle. Due to the small overall sizes of Ag/TC and Au/TC nanospheres in experiments~\cite{Kometani2001}, the behavior of their absorption spectra is almost entirely determined by the near-field coupling of the Frenkel exciton with the electric dipole plasmon, while the role of higher order multipoles ($n > 1$) is insignificant.

\begin{figure}[t]
	\centering\includegraphics[width=0.94\linewidth]{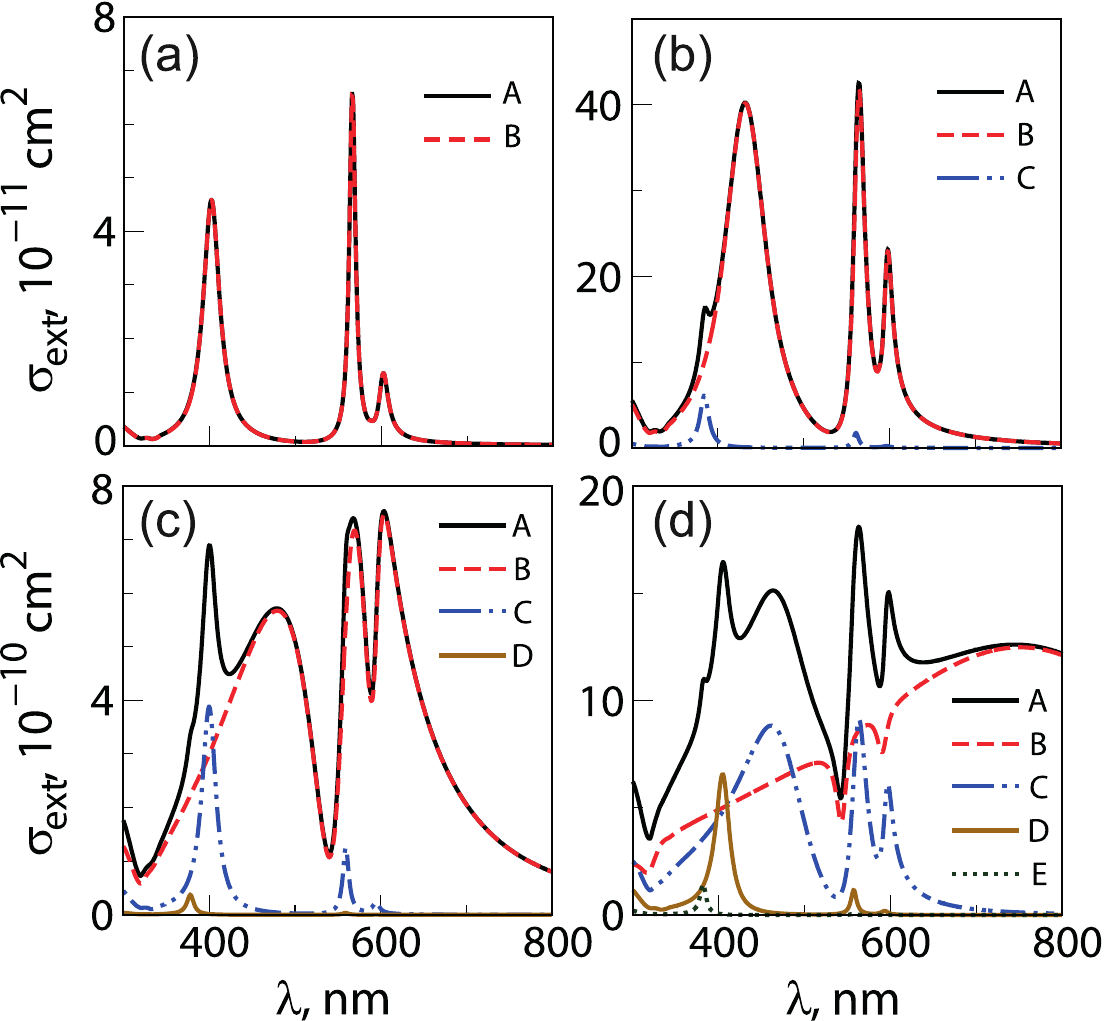}
	\caption{Spectral dependences of light extinction cross sections, $\sigma_{\text{ext}}$, by two-layer nanospheres consisting of a silver core and an outer J-aggregate shell of TDBC dye. Calculations were performed using the extended Mie theory \cite{Bohren1998} for four sets of inner and outer particle radii: $r_1=12$ nm, $r_2=20$ nm (a); $r_1=32$ nm, $r_2=40$ nm (b); $r_1=52$ nm, $r_2=60$ nm (c); $r_1=92$ nm, $r_2=100$ nm (d). The red, blue, dark yellow and green curves are the contributions of dipole ($n=1$, B), quadrupole ($n=2$, C), octupole ($n=3$, D) and hexadecapole ($n=4$, E) terms into the general formula \eqref{sigma-ext-sum} of multipole expansion, respectively. The solid black curve (A) is their sum.}
	\label{fig:plexciton-multipole}
\end{figure}

Calculations \cite{Leb-Medv2012, Moritaka2020} for two-layer Ag/TC and Au/TDBC nanospheres show that the situation changes drastically with a significant increase in the total particle size. Then, the near-field coupling of the Frenkel exciton not only with the dipole, but also with the multipole plasmons of higher order $n>1$ (especially with the quadrupole and octupole) plays a significant role in the formation of the optical spectra of hybrid plexcitonic nanoparticles. We demonstrate this effect by calculating light extinction spectra within the framework of the Mie theory using Ag/TDBC hybrid particles as an example (see Fig. \ref{fig:plexciton-multipole}). The thickness of the excitonic shell is chosen to be fairly large, $l_{\text{J}}=8$ nm, as sufficiently strong grounds to justify the scalar ''isotropic'' model (\ref{eps-J}) used to describe dielectric function, $\varepsilon _{\text{J}}(\omega)$, of the excitonic J-aggregate shell of the particle. The following parameters of the TDBC J-aggregate were used in the calculations: $\hbar \omega_{\text{ex}}=2.12 $ eV ($\lambda_{\text{ex}}=585$ nm), $\gamma_{\text{ ex}}=48$ meV, $f_{\text{J}}=0.44$ and $\varepsilon_{\infty}^{\text{J}}=2.3$ \cite{Bellessa2014}. Following \cite{Leb-Medv2012, Kond-Leb_OE2019, KML_OE2022}, the dielectric function of the metal core was calculated taking into account the size effect (see Sect. 2 and formula \eqref{eps-size-dep}). The values chosen for the radius of the silver core were $r_{1} = 12$, $32$, $52$ and $92$ nm.

The formation of the light absorption spectrum of an Ag/TDBC particle with a small radius is almost entirely determined by the interaction of a Frenkel exciton with an electric dipole plasmon, $n=1$. This is shown in Fig. \ref{fig:plexciton-multipole}a for a particle with outer radius $r_2 = 20$ nm. In accordance with the analytical model (see Sect. \ref{Sect5-analytical-hybrid-freqs}), with isotropic orientation of molecular aggregates in the organic shell, three pronounced maxima are formed in the extinction spectrum. The left plexcitonic peak appears at a wavelength of $\lambda = 402$ nm near the position of the dipole plasmon resonance of the bare silver core, $\lambda_{\text{pl}} = 389$ nm, shifted relative to it towards longer wavelengths by $\Delta \lambda = 13$ nm. In the vicinity of the excitonic transition frequency of the TDBC J-aggregate ($\lambda_{\text{ex}} = 585$ nm), two split plexcitonic peaks appear in the cross section at $\lambda= 566$ and $603$ nm, whose maxima differ substantially from each other in intensity,
\begin{equation*}
	\sigma^{(\text{left})}_{\text{ext}}/\sigma^{(\text{right})}_{\text{ext}} = 4.86.
\end{equation*}

As the outer particle radius increases to $r_2=40$ nm (see Fig. \ref{fig:plexciton-multipole}b), a small contribution to the extinction comes from the quadrupole term (blue curve), which leads to the appearance of a weak peak in the total cross section, $\sigma_{\text{ext}}$ (black curve), in the region of $350 \lesssim \lambda \lesssim 400$ nm. The plexcitonic effect manifests itself as a deep dip (at $\lambda=586$ nm) between two peaks of different intensities 
\begin{equation*}
	\sigma^{(\text{left})}_{\text{ext}}/\sigma^{( \text{right})}_{\text{ext}} = 1.84,
\end{equation*}
near the position of the TDBC J-band maximum, $\lambda_{\text{ex}} = 588$ nm. Further increase in the outer radius to $r_{2}=60$ nm leads in the range of $350 \lesssim \lambda \lesssim 450$ nm to a very noticeable modification of the extinction spectrum as a result of the contribution of the quadrupole term (see Fig. \ref{fig:plexciton-multipole}c). Here, compared to the previous case (Fig. \ref{fig:plexciton-multipole}b), the intensities of the middle and right dipole plexcitonic peaks equalize, and the relative depth of the dip between the left and middle dipole peaks decreases. In addition, in the vicinity of $\lambda = 380$ nm, the contribution of the octupole term (green curve) becomes noticeable, which, however, does not lead to any significant change in the behavior of the total extinction cross section.

A drastic change in the behavior of the spectrum of the plexcitonic Ag/TDBC particle occurs at large values of $r_2 = 100$ nm (see Fig. \ref{fig:plexciton-multipole}d). First of all, attention is drawn to the increase in the total number (six) of spectral peaks due to the increase in the contribution of quadrupole and octupole resonances to the cross section. In this case, the peak position is strongly shifted to longer wavelengths and equals to $\lambda_{\max} = 751$ nm due to the interaction of the exciton with the dipole plasmon. The cross section in this region ($\lambda\gtrsim 650$ nm) can be reproduced quite well taking into account only the dipole term in the Mie expansion. By contrast, the behavior of the spectrum in the region of $\lambda \lesssim 650$ nm is mainly determined by the contribution of multipole resonances.

\subsection{Role of size effects}

Below is a summary of the main results clarifying the influence of size effects on the optical properties of metal nanoparticles coated with a thin layer of a molecular dye aggregate. The specific manifestation of these effects is determined by several factors. Previously, in Section \ref{Sect-SPP} we have demonstrated the importance of the size effect for the dielectric function of a noble metal core in a hybrid nanoparticle associated with the scattering of free electrons at the Metal/J-aggregate interface. This effect has been shown to be particularly strong if the particle size is significantly smaller than the electron mean free path, $l_{\infty }$, in bulk silver or gold samples. In addition, as follows from the formulas of the quasistatic approximation (\ref{sigma-abs-static}) and (\ref{alpha}), the dipole polarizability $\alpha$ of the hybrid particle, and therefore the absorption, $\sigma_{\text{abs}}$, and scattering, $\sigma_{\text{scat}}$, cross sections strongly depend on this particle's overall radius (since $\alpha\propto r^{3}_2$), and also on the ratio $r_1/r_2$ of the inner and outer radii of the concentric spheres due to the near-field coupling between the core and the shell. Beyond the applicability of the quasistatic approximation (i.e. at $kr\gtrsim 1$), a further significant increase in the outer radius of the hybrid nanoparticle leads to more complex dependences of the spectra on its geometrical parameters than those predicted by simple expressions (\ref{sigma-abs-static}) and (\ref {alpha}). It was demonstrated for two-layer Ag/J-aggregate nanoparticles by direct calculations \cite{Leb-Medv2013a} of the cross sections $\sigma_{\text{abs}}$ and $\sigma_{\text{scat}}$ within the framework of extended Mie theory using the size- and frequency-dependent dielectric function of the silver core and the standard isotropic model \eqref{eps-J} for the dielectric function of the J-aggregate shell. 

For particles with core radii $r_1=10$ nm and $r_1=30$ nm, when the thickness of the J-aggregate shell, $l_{\text{J}}$, varies from $1$ to $12$ nm, significant changes have been shown to occur in the positions and intensities of spectral photoabsorption peaks even with relatively small changes in the thickness of the outer excitonic layer.

It was also shown that the spectral peaks intensities increase substantially for the bigger external particle size, mostly due to an increase in its total volume $\mathcal{V}=4\pi r^3_2/3$. In addition, with a significant increase in the particle size, the behavior of the absorption spectra becomes more complex due to an increase in the total number of spectral peaks (see Fig. \ref{fig:plexciton-multipole}). It is caused by the effects of interaction of the Frenkel exciton in the organic shell not only with the electric dipole, but also with multipole (quadrupole, octupole, etc.) localized plasmons in the metal core of the particle \cite{Leb-Medv2012}. In a certain range of nanoparticle sizes, when the contributions of light absorption and scattering to the resulting total extinction cross section become of the same order, the spectral behavior of $\sigma_{\text{ext}}\left(\lambda \right)$ is additionally affected by the competition of contributions from these processes. For silver particles coated with a thin layer of a J-aggregate, the extinction cross section is mainly determined by the light absorption at an outer radius less than $25$--$35$ nm, while at larger values of the radius the contribution of the light scattering begins to dominate. For hybrid particles with a gold core, the scattering process becomes predominant when the outer radius of the particle is $r_2\gtrsim 55$--$60$ nm.

\section{Optical spectra of three-layer metalorganic nanospheres}

\subsection{Absorption and luminescence spectra}

In a number of works \cite{Yoshida2009a, Yoshida2010, Wurtz2007, Yoshida2009b, Yoshida2009b, Shapiro2015} three-layer nanoparticles of various geometrical shapes and sizes were synthesized consisting of a gold or silver core, an outer J-aggregate shell of a cyanine dye (TC, PIC, TDBC, etc.), and an intermediate organic layer (spacer) between them. Interacting with an external electromagnetic field, this layer acts as a passive dielectric layer between two active layers of the nanoparticle: the core, where localized surface plasmons are excited, and the outer excitonic shell \cite{Leb-Medv2013b, Moritaka2023, KML_OE2022, Kond-Leb_OE2019}. By varying the thickness of the spacer layer it is also possible to affect the strength and features of the plexcitonic coupling between the core and shell and, thereby, modify the spectral characteristics of the hybrid system compared to two-layer Metal/J-aggregate particles.

\begin{figure}[t]
	\centering\includegraphics[width=0.94\linewidth]{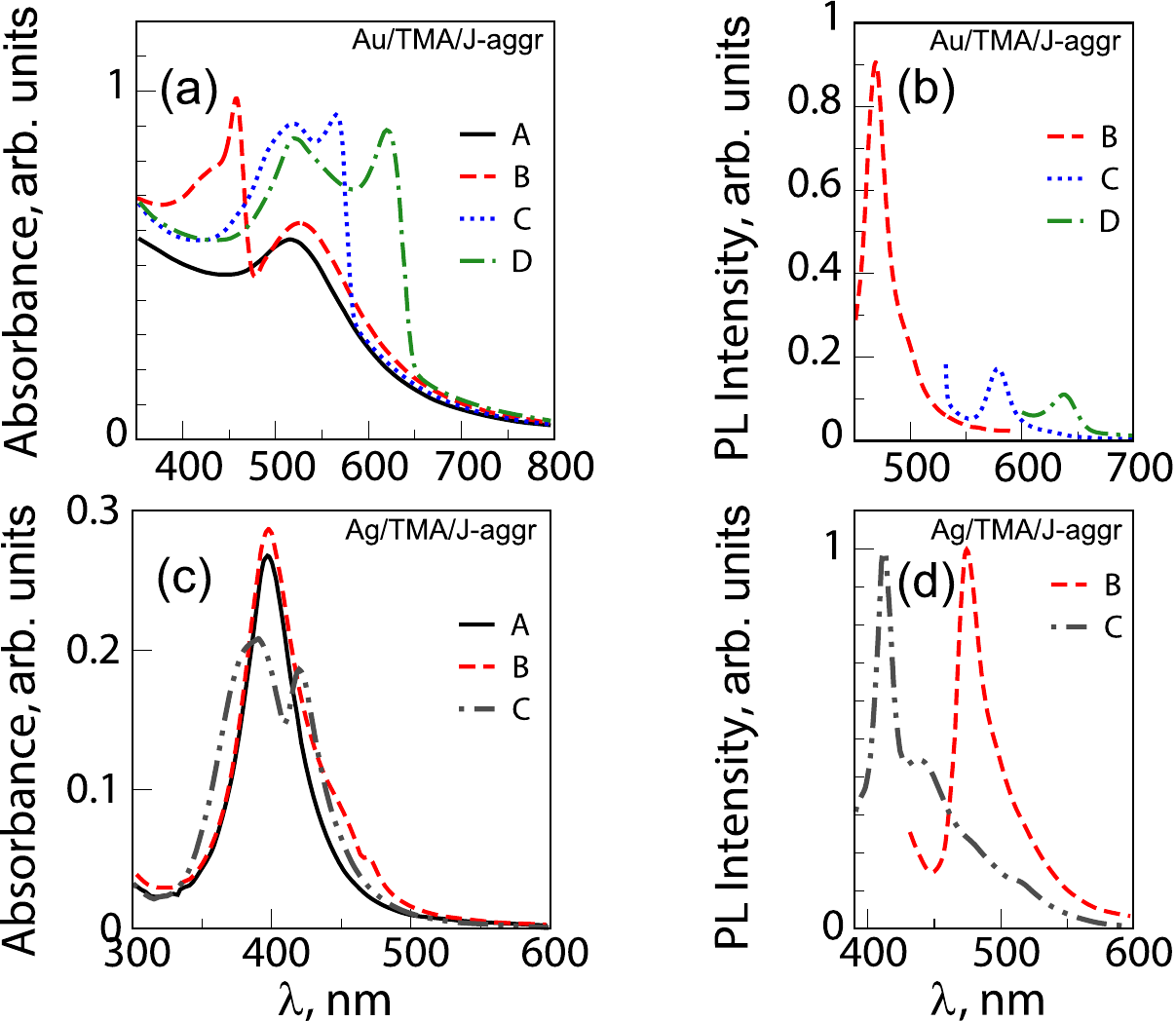}
	\caption{Experimental absorption (a, c) and photoluminescence (b, d) spectra of colloidal three-layer Metal/Spacer/J-aggregate nanospheres with gold (a, b) \cite{Yoshida2009a} and silver (c, d) \cite{Yoshida2010} cores and TMA spacer. For particles with a gold core, the J-aggregate shell consisted of TC (red dashed curves, B), PIC (blue dotted curves, C), Thia(Et) (green dash-dotted curves, D) dyes. For particles with a silver core, J-aggregates of TC (red dashed curves, B) and OC (gray dash-dotted curves, C) dyes were used. Panels (a) and (c) show absorption spectra of Au/TMA and Ag/TMA nanospheres without a J-aggregate layer (solid black curves, A). Core diameters ($D$), spacer thicknesses ($\ell_{\text{s}}$), and outer shell thicknesses ($l_{\text{J}}$): Au/TMA/J-aggregate -- $D = 5.2$~nm, $\ell_{\text{s}} = 1.1$~nm, $l_{\text{J}} = 1.7$~nm; Ag/TMA/J-aggregate -- $D = 14.1$~nm, $\ell_{\text{s}} = 1$~nm, $l_{\text{J}} = 4$~nm. Photoluminescence excitation wavelengths: $\lambda = 430$ nm (Au/TMA/TC), $\lambda = 520$ nm (Au/TMA/PIC), $\lambda = 550$ nm (Au/TMA/Thia(Et)), $\lambda = 380$ nm (Ag/TMA/TC), and $\lambda = 405$ nm (Ag/TMA/OC).}
	\label{fig:3layer-luminescence}
\end{figure}

The key difference between three-layer metalorganic nanoparticles with an intermediate passive layer and two-layer particles is that when they interact with light, it is not only absorption and scattering processes possible but photoluminescence too. This is demonstrated in Fig. \ref{fig:3layer-luminescence}, with the experimental data \cite{Yoshida2009a} on the absorption and photoluminescence spectra of a colloidal solution of hybrid three-layer Au/TMA/TC nanospheres (solid red curves), as well as the absorption spectrum of a two-layer Au/TMA particle without the outer J-aggregate layer (dashed curve). Note that in the absorption spectrum of Au/TMA/TC nanoparticles there are two plexcitonic peaks: a sharp one at $\lambda=459$ nm and a broader one at $\lambda=526$ nm. It is important to stress, that the experiment demonstrated fairly intense photoluminescence of three-layer Au/TMA/TC nanospheres with a maximum emission spectrum at $\lambda = 469$ nm (see Fig. \ref{fig:3layer-luminescence}) attributed by the paper's authors to the luminescence of the TC J-aggregate \cite{Yoshida2009a}. In contrast, for bilayer Au/TC nanoparticles in which the gold cores were directly coated with the J-aggregate, no corresponding luminescence was observed. This fundamental difference was explained  \cite{Yoshida2009a} by the quenching of luminescence due to the excitation energy transfer from the J-aggregate shell to the metal core. Such quenching can be very effective in two-layer nanoparticles, but is significantly suppressed in three-layer nanoparticles due to the presence of a passive TMA spacer. The presence of photoluminescence, in addition to the light absorption by three-layer nanoparticles, was also observed for a number of other dyes forming the outer J-aggregate shell. The corresponding examples are Au/TMA/PIC and Au/TMA/Thia(Et) nanoparticles (see the structural formulas of the dyes in Table \ref{table-Jaggr}) with a spherical \cite{Yoshida2009a} and rod-shaped \cite{Yoshida2009b} gold core, as well as three-layer Ag/TMA/TC and Ag/TMA/OC nanospheres with a silver core \cite{Yoshida2010}.

\subsection{Influence of thickness of spacer layer on optical spectra}

In a number of problems in optics and spectroscopy of nanoparticles with a metal core and a double shell consisting of a passive organic intermediate layer and an active J-aggregate outer layer, what stands out is that changing the thickness of the passive layer provides for additional means of altering the coupling strength between the core and the shell. This offers some new possibilities for the modification of the spectral characteristics of the hybrid nanosystem compared to metal/J-aggregate two-layer particles. This has been demonstrated in \cite{Leb-Medv2013b} using spherical Ag/TMA/TC nanoparticles with a silver core, an outer J-aggregate shell of the cyanine TC dye and an organic spacer of TMA [N,N,N-trimethyl(11-mercaptoundecyl)ammonium chloride]. Similar calculations of light absorption spectra were performed in \cite{Moritaka2023} for the Au/TMA/J-aggregate composite nanospheres with a gold core.

\begin{figure}[t]
	\centering\includegraphics[width=0.94\linewidth]{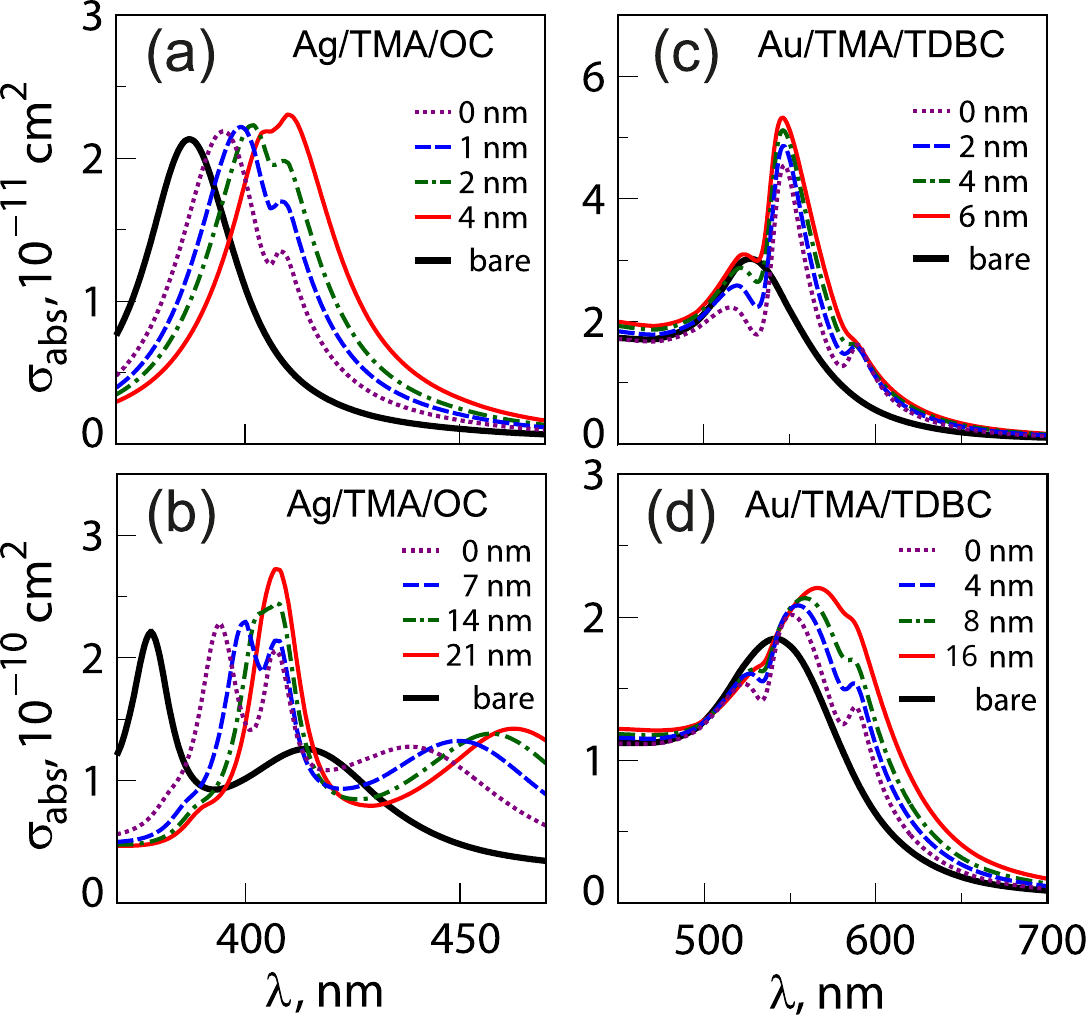}
	\caption{Light absorption cross sections, $\sigma_{\text{abs}}$, of three-layer Ag/TMA/OC (a, b) and Au/TMA/TDBC (c, d) nanospheres in water as functions of the light wavelength, $\lambda$, in vacuum at different thicknesses, $\ell_{\text{s}}$, of the spacer layer. The calculations have been performed by using the modified Mie theory for the following values of the metal core radius, $r_{1}$, the outer J-aggregate shell thickness, $l_{\text{J}}$, and sets of $\ell_{\text{s}}$ values: (a) $r_{1} = 10$ nm, $l_{\text{J}} =1$ nm, $\ell_{\text{s}} = 0$, $1$, $2$, $4$ nm; (b) $r_{1} = 70$ nm, $l_{\text{J}} =7$ nm, $\ell_{\text{s}} = 0$, $7$, $14$, $21$ nm; (c) $r_{1} = 20$ nm, $l_{\text{J}} =1$ nm, $\ell_{\text{s}} = 0$, $2$, $4$, $6$ nm; (d) $r_{1} = 40$ nm, $l_{\text{J}} = 1$ nm, $\ell_{\text{s}} = 0$, $4$, $8$, $16$ nm. Values of the $\ell_{\text{s}}$ spacer thickness are specified in the legends for each colored spectral curve. Black curves show the absorption spectra of the bare metal cores.}
	\label{fig:spheres-spacer-effect}
\end{figure}

To demonstrate the near-field nature of the plexcitonic coupling and the strong dependence of the results on the distance between the plasmonic and excitonic subsystems, we plot in Figure \ref{fig:spheres-spacer-effect} the light absorption cross sections by three-layer nanospheres of Ag/TMA/OC (panels \ref{fig:spheres-spacer-effect}a and \ref{fig:spheres-spacer-effect}b) and Au/TMA/TDBC (panels \ref{fig:spheres-spacer-effect}c and \ref{fig:spheres-spacer-effect}d). These series of calculations have been performed using the extended Mie theory for concentric spheres at four different values of the spacer layer thicknesses $\ell_{\text{J}}$ and at different radii $r_{1}$ of the metal core and thicknesses $l_{\text{J}}$ of the outer J-aggregate shell. 

Both series of calculations demonstrate strong dependences of the results for absorption cross sections on the distance $\ell_{\text{s}}$ between two interacting active subsystems, plasmonic and excitonic. In particular, Fig. \ref{fig:spheres-spacer-effect}a illustrates a significant weakening of the near-field plexcitonic coupling efficiency with the increasing thickness of $\ell_{\text{s}}$ for a silver core. This is manifested  as a decrease in the depth of the dip in the vicinity of $\lambda = 406$ nm, with increasing $\ell_{\text{s}}$. The weakening of plexcitonic coupling with an increase in the passive spacer thickness also takes place for composite nanospheres of a larger size, as shown in Fig. \ref{fig:spheres-spacer-effect}b for a core radius of $r_{1} = 70$ nm. For the geometrical parameters chosen, the splitting of spectral peaks located at wavelengths of $\lambda = 394$ and $407$ nm is pronounced in the absence of an intermediate layer ($\ell_{\text{s}} = 0$). However, the splitting significantly decreases in the presence of passive spacer with thicknesses of $\ell_{\text{s}} = 7$ and $14$ nm, and completely disappears at $\ell_{\text{s}} = 21$ nm. Note also the distinctions in the overall behavior of the absorption spectra shown in Figs. \ref{fig:spheres-spacer-effect}a and \ref{fig:spheres-spacer-effect}b for some very different particle sizes. This is to be expected, since an increase in the particle size modifies significantly its effective polarizability and essentially enhances the cross section contribution of the quadrupole and octupole plexcitonic resonances.

\begin{figure*}[t]
	\centering\includegraphics[width=0.98\textwidth]{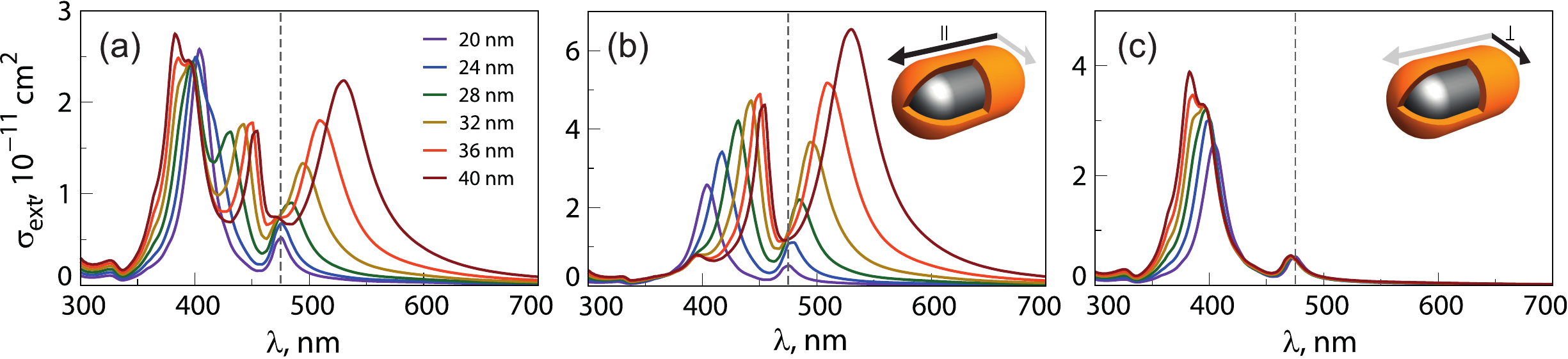}
	\caption{Extinction cross sections of round-ended Ag/TC nanorods in an aqueous solution. The length, $L$, of silver core varies from $20$ to $40$ nm as indicated in the legend; the core diameter is $D = 20$ nm; the J-aggregate shell thickness is $l_{\text{J}} = 1$ nm. (a) Extinction cross sections of the rod averaged over its possible orientations. Panels (b) and (c) show extinction cross sections for light polarized (b) parallel and (c) perpendicular to the axis of the rod. At \textit{L} = 20 nm, the rod transforms into a spherical nanoparticle. Dashed vertical line shows the peak position of TC J-aggregate.}
	\label{fig:AgTCSphRod}
\end{figure*}

The general trend of the plexcitonic coupling, consisting in the rapid fall of the efficiency with the bigger distance between the plasmonic and excitonic subsystems, also holds true for three-layer nanospheres consisting of a gold core, a passive dielectric spacer, and an outer J-aggregate shell. This is clearly demonstrated in Fig. \ref{fig:spheres-spacer-effect}c and \ref{fig:spheres-spacer-effect}d for particles with the outer layer of J-aggregates of the TDBC dye. As $\ell_{\text{s}}$ thickness of the organic TMA spacer increases, the spectral dips located at wavelengths of $\lambda = 530$ and $580$ nm gradually diminish, while a dip near $\lambda = 580$ nm practically vanishes. It is seen that the behavior of the spectral curves in Figs. \ref{fig:spheres-spacer-effect}c and \ref{fig:spheres-spacer-effect}d differs substantially from the case shown in Figs. \ref{fig:spheres-spacer-effect}a and \ref{fig:spheres-spacer-effect}b for the metalorganic Ag/TMA/OC system. This stems from the considerable difference in the dielectric properties of the Ag and Au metal cores, as well as of the OC and TDBC dye J-aggregates, so that plexcitonic resonances of Ag/TMA/OC and Au/TMA/TDBC systems appear in different parts of the visible spectrum.

Note also that all our calculations in this section have been performed using the standard isotropic model \eqref{eps-J} for the description of the permittivity of molecular dye aggregates. Taking into account the anisotropy effects of the outer excitonic shell of a three-layer metalorganic nanoparticle can change the specific spectral behavior of the absorption cross section for the given geometrical parameters of a particle. Nevertheless, the outlined general trend, where the near-field plexcitonic coupling diminishes when the excitonic and plasmonic components are spatially separated, certainly holds.

\section{Effects of plexcitonic coupling in nanorods, nanoplatelets and their dimers}

\subsection{Spectral behavior of plexcitonic nanorods with varying their lengths and number of spacer layers}

Due to the presence of longitudinal and transverse plasmon resonances in elongated metal nanoparticles, spectra of hybrid metalorganic nanorods have their own significant features compared to plexcitonic nanospheres. In a general case, both transverse and longitudinal plasmon-polaritons can interact with the Frenkel exciton. The positions of the maxima of the longitudinal and transverse resonances are determined by the ratio of the longitudinal and transverse rod dimensions, $L_{\parallel}/L_{\perp}$, and can be very far apart from each other on the frequency (or wavelength) scales. Meanwhile, the strongest coupling usually occurs between an exciton and a plasmon with the closest resonance energies. Accordingly, in the case of exact resonance $\omega^{\parallel}_{\text{pl}} = \omega_{\text{ex}}$ or $\omega^{\perp}_{\text{pl}} = \omega_{\text{ex}}$ the effects of strong coupling of an exciton with one of the plasmonic modes (such as, for example, induced transparency) can occur similarly to the plexcitonic nanospheres considered above. This directly follows from the available experimental data \cite{Wurtz2007, Yoshida2009b, Shapiro2015, Ni2010, Melnikau2016, Wu2021, Guo-Wu2021, Dey2023} and calculations \cite{Shapiro2015, Kondorskiy2015} carried out by the FDTD method.

Figure \ref{fig:AgTCSphRod} shows the results of calculations of the extinction spectra of an Ag/TC nanorod. As its length $L$ increases, field modes excited by the incident light polarized along the axis of the nanorod evolve from the weak plexcitonic coupling regime to the strong coupling regime. Modes excited by polarization of external radiation perpendicular to the nanorod axis stay in the weak coupling regime in all cases considered here. A characteristic feature of such spectra is that when the ratio, $L_{\parallel}/L_{\perp}$, of length to diameter of the rod changes, the positions of the maxima of plexcitonic resonances corresponding to the longitudinal and transverse modes shift in significantly different ways (see Figs. \ref{fig:AgTCSphRod}b and \ref{fig:AgTCSphRod}c). As the ratio $L_{\parallel}/L_{\perp}$ increases, the peaks of plexcitonic resonances for the longitudinal mode shift towards the long-wavelength region of the spectrum, and for the transverse mode this shift is directed towards short wavelengths. Meanwhile, the plexcitonic peaks for the longitudinal mode are shifted much more than the peaks corresponding to the transverse mode. As can be seen from Figures \ref{fig:AgTCSphRod}b and \ref{fig:AgTCSphRod}c, the magnitudes of the maxima of these peaks also change differently. These features are directly reflected in the spectral behavior of extinction cross sections averaged over the orientations of the rod in space.

\begin{figure}[b]
	\centering\includegraphics[width=0.9\linewidth]{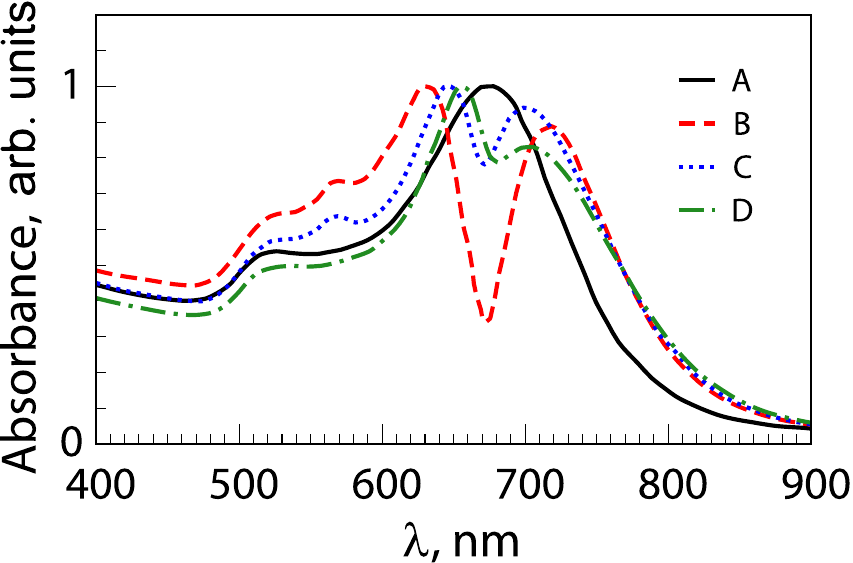}
	\caption{Absorption spectrum \cite{Yoshida2009b} of a colloidal solution containing multilayer nanorods, Au/TMA/ (PSS/PDDA)$_n$/Thia(Ph), with an outer J-aggregate shell of the Thia(Ph) dye (see Table \ref{table-Jaggr}). The black solid curve (A) is a bilayer Au/TMA particle. The colored curves correspond to multilayer particles with different numbers of intermediate layers of PSS/PDDA polyelectrolyte: $n=0$ (red dashed curve B), $n=1$ (blue dotted curve C), $n=2$ (green dash-dotted curve D).}
	\label{fig:multilayered_nanorods}
\end{figure}

Further we demonstrate in Fig. \ref{fig:multilayered_nanorods} one more important feature of plexcitonic coupling for hybrid multilayer nanorods synthesized in \cite{Yoshida2009b} with a spacer layer representing a monolayer of TMA precipitate and several  layers ($n=0$, $1$, $2$) of PSS/PDDA [poly(sodium 4-styrenesulfonate) / poly(diallyldimethylammonium)chloride] polyelectrolyte. It consists of a significant dependence of the absorption efficiency of such multilayer system on the number of intermediate passive dielectric layers between the plasmonic component (rod core) and its outer excitonic shell. As follows from the Fig. \ref{fig:multilayered_nanorods}, the spectral behavior is distinguished by  a decrease in the depth of the spectral dip near the resonance, $\lambda = 680$ nm, observed with an increase in the number $n$ of intermediate layers of the PSS/PDDA polyelectrolyte. Such a behavior is due to the weakening of the efficiency of the near-field coupling between the plasmon localized in the core and the Frenkel exciton in the outer shell of a hybrid nanorod.

\subsection{Manifestation of optical chirality in extinction and circular dichroism spectra}

There have been a number of recent papers on the optical chirality effects in metalorganic nanostructures. These studies are motivated to a large extent by the considerable applied medical and pharmaceutical interest in the effects of light interaction with the so-called enantiomers, i.e. isomers that are mirror symmetrical with respect to each other. The great practical importance of such isomers is due to the fact that they perform different functions and carry out essentially different biological activities. Optical detection and separation of enantiomers, as well as control of photochemical reactions involving them, are topical problems in modern nanophotonics (see \cite{Solomon2020} and references therein). On the other hand, in most of the past works on plexcitonic nanoparticles there has been no discussion of their chiral properties, and studies of the fundamental features of the plexcitonic coupling in chiral composite particles have intensified only in the past few years.

In particular, a series of experimental studies (see, e.g., \cite{Wu2021, ZhuWu2021, He-Guo2023, Kumar2023, Cheng-Yang2023}) published recently have been devoted to the studies of the effects of plexcitonic coupling in chiral Metal/J-aggregate nanoparticles. Note that the chirality of a metalorganic nanosystem stem from certain fundamentally different factors: the chirality of the plasmonic core \cite{Cheng-Yang2023}; the chirality of aggregates in the organic shell \cite{Wu2021}; and the chiral geometry of a complex system consisting of several achiral particles \cite{ZhuWu2021}. Whenever a chiral molecular aggregate is located near the surface of a metal nanostructure, near-field plasmon--exciton interaction leads to the appearance of circular dichroism in the vicinity of the plasmon resonance frequency, even if the metal nanostructure itself is achiral. This plexcitonic phenomenon creates an opportunity for endowing achiral plasmonic nanostructures with optical activity.

\begin{figure}[h]
	\centering\includegraphics[width=0.94\linewidth]{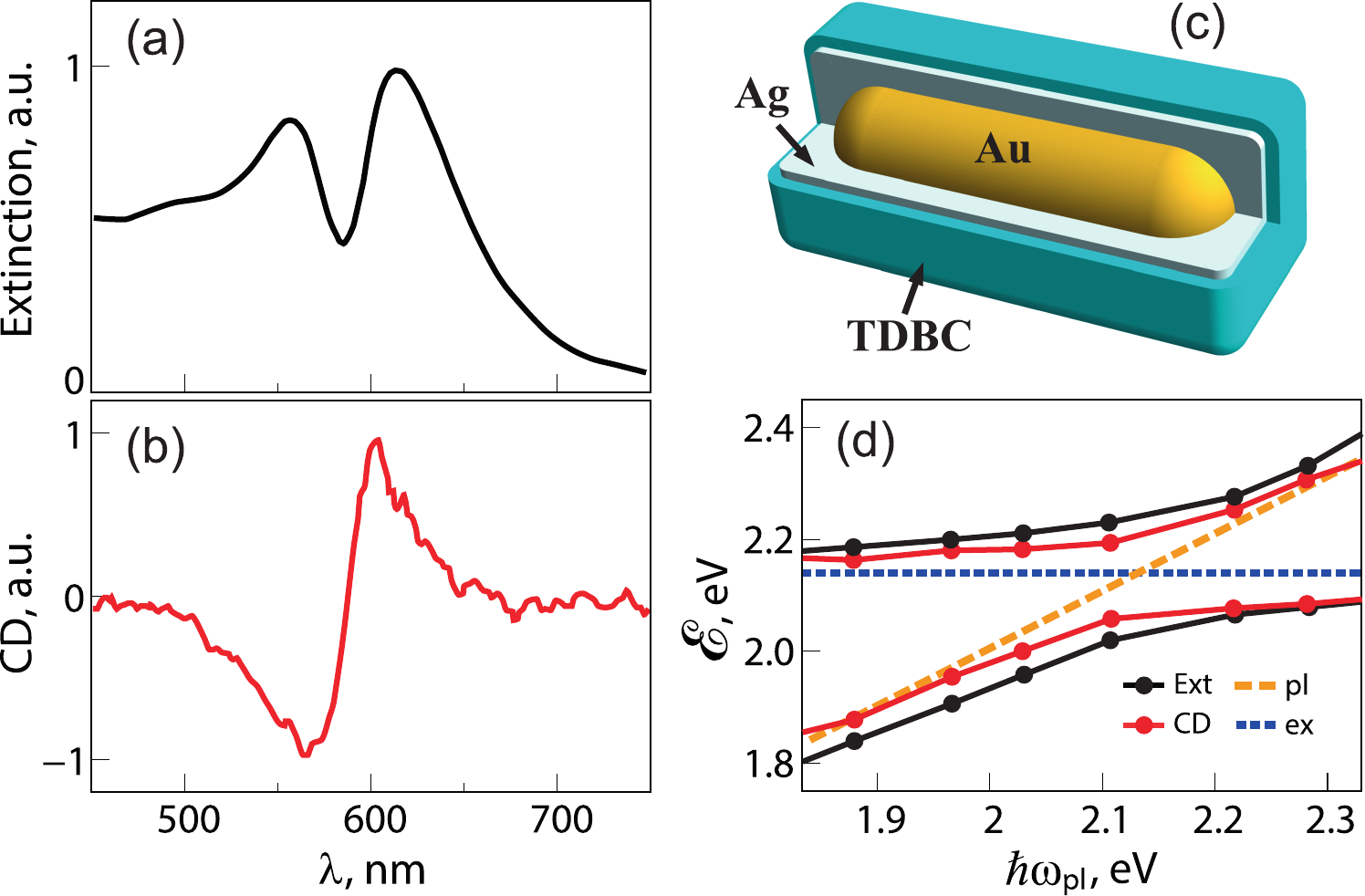}
	\caption{Extinction (a) and circular dichroism (b) spectra of Au/Ag nanocuboid coated with the chiral J-aggregate of the TDBC dye \cite{Wu2021}. (c) Schematic representation of a particle. (d) Energies of hybrid modes of a composite nanoparticle (black and red dots connected by solid curves) depending on the spectral position, $\hbar\omega_{\text{pl}}$, of plasmon resonance of the Au/Ag core (orange dashed line). The blue dashed line is the energy of the J-band maximum, $\hbar\omega_{\text{ex}}$, of the TDBC dye.}
	\label{fig:Plexcitonic_Chirality}
\end{figure}

In this context, the authors of \cite{Wu2021} have recently studied circular dichroism and light extinction spectra for Au/Ag composite nanocuboids coated with chiral J-aggregates of the TDBC dye, the monomers of which do not exhibit optical activity (see Fig. \ref{fig:Plexcitonic_Chirality}a and \ref{fig:Plexcitonic_Chirality}b). The plasmon resonance position of the two-component metal core of the Au/Ag nanocuboid was controlled by changing the thickness of the silver shell of the gold rod (see Fig. \ref{fig:Plexcitonic_Chirality}c). This allowed the authors of \cite{Wu2021} to determine experimentally the dependence of the energy of hybrid plexcitonic modes on the magnitude of the spectral detuning of the resonances of the isolated metal and J-aggregate subsystems. The dependences obtained in \cite{Wu2021} are presented in Fig. \ref{fig:Plexcitonic_Chirality}d. Formation of hybrid modes was confirmed by the measurements of the positions of peak maxima not only in the extinction spectra, but also in circular dichroism. Note that the circular dichroism of the hybrid particle was not owed just to an increase in the intensity of the dichroic response of the chiral TDBC J-aggregate due to local field enhancement, but was caused specifically by the formation of new hybrid states in the strong coupling regime. The observed splitting energy of the plexcitonic peaks at the quasi-crossing point differed for the extinction and circular dichroism spectra and, according to data \cite{Wu2021} are equal to $\Delta \mathcal{E} = 214$ meV and $\Delta \mathcal{E} = 136$ meV, respectively. This difference in the spectral profiles of the two types (see Fig. \ref{fig:Plexcitonic_Chirality}a and \ref{fig:Plexcitonic_Chirality}b) was explained \cite{Wu2021} within the framework of the quasistatic approximation. Note that, compared to extinction spectra, circular dichroism spectra contain narrower bands, and the different signs of the signals (positive and negative) of the two hybrid modes ensure their simpler and more reliable experimental registration.

\subsection{Induced transparency of plexcitonic nanoplatelets}

\begin{figure}[h]
	\centering\includegraphics[width=0.94\linewidth]{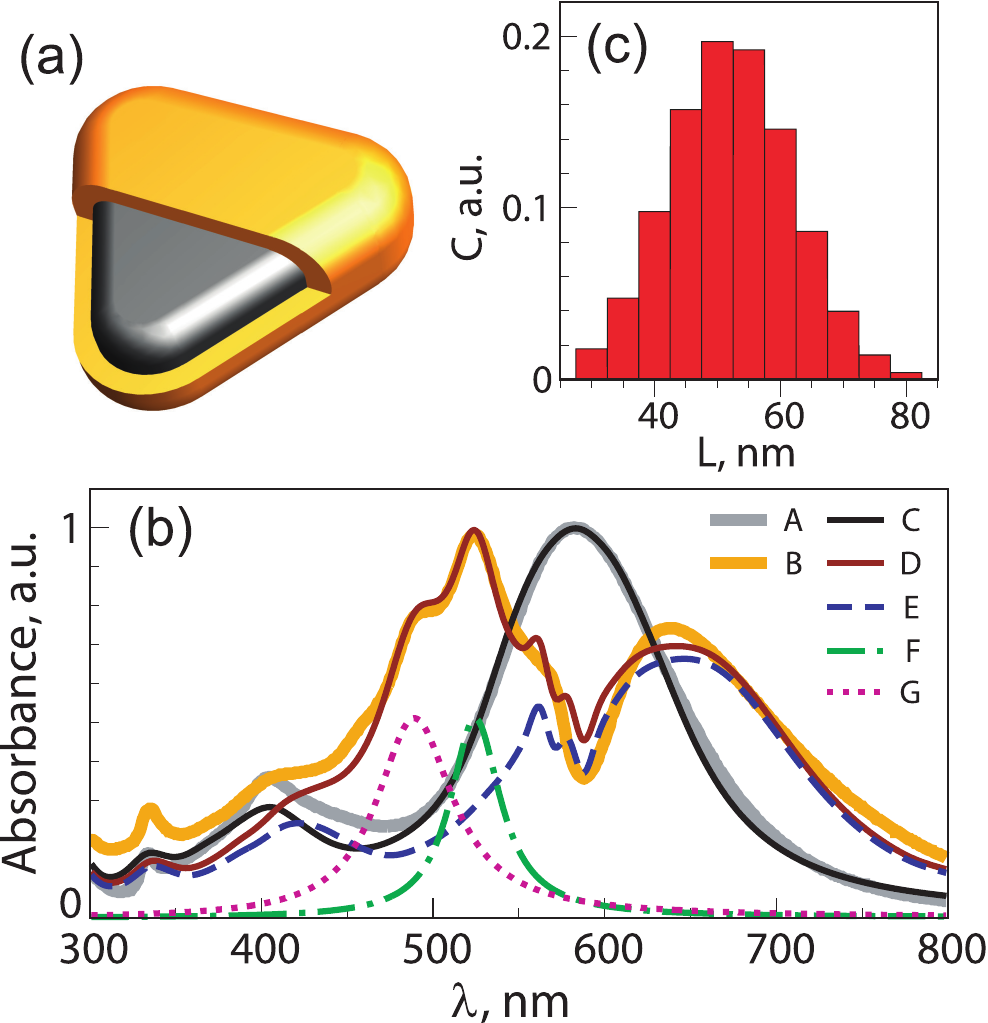}
	\caption{(a) Schematic view of a bilayer metalorganic nanoprism. (b) Experimental data \cite{DeLacy2015} and results of FDTD calculations of absorption spectra of bare silver nanoprisms and two-layer Ag/PIC nanoprisms. Thick gray (A) and orange (B) curves are experimental data for the spectra of bare and coated nanoprisms, respectively. Black (C) and wine (D) curves show the calculation results for bare and coated nanoprisms. The wine curve is obtained taking into account the absorption contributions of the PIC dye monomer and dimer, as well as the size dispersion of nanoprisms. Blue dashed curve (E) is the calculation of absorption by two-layer nanoprisms. The green dash-dotted curve (F) and the pink dotted curve (G) are contributions to the absorption from the the monomer and dimer. (c) Histogram of the concentration distribution, $C$, of nanoprisms versus their side length, $L$.}
	\label{fig:DeLacyPrisms}
\end{figure}

The effect of induced transparency described in section \ref{Sect-Induced-Transparency} is most prominently manifested in a number of experimental works on the optical properties of plexcitonic nanoplatelets: nanoprisms and nanodisks. Figure \ref{fig:DeLacyPrisms} shows experimental data \cite{DeLacy2015} on the light absorption spectra of silver triangular nanoprisms coated with a thick layer of J-aggregates of the PIC dye. The choice of a nanoprism as a core makes it possible to satisfy the resonance condition between the plasmon of the core and the exciton of the J-aggregate shell by varying the length of the edge of the prism base. A quantitative measure of the efficiency of the plexcitonic coupling is the experimentally determined energy splitting between the upper and lower plexcitonic branches. The magnitude of this splitting, $\hbar\Omega_{\text{R}} = 207$ meV, was determined for such a length of the prism base edge that led to the coincidence (resonance) at $\lambda = 582$ nm of unperturbed plasmon, $\hbar\omega_{\text{pl}}$, and exciton, $\hbar\omega_{\text{ex}}$, energies. In the theory of nonadiabatic transitions between electronic terms of a diatomic molecule, the fulfillment of the resonance condition ($\hbar \omega_{\text{ex}} = \hbar \omega_{\text{pl}}$) corresponds to the crossing point of diabatic potential energy curves. The splitting of hybrid modes of such composite nanoplatelets indicates the presence of a strong plexcitonic coupling regime (see section \ref{Sect-coupling-regimes}) in the phenomenon of induced transparency.

In the solution prepared in the experimental work \cite{DeLacy2015}, apart from the silver nanoprisms with a fairly wide range of sizes coated with J-aggregates, there were also monomers and dimers of the dye. Therefore, the resulting absorption spectra measured in a colloidal solution are determined as a sum, $\sigma(\omega )=\sum\limits_{j}c_{j}\sigma _{j}(\omega)$, of absorption cross sections $\sigma_{j}(\omega)$, of its individual components $j$ with weighting factors proportional to their concentrations $c_{j}$ in solution. We calculate the absorption cross sections of nanoprisms with sizes varying from $L = 30$ nm to $L = 75$ nm with a step of 5 nm. The height of the prism is $h = 10$ nm, the thickness of the organic shell is $l_{\text{J}} = 8$ nm. Parameters associated with the J-aggregate of the PIC dye are given in Table \ref{table-Jaggr}.

\begin{figure}[b]
	\centering\includegraphics[width=0.98\linewidth]{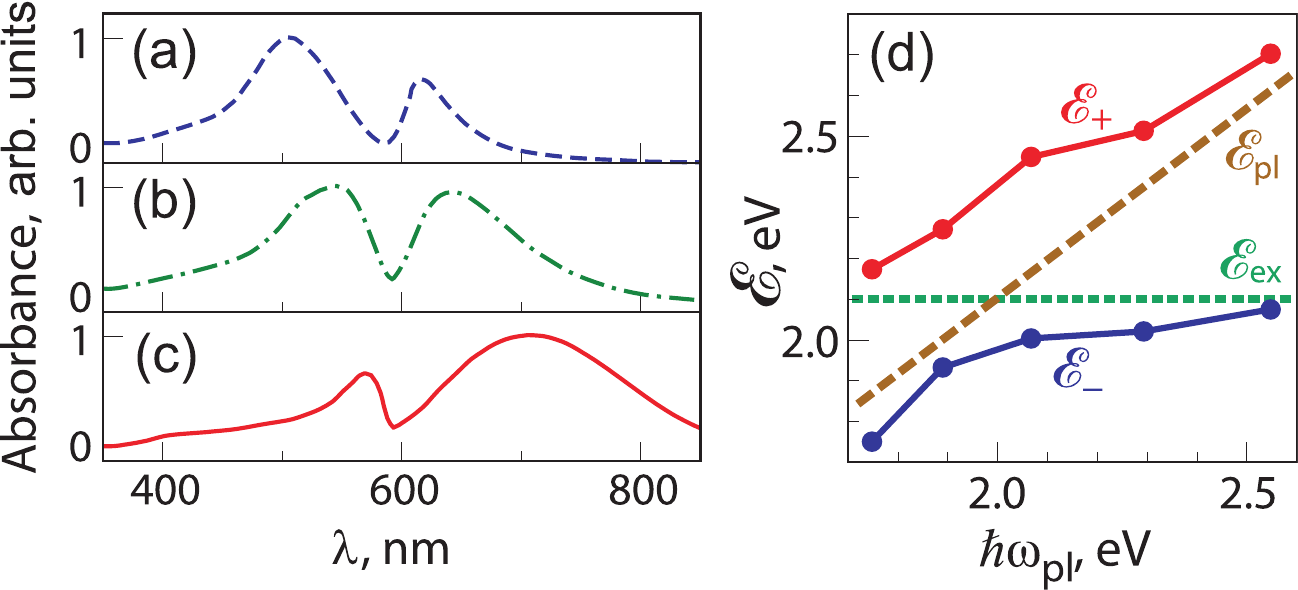}
	\caption{(a)--(c) Absorption spectra of silver nanoprisms coated with the J-aggregate of the TDBC dye in an aqueous solution at different edge lengths: (a) 60 nm, (b) 100 nm, (c) 140 nm \cite{Balci2013}. (d) Behavior of the energy branches $\mathcal{E}_{+}$ (red curve) and $\mathcal{E}_{-}$ (blue curve) of the Ag/TDBC hybrid modes depending on the plasmon energy, $\mathcal{E}_{\text{pl}}=\hbar \omega_ {\text{pl}}$, of a bare prism (dark yellow line). The green horizontal line is the energy, $\mathcal{E}_{\text{ex}}=\hbar \omega_{\text{ex}}$, of the excitonic transition in the J-band of the TDBC dye aggregate.}
	\label{Fig22_prism_ultrastrong}
\end{figure}

The induced transparency (i.e. the presence of a dip in the light absorption spectrum) was also observed in \cite{Balci2013}, where the object demonstrating this phenomenon was a triangular silver nanoprism coated with a J-aggregate of the TDBC dye (see Fig. \ref{Fig22_prism_ultrastrong}a--\ref{Fig22_prism_ultrastrong}c). In the experiment, the wavelength of the localized plasmon was tunable in the range from 400 to 1100 nm. Reconstruction of the energy branches $\mathcal{E}_{+}$ and $\mathcal{E}_{-}$ of the hybrid modes of the Ag/TDBC plexcitonic nanoprism based on measurements of the dependence of the spectral maxima positions made it possible to determine the characteristic value of the constant $V = \hbar g$ of the plexcitonic coupling for this system. Under conditions of the resonance of the plasmonic and excitonic subsystems, it corresponds to the energy splitting of hybrid modes  $\Delta \mathcal{E} = \operatorname{Re}\{\mathcal{E}_{+} - \mathcal{E}_{-}\} = \hbar \Omega_{\text{R}} = 2 V = 400$ meV (see Fig. \ref{Fig22_prism_ultrastrong}d). This energy splitting was equal to 19 \% of the excitonic transition energy in the J-band of the TDBC aggregate. Given the semiempirical criterion in section \ref{Sect-coupling-regimes}, this corresponds to the ultrastrong plexcitonic coupling regime. Such a significant value of coupling energy, $V = \hbar g$, is the result of strong localization of the electric field near the sharp corners of the nanoprism. A comparison of the results of spectral peak splitting energies presented in Table \ref{tab:RabiSplitting}, obtained by different authors in studies of the strong and ultrastrong plexcitonic coupling regimes for metalorganic nanosystems, shows that the author of Ref. \cite{Balci2013} had obtained one of the highest values of $\hbar \Omega_{\text{R}}$.

In order to study possible plexcitonic coupling regimes, we have performed a computer simulation of the optical spectra of silver and gold nanoprisms and nanostars coated with dye aggregates with different optical properties (TC, PIC, TDBC) \cite{Lam-Kond-Leb2019}. The calculations were carried out using the FDTD method combined with the standard scalar isotropic model \eqref{eps-J} for the description of the excitonic shell. At the same time, typical dips were reproduced in the absorption spectra of the systems under study, consistent with the effect of induced transparency. 

\begin{figure}[t]
	\centering\includegraphics[width=0.94\linewidth]{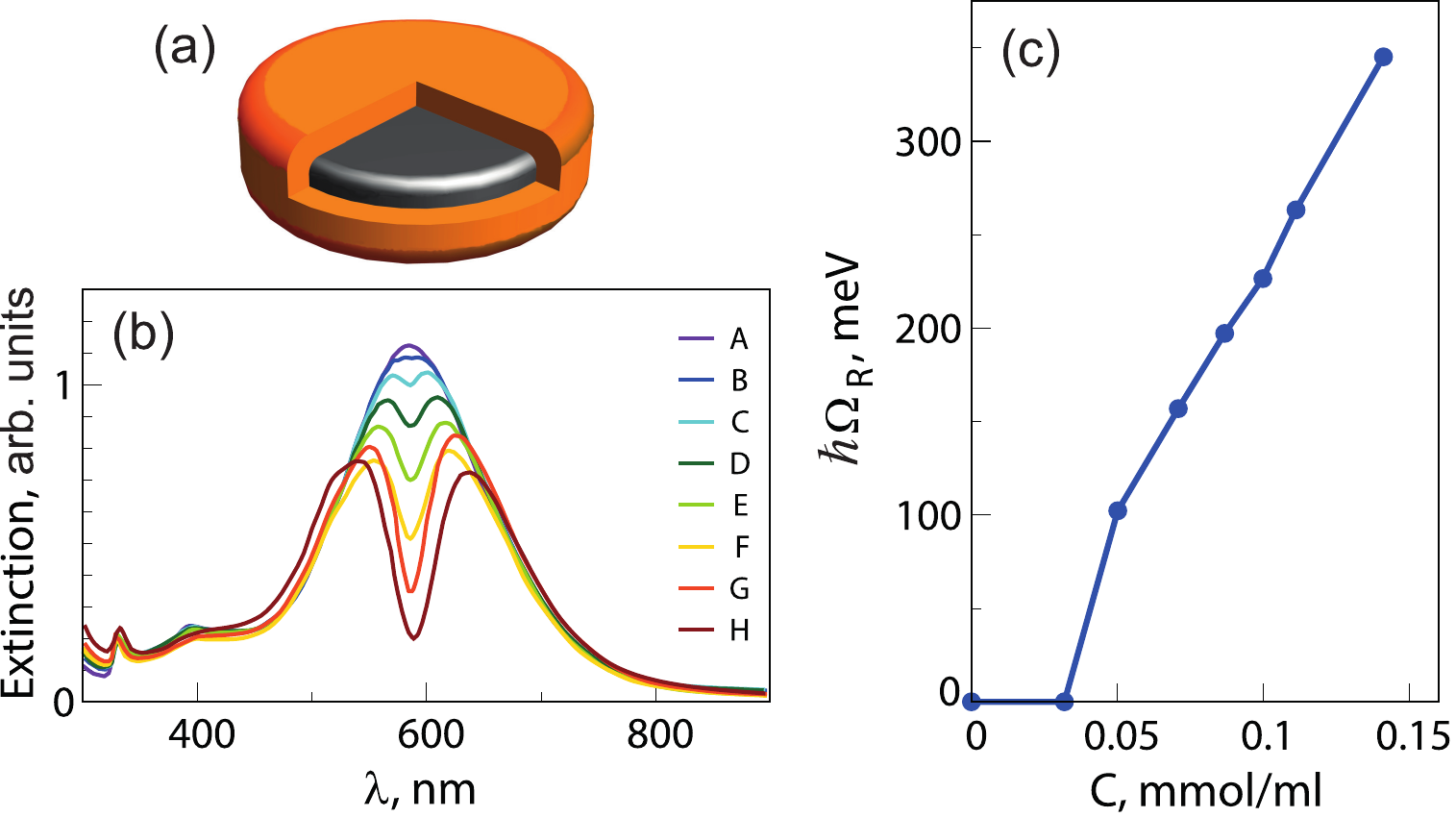}
	\caption{Demonstration of the possibility of controlling the plexcitonic coupling energy in the Ag/TDBC hybrid system by changing the amount of aggregated dye in the outer excitonic layer. (a) Schematic illustration of a bilayer metalorganic nanodisk. (b) Extinction spectra of plexcitonic nanodisks at different dye concentrations in a colloidal solution of Ag/TDBC nanoparticles. The order of the curves from top to bottom, designated A, B, C,..., H, corresponds to the increasing dye concentration. Curve A shows spectrum of a bare silver disk. (c) The magnitude of the energy splitting, $\hbar \Omega_{\text{R}}=2\hbar g$, of the upper and lower states of the hybrid system depending on the concentration $C$ of the TDBC dye in the solution.}
	\label{fig:balci2019}
\end{figure}

The effect of induced transparency was also studied in \cite{Balci2019} using silver nanodisks coated with a J-aggregate layer of TDBC-dye (see Fig. \ref{fig:balci2019}a). During the experiments, the plexcitonic coupling strength was varied and a very high value of the hybrid modes splitting energy, $\hbar\Omega_{\text{R}} > 350$ meV, was achieved indicating a transition to the ultrastrong plexcitonic coupling regime (see Sect. \ref{Sect-coupling-regimes}). The disk diameter was selected in such a way that the frequencies of the plasmonic and excitonic resonances coincided ($\lambda_{\text{ex}}^{\text{TDBC}}=585$ nm). A clear experimental sign of the formation of plexcitonic nanoparticles was a pronounced dip in their extinction spectra.   

Experimental data \cite{Balci2019} on extinction spectra are shown in Fig. \ref{fig:balci2019}b for different dye concentrations. The authors controlled the amount of the aggregated dye molecules attached to silver nanodisks by controlling the concentration of dye molecules in the colloidal solution. It can be seen from Fig. \ref{fig:balci2019}b that such a variation of concentration  significantly affects the depth of the dip in the extinction spectrum of Ag/TDBC nanodisks and causes a transition from the weak coupling regime (at low concentrations) to the strong, and then ultrastrong coupling regimes (at high concentrations). The experimental results displayed in Fig. \ref{fig:balci2019}c show an increase in the Rabi splitting energy with the increasing dye concentration after reaching the strong coupling regime. The above is in accordance with the theory described in Sect. \ref{Sect-Coupl-Ocs-Model}. Thus, the main result of the work \cite{Balci2019} was to demonstrate the possibility of effectively controlling the process of a J-aggregate layer formation on the surface of a nanodisk and controlling the amount of plexcitonic coupling energy and, accordingly, the extinction and transmission spectra of hybrid nanoplatelets.

\begin{figure*}[t]
	\centering\includegraphics[width=0.98\textwidth]{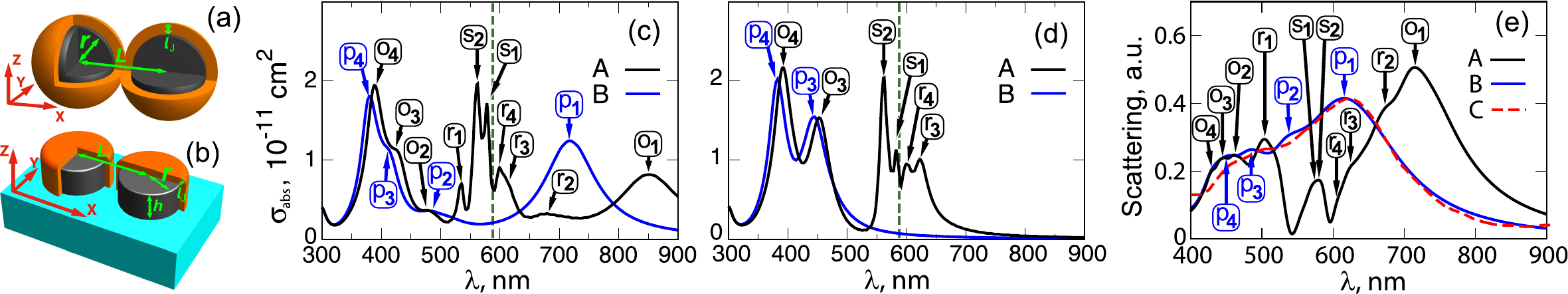}
	\caption{Demonstration of the spectral band replication effect using plexcitonic dimers consisting of a pair of silver-core nanoparticles coated with a TDBC dye J-aggregate shell. (a, b) Schematic view of the dimers consisting of (a) two hybrid nanospheres and (b) two hybrid nanodisks on a glass substrate. (c)--(e) Results of calculations~\cite{Kond-Leb_OE2019} of the optical spectra of plexcitonic dimers in an aqueous solution (black curve, A) in comparison with that for bare silver dimers (blue curve, B). (c, d) Light absorption spectra of Ag/TDBC nanospheres for two values of the distance between their centers: (c) $L=18$ nm and (d) $L=22$ nm; the core radius and the shell thickness were $r=10$ nm and $l_{\text{J}}=3$ nm. (e) Dark-field spectra for Ag/TDBC nanodisks compared with the experimental data~\cite{Gunnarsson2005} for bare silver disks (red dashed curves, C). The radius and height of the silver disks are $r=40$ nm and $h=40$ nm; the distance between their centers is $L=110$ nm, the shell thickness is $l_{\text{J}}=12$ nm. Vertical green dashed lines indicate the position of the J-band center. Arrows indicate: $\mathsf{p_i}$ are ''plasmonic'' peaks of the bare dimer; $\mathsf{o_i}$ and $\mathsf{r_i}$ are plexcitonic peaks associated with the ''original'' and ''replication'' bands of the dimer; $\mathsf{s_i}$ are resonances of the J-aggregate shell of the dimer.}
	\label{fig:replication_dimers}
\end{figure*}

\subsection{Spectral band replication of plexcitonic nanoparticle dimers}

An unusual manifestation of the strong plexcitonic coupling regime is the effect of replication of spectral bands. As was shown in \cite{Kond-Leb_QE2018,Kond-Leb_OE2019}, the spectra of pairs of closely spaced and partially overlapping bilayer metalorganic nanostructures contain twice the number of plexcitonic spectral bands compared to cases when the organic shell is absent. These plexcitonic bands can be divided into two groups: the ''original'' bands ($\mathsf{o_i}$), which accurately reproduce the plasmonic peaks ($\mathsf{p_i}$), and their ''replicas'' ($\mathsf{r_i}$), with a specific relative position and intensity distribution. Figure~\ref{fig:replication_dimers} explains the essence of the phenomenon using pairs of two-layer Ag/TDBC nanoparticles of spherical (Figs.~\ref{fig:replication_dimers}c--\ref{fig:replication_dimers}d) and disk-like (Fig.~\ref{fig:replication_dimers}e) shapes by way of an example.

The positions of the $\mathsf{o_i}$ bands in the spectra of bilayer dimers are close to the positions of the $\mathsf{p_i}$ bands of plasmonic dimers. There are also narrow $\mathsf{s_i}$ spectral peaks, which should be attributed to excitonic resonances in the J-aggregate shell. Significantly, plexcitonic dimers have an additional group of spectral bands, $\mathsf{r_i}$ (see Figs.~\ref{fig:replication_dimers}c--\ref{fig:replication_dimers}d). These bands are always located on the opposite side of the $\mathsf{o_i}$ bands relative to the excitonic resonance of the dye J-aggregate. The relative positions, widths and intensities of the $\mathsf{p_i}$ plasmonic bands are quite accurately reproduced by the $\mathsf{o_i}$ bands and, with some distortions, are repeated by the $\mathsf{r_i}$ bands on the other side of the excitonic resonance. This becomes particularly noticeable if we trace how changes in the positions of the $\mathsf{p_i}$ bands, which occur when the distance between nanostructures is varied, are reflected in the positions of the $\mathsf{o_i}$ and $\mathsf{r_i}$ bands. Such a behavior in the optical spectra of plexcitonic dimers composed of two hybrid Ag/TDBC metalorganic nanospheres and nanodisks of the core-shell type were demostrated in~\cite{Kond-Leb_OE2019} by the FDTD simulations. Spectral band replication phenomenon was also explained qualitatively and quantitatively by using the effective multilevel Hamiltonian model (see Sect. \ref{Sect-multilevel-hamiltonian}). This Hamiltonian includes the interaction of quasi-degenerate excitonic modes of the organic shells with multiple modes of a plasmonic dimer composed of two bare metal nanospheres or nanodisks.

It should be noted that the relationship between the occurrence of the replication phenomenon and the emergence of a strong plasmon--exciton interaction regime was studied in~\cite{KondorMek2022} by analyzing the change in the spectra of metalorganic nanoparticle dimers upon the introduction of a passive dielectric spacer between their cores and shells. With an increase in the thickness of this spacer, the ''replicas'' of plasmon resonances approach the position of the J-aggregate band and disappear. This confirms the validity of the spectral bands classification in terms of plexcitonic ''originals'' and ''replicas'' of plasmon resonances. A peculiar feature of the spectral band replication is that it is observed when the molecular aggregate J-band lies at some distance relative to the plasmon resonance peak toward the long-wavelength side. This distinguishes it from the traditional effects of the dip in the absorption spectra of hybrid metalorganic nanoparticles that occurs when the resonance frequencies of the plasmonic and excitonic subsystems are close or coincide.

\section{Effects of anisotropy of excitonic shell and their influence on light absorption and scattering spectra}

\subsection{Optical anisotropy of molecular J-aggregates}

Theoretical description of light absorption and scattering spectra, as well as analysis of plexcitonic coupling regimes in Metal/J-aggregate and Metal/Spacer/J-aggregate hybrid nanostructures were previously performed almost always within the framework of various isotropic models. In the isotropic models, the outer layer of an organic dye is usually treated as an isotropic medium with an effective scalar local dielectric function of the resonance type (\ref{eps-J}). In many cases, such theoretical approaches make it possible to describe the optical spectra of plexcitonic nanosystems and adequately interpret one or another coupling regime. However, isotropic approaches turn out to be inapplicable for analyzing cases when the outer organic shell of the system has pronounced anisotropic dielectric properties.

It is well known that nano- and micrometer-sized structures consisting of large extended molecules are, as a rule, anisotropic. A striking example of dyes capable of forming molecular J-aggregates with pronounced anisotropic dielectric properties is pseudoisocyanine dye (PIC). The synthesis of molecular aggregates of various dyes including PIC, as well as studies of their morphology have been the subject of intense research for many years~\cite{Scherer1984, Misawa1994, Tani2007, Tani2012, Haverkort2014}. In particular, to clarify the location and orientation of the molecules composing the aggregate, the linear dichroism of the film of a highly oriented J-aggregate of pseudoisocyanine bromide was studied in \cite{Misawa1994}. Anisotropic behavior of the absorption and fluorescence spectra of fibril-like J-aggregates of pseudoisocyanine dyes in thin-film matrices was observed in \cite{Tani2007}. Experimental results for oriented PIC dye J-aggregates demonstrated a significant difference in the absorption spectra obtained for polarizations parallel and perpendicular to the J-aggregate orientation axis. Pronounced anisotropic absorption has also been demonstrated for J-aggregates of a number of carbocyanine dyes~\cite{Didraga2004,Pugzlys2006}. It has been shown that when light is polarized parallel to the alignment direction of these molecular aggregates, intense narrow absorption bands appear in their spectra.

Anisotropic effects in the processes of absorption and scattering of light by various nanoparticles have been studied extensively for many years given the important role these effects play in some technological and biological applications~\cite{Joannopoulos2010}. A modification of the Mie theory for the description of the scattering and extinction spectra of spherical nanoparticles coated with an optically anisotropic outer shell was proposed in \cite{Roth1973}. Within the framework of the quasistatic approximation, an analytical description of the optical spectra of ellipsoidal plasmonic particles covered with an anisotropic shell was developed in \cite{Ambjornsson2006}. Orientational effects in dipole-dipole interactions between neighboring dye molecules was considered recently using a coupled dipole model in a spherical metalorganic core--shell systems~\cite{Ru2018a}. The same work proposed a theoretical study of the impact of the concentration of dye molecules in the shell and coating uniformity on the process of light scattering. A thin-shell approximation of the Mie scattering problem for spherical core--shell and core--double-shell structures with radial anisotropy in the outer layer was offered in \cite{Ru2018b, Tang2021}. This approximation describes some orientational and anisotropic effects resulting from resonant dye-plasmon interactions. In addition, the authors in \cite{Ru2019, Tang2020} developed a model to estimate the effective dielectric function of an anisotropic layer of dye molecules adsorbed on a metal surface, and to describe the electromagnetic core--shell interaction in such systems. Recently we have clarified \cite{KML_OE2022,Kondorskiy2024} the role of orientational effects in the molecular self-assembling of the J-aggregate excitonic shell in the formation of the optical spectra of some core-shell and core--double-shell plexcitonic nanoparticles of various sizes, shapes and compositions.

\subsection{Features of spectra of plexcitonic systems with isotropic and anisotropic outer shells}

We now discuss the role of the anisotropy of the outer shell of ordered molecular J-aggregates in the formation of plexcitonic coupling and optical spectra of metalorganic nanoparticles. In \cite{KML_OE2022,Kondorskiy2024} we have performed theoretical studies of the effects of plasmon--exciton coupling and the behavior of light absorption and scattering spectra by hybrid nanoparticles of various shapes containing a metal core and an outer shell of molecular J-aggregates of organic dyes. As shown in \cite{KML_OE2022}, taking into account the anisotropy of the J-aggregate shell of a composite system makes it possible to reliably explain the available experimental data on the optical spectra of a number of studied metalorganic nanoparticles, which cannot be accurately described within the framework of the standard isotropic model (\ref{eps-J}) of the outer J-aggregate shell.

\begin{figure}[h]
	\centering\includegraphics[width=0.94\linewidth]{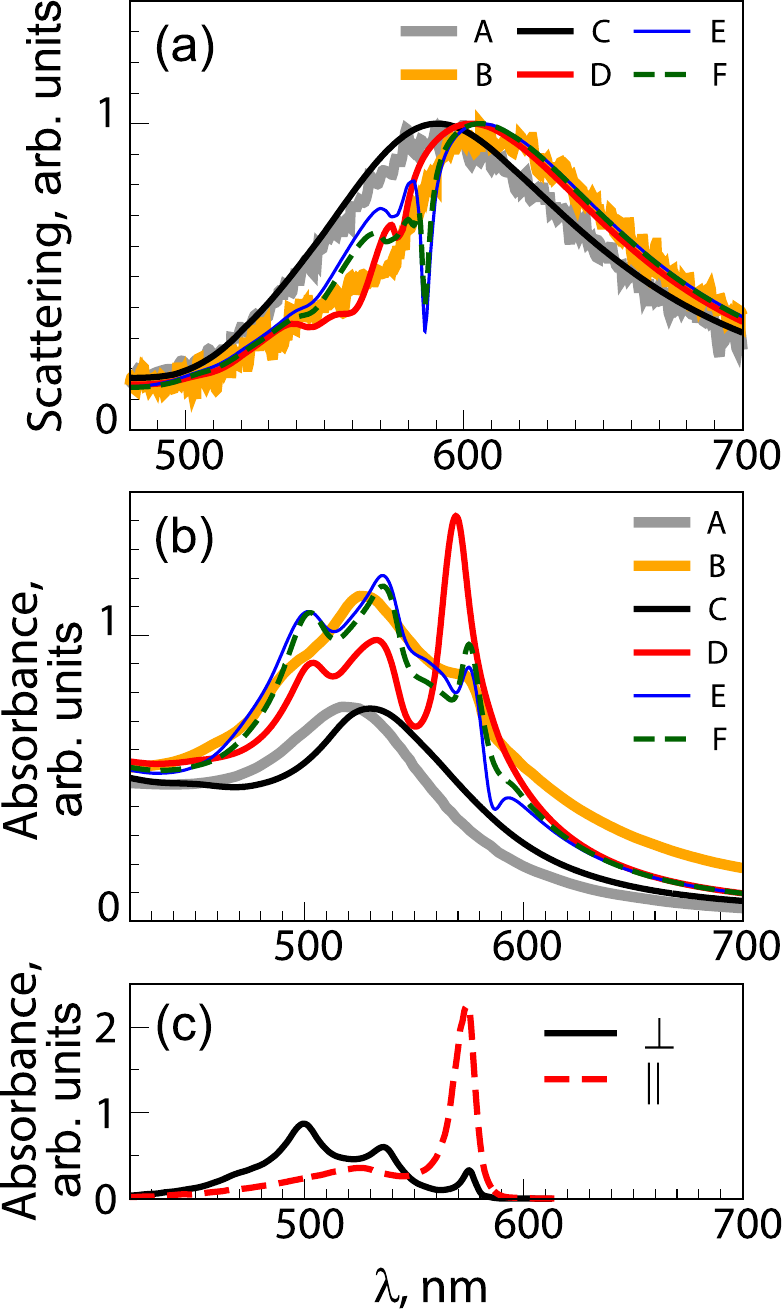}
	\caption{(a) Comparison of experimental and theoretical data on the scattering spectra of two-layer, Au/MUA, and three-layer, Au/MUA/PIC, nanospheres with a diameter of $D = 100$ nm. Thick gray (A) and orange (B) curves are experimental data for two-layer and three-layer particles. Results of calculations: black curve (C) is for two-layer particles; while red (D), thin blue (E), and green dashed (F) curves are for three-layer particles. Calculations have been performed for J-aggregates of the PIC dye in the outer shell which have (D) normal (D) and tangential (E) orientation relative to the surface, as well as using an isotropic (F) dielectric function (\ref{eps-J}). (b) Comparison of experimental~\cite{Yoshida2009a} and theoretical~\cite{KML_OE2022} data on the absorption spectra of two-layer, Au/TMA, and three-layer, Au/TMA/PIC nanospheres with a diameter of $D = 9.1$ nm. The curve symbols are the same as in panel (a). (c) Absorption spectra of a PIC dye J-aggregate with incident light polarized parallel (dashed red curve) and perpendicular (solid black curve) to the aggregate axis~\cite{Misawa1994}.}
	\label{fig:aniso_oe}
\end{figure}

For the illustration see Figs~\ref{fig:aniso_oe}a and \ref{fig:aniso_oe}b which demonstrate the optical spectra of hybrid three-layer Au/MUA/PIC and Au/TMA/PIC nanoparticles with gold cores and the same molecular aggregate of the PIC dye but with different passive dielectric MUA and TMA layers. The effect manifests itself in a qualitatively different shape of these spectra for the cases of normal and tangential orientation of aggregated molecules in the dye shell relative to the surface of the nanostructure. It was shown that the results of theoretical calculations are in good agreement with the available experimental data provided only that one takes into account the optical anisotropy of the shell with one clearly defined spatial orientation of the aggregates. The agreement was achieved for Au/MUA/PIC nanoparticles (Fig.~\ref{fig:aniso_oe}a) provided the PIC aggregate is oriented along the normal to the surface, and for Au/TMA/PIC particles (Fig.~\ref{fig:aniso_oe}b) -- when the PIC aggregate is oriented along the surface. Note here that according to the experimental results \cite{Yoshida2009a, Yoshida2009b, Uwada2007} the use of a number of optically passive organic substances (for example, TMA or MUA) as dielectric layers promotes the aggregation of dye molecules on the surface of nanoparticles.

A comparison of the results for Au/MUA/PIC and Au/TMA/PIC nanoparticles, which have intermediate MUA and TMA layers that are completely different in nature, indicates a strong dependence of the orientation of the molecular aggregate of a hybrid nanoparticle on the material of the surface on which the aggregate is assembled. This conclusion is consistent with modern chemical concepts that metal nanoparticles coated with MUA and TMA form oppositely charged particles~\cite{Kalsin2006,Kowalczyk2012}. Thus, it is possible to effectively control the optical properties of hybrid metalorganic nanostructures by selecting the appropriate optically passive spacer between the plasmonic core and the outer excitonic shell. Therefore the influence of the passive dielectric layer on the resulting spectra of the hybrid system depends not only on the specific value of its permittivity, $\varepsilon_{\text{s}}$, and thickness, $\ell_{\text{s}}$. In addition, these spectra are strongly affected by the physicochemical properties of the passive spacer material, since they determine the orientation of molecular aggregates in the process of the outer J-aggregate shell formation.

Note that the use of the isotropic scalar dielectric function (\ref{eps-J}) of the outer organic shell in the calculations can provide a reasonable explanation for some of the available experimental data. Analysis shows that if the effective oscillator strength of the excitonic transition in the J-band of a molecular aggregate is sufficiently small, then the optical spectra of metalorganic nanoparticles with a tangential orientation of the J-aggregate in the outer shell of the dye turn out to be quite close to the spectra of hybrid nanoparticles with an optically isotropic shell. In this case, the behavior of these spectra will differ significantly from the calculated spectra of the same hybrid nanoparticles, but with a normal orientation of the J-aggregate in the excitonic shell.

\begin{figure}[h]
	\centering\includegraphics[width=0.94\linewidth]{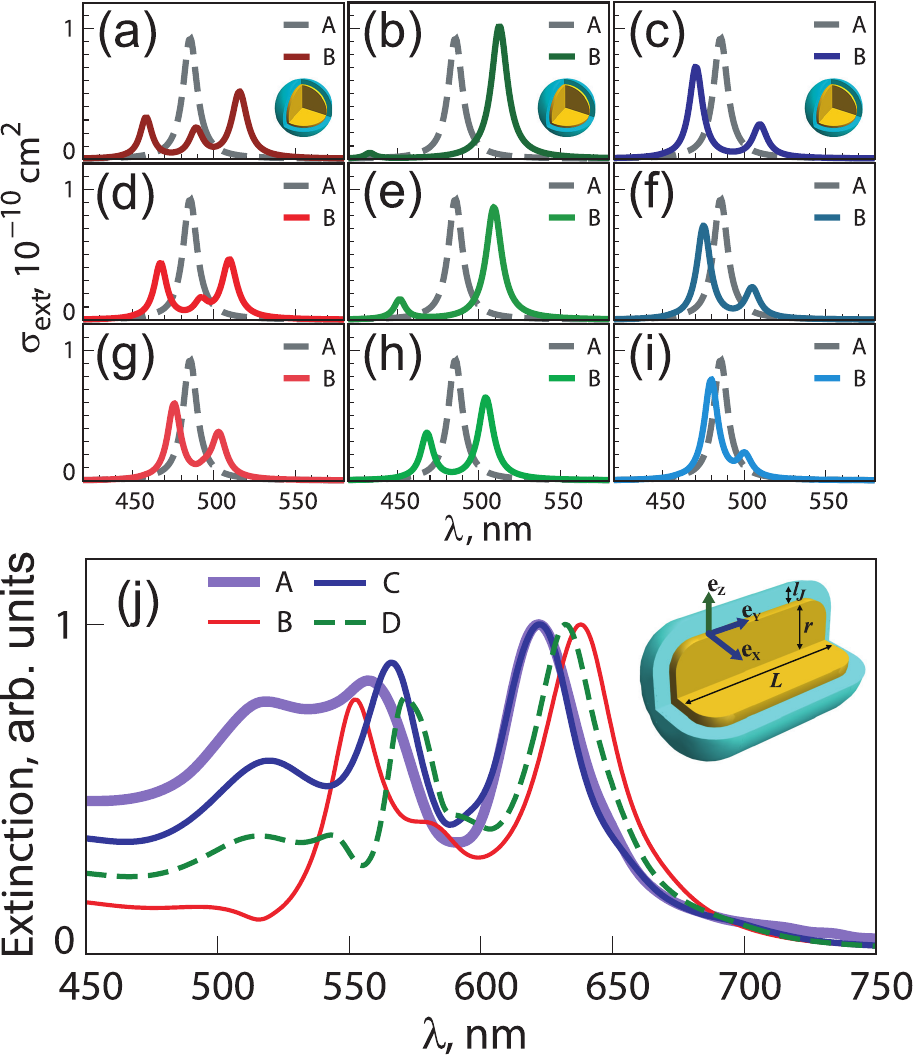}
	\caption{(a)--(i) Extinction spectra of a two-layer nanosphere with a gold core and a J-aggregate shell with different values of the effective oscillator strength, $f_{\text{J}}$, and different orientations of the J-aggregate relative to the metal surface~\cite{ Kondorskiy2024}. (a)--(c) Calculations performed at $f_{\text{J}} = 0.15$ using the dielectric functions for (a) isotropic shell, (b) normal, and (c) tangential orientations of J-aggregates in the dye shell. (d)--(f) and (g)--(i) are similar calculations at $f_{\text{J}} = 0.08$ and $f_{\text{J}} = 0.03$, respectively. Gray dashed curves (A) are calculations for a bare particle. Colored solid curves (B) are results for bilayer particles with a core diameter of $D=25$ nm and a shell thickness of $l_{\text{J}}=3$ nm. The dielectric function of a metal is described by the Drude formula with parameters corresponding to the plasmon resonance of gold~\cite{Stete2023}. Parameters of the J-aggregate: $\varepsilon^{\infty}_{\text{J}}$ = 1.7, $\omega_{\text{ex}}$ = 2.5 eV, $\gamma_{\text{ex} }$ = 0.05 eV. (j) Extinction spectra of TDBC-coated gold nanorod. Thick lavender curve (A) shows experimental data~\cite{Stete2023}. Red thin (B), blue (C), and green dashed (D) curves are calculations~\cite{Kondorskiy2024}. Calculations have been performed for different orientations of J-aggregates in the outer shell: normal (B) and tangential (C) to the surface, as well as using an isotropic (D) dielectric function (\ref{eps-J}). The nanorod radius is $r = 9.8$ nm, the length is $L=32$ nm, and the dye shell thickness is $l_{\text{J}}=3$ nm.}
	\label{fig:COL}
\end{figure}

The nature of the differences in the spectra of metalorganic nanoparticles with isotropic and anisotropic shells in the regimes of strong and ultrastrong plexcitonic coupling was studied in~\cite{Kondorskiy2024}. It was shown that the use of the isotropic dielectric function of the J-aggregate can lead to significant discrepancies between theoretical predictions and experimental spectra of hybrid nanoparticles. This discrepancy is especially noticeable when studying metalorganic structures with a shell of the J-aggregate of the TDBC dye, since the value of the effective oscillator strength of its excitonic band is quite high ($f\approx 0.22$). The energy of plexcitonic coupling (Rabi splitting) in such a system reaches hundreds of meV. In this case, calculations carried out using the isotropic model (\ref{eps-J}) for the dielectric function $\varepsilon_{\text{J}}$ predict a third resonance between the two plexcitonic bands~\cite{Leb-Medv2012,Antosiewicz2014} (see Fig.~\ref{fig:COL}a and Fig.~\ref{fig:COL}d). However, this resonance was experimentally observed only in cases of large thickness of J-aggregate layers ($\gtrsim$ 10 nm) (see, for example, ~\cite{Bellessa2009}). At such thicknesses of the organic excitonic shell, a significant part of the shell is far enough from the metal surface, so that this shell part is not coupled strongly with the plasmonic core. Then the anisotropy effects associated with the strong plexcitonic coupling regime do not manifest themselves notably. In experiments with nanoparticles coated with an organic shell with a thickness of several nanometers, such a resonance was not observed (see discussion in \cite{Stete2023}), although its appearance was predicted by theoretical calculations within the framework of the isotropic model.

The results of calculations presented in Figures~\ref{fig:COL}a--~\ref{fig:COL}i show that the use of isotropic dielectric function (\ref{eps-J}) to describe the outer excitonic shell of a metalorganic nanoparticle leads to the appearance of an additional band in the spectrum near $\lambda = 490$ nm, with the intensity which increases with the increasing effective oscillator strength, $f_{\text{J}}$. However, in the spectra obtained using the tensor model of an anisotropic shell, such a band is absent for both (tangential and normal) orientations of J-aggregates in the organic shell. The figure demonstrates that the assumption of the tangential orientation of J-aggregates of the TDBC dye in the outer shell allows one to obtain good agreement between theoretical and experimental data. In this case, the use of isotropic dielectric function does not allow for achieving a satisfactory description of the experiment.

\section{Conclusions}

The review provides an insight into the physical phenomena in light interaction with hybrid nanostructures due to the near-field coupling of Frenkel excitons in the outer organic shell with surface plasmon-polaritons localized in the metal core. This rapidly emerging research field extends some general ideas in the subwavelength optics, plasmonics and organic photonics. 
We have briefly discussed key properties of localized plasmon-polaritons in metal nanoparticles and molecular excitons in ordered dye aggregates, something which is believed to be necessary for understanding the effects of plexcitonic coupling, and main features in the behavior of the absorption, scattering and luminescence spectra of metalorganic nanostructures. We have outlined basic physical approaches and methods for the description of the optical phenomena induced by light interaction with such structures and for the formulation of simple models aimed at the interpretation of various plexcitonic coupling regimes, including that of the weak, strong, and ultrastrong coupling. In addition to traditionally discussed topics, considerable attention has been paid to  the investigation of some novel points in the optics of plexcitonic nanostructures. In particular, we have explored the influence of size effects on the behavior of the optical spectra of metalorganic nanoparticles and analyzed the role of electric dipole, quadrupole and octupole plasmons in the plexcitonic coupling, as well as studied the role of the passive dielectric organic layer (the spacer layer between the plasmonic core and the excitonic shell) in the formation of nanostructures with qualitatively different optical properties. In addition, we have overviewed a number of available experimental and theoretical results on the light absorption, scattering, extinction, photoluminescence, and circular dichroism spectra of metalorganic nanoparticles of various compositions, sizes, and geometrical shapes (spheres, disks, rods, prisms).

Importantly, our theoretical analysis of plexcitonic effects has not been limited to the presentation of the standard scalar ''isotropic'' model of the excitonic shell of a composite nanoparticle; we have also covered a new physical approach based on the use of a tensor model for describing molecular aggregates with tangential and normal components of the dielectric function significantly different from each other. Within the framework of this approach, we have provided recently a successful quantitative explanation of the available experimental data on the optical spectra of plexcitonic nanospheres and nanorods, which adequately takes into account the anisotropy of the system's excitonic shell. Among other phenomena studied recently by means of new approaches is the replication of spectral bands of plexcitonic dimers of bilayer nanodisks and nanospheres. The phenomenon has been explained with numerical calculations using the FDTD method yielding a clear physical interpretation within the framework of the effective multilevel Hamiltonian model. Another example of a new phenomenon studied recently and discussed in our review is the optical chirality of plexcitonic systems manifesting itself in the circular dichroism and extinction spectra of a metal nanorod coated with a chiral dye J-aggregate.

The experimental and theoretical data presented in the review clearly demonstrate that the choice of shape, size, and composition of plasmonic and excitonic subsystems can affect dramatically the efficiency and the regime of plexcitonic coupling in a variety of hybrid nanosystems. Thus, using modern synthesis methods, it is possible to create hybrid nanoparticles, multilayer nanostructures and their arrays with new unique optical properties. As such, metalorganic nanostructures and nanomaterials are considered to be promising objects of nanophotonics and molecular optoelectronics. Such nanostructures can be used in highly sensitive biological and chemical sensors, photodetectors with high spectral selectivity, efficient converters of light energy in the visible and near-IR ranges, and novel fluorescent materials. Hybrid metalorganic nanostructures seem to be promising for achieving efficient coherent transmission of near-field interaction in nanophotonic devices. Such plexcitonic structures allow for efficient control and manipulation of light fields on nanometer spatial and ultrashort time scales. They are also of interest for a number of photonic applications due to the enormous values of nonlinear optical susceptibility of molecular aggregates forming the organic component of hybrid metalorganic nanosystems. In the past few years, the possibility of using them to enhance the effects of optical chirality has been discussed a lot as applied to problems of selective detection of chiral organic molecules and their complexes at low concentrations.

In conclusion, we note that physical approaches and methods for describing plexcitonic phenomena in metalorganic nanosystems have much in common with the consideration of similar effects caused by the coupling between plasmon-polaritons and excitons in inorganic composite materials, for example, in systems containing metal nanoparticles and semiconductor quantum dots, quantum wells or quantum wires. To a certain extent, this also applies to theoretical models instrumental in describing optical effects caused by the electromagnetic coupling of excitons with polaritons in microcavities. So, the insights on the optics of metalorganic nanostructures presented in the review might help our understanding of the essence of plexcitonic phenomena in the systems under consideration.

The work was supported by the Russian Science Foundation (grant \# 19-79-30086). We are very grateful to our colleague Sergei Moritaka for discussing the results, constructive recommendations and valuable comments, as well as for the great assistance provided to the authors in preparing a number of materials for the review. We also thank Alexander Narits for his careful reading of the manuscript and valuable comments.


\begin{thebibliography}{999}

\bibitem{Gallop2024} Gallop N P et al. \textit{Nat. Mater.} \textbf{23} 88 (2024)

\bibitem{Perego2022} Perego J et al. \textit{Nat. Commun.} \textbf{13} 3504 (2022)

\bibitem{Koduru2023} \textit{Hybrid Nanomaterials for Sustainable Applications: Case Studies and Applications} (Eds J R Koduru, R R Carri, N M Mubarak) (Elsevier, 2023)

\bibitem{Beddoes2023} Beddoes B et al. \textit{Opt. Express} \textbf{31} 18336 (2023)

\bibitem{Xie2023} Xie Y et al. \textit{Polymers} \textbf{15} 3721 (2023)

\bibitem{Vats2023} Vats G et al. \textit{Adv. Mater.} \textbf{35} 2205459 (2023) 

\bibitem{Youngblood2023} Youngblood N et al. \textit{Nat. Photon.} \textbf{17} 561 (2023)

\bibitem{Kim2022} Kim J H et al. \textit{Adv. Mater.} \textbf{34} 2104678 (2022)

\bibitem{Manzhos2021} Manzhos S et al. \textit{Adv. Phys. X} \textbf{6} 1908848 (2021)

\bibitem{Milichko2016} Milichko V A et al. \textit{Phys. Usp.} \textbf{59} 727 (2016)

\bibitem{Reus2024} Reus M A et al. \textit{Adv. Opt. Mater} \textbf{12} 2301008 (2024)

\bibitem{Yang2024} Yang S et al. \textit{J. Lumin.} \textbf{270} 120560 (2024)

\bibitem{Trapani2022} Trapani D et al. \textit{Materials} \textbf{15} 5450 (2022)

\bibitem{Narayan2023} \textit{Encyclopedia of Sensors and Biosensors} (Ed. R Narayan) (Elsevier, 2022)

\bibitem{Firoozi2023} Firoozi A et al. \textit{Sci. Rep.} \textbf{13} 11325 (2023)

\bibitem{Ates2022} Ates H C et al. \textit{Nat. Rev. Mater.} \textbf{7} 887 (2022)

\bibitem{Rodrigues2023} Rodrigues F et al. \textit{Photonics} \textbf{10} 182 (2023)

\bibitem{Chandrasekar2022} Chandrasekar R \textit{Chem. Commun.} \textbf{58} 3415 (2022)

\bibitem{Davis2017} Davis T J, G{\'o}mez D E, Roberts A \textit{Nanophotonics} \textbf{6} 543 (2017)

\bibitem{Fernandez-Bravo2019} Fernandez-Bravo A et al. \textit{Nat. Mater.} \textbf{18} 1172 (2019)

\bibitem{Shalaev2017} Wang Z et al. \textit{Laser Photon. Rev.} \textbf{11} 1700212 (2017) 

\bibitem{Balykin2018} Balykin V I \textit{Phys. Usp.} \textbf{61} 846 (2018); \textit{Usp. Fiz. Nauk.} \textbf{188} 935 (2018)

\bibitem{Park2014} Park C et al. \textit{Phys. Rev. Lett.} \textbf{113} 113901 (2014)

\bibitem{Kazantsev2017} Kazantsev D V et al. \textit{Phys. Usp.} \textbf{60} 259 (2017); \textit{Usp. Fiz. Nauk.} \textbf{187} 277 (2017) 

\bibitem{Khodadadi2020} Khodadadi M, Nozhat N, Moshiri S M M \textit{Opt. Express} \textbf{28} 3305 (2020)

\bibitem{Agranovich2011} Agranovich V M, Gartstein Y N, Litinskaya M \textit{Chem. Rev.} \textbf{111}, 51795214 (2011)

\bibitem{Will2019} Will P-A, Reineke S, in \textit{Handbook of Organic Materials for Electronic and Photonic Devices} (Ed. O. Ostroverkhova) (Elsevier, 2019) p. 695

\bibitem{Zhou2021} \textit{Optoelectronic Organic-Inorganic Semiconductor Heterojunctions} (Ed. E Zhou) (CRC Press, 2021)

\bibitem{Klimov2009} Klimov V \textit{Nanoplasmonics} (New York: Jenny Stanford Publishing, 2014); Translated from:
\textit{Nanoplazmonika} (Moscow: Fizmatlit, 2009) 

\bibitem{H-Chang2021} Chang H et al. \textit{Plasmonic Nanoparticles: Basics to Applications (I)} in: \textit{Nanotechnology for Bioapplications} pp 133-159 (Springer Nature Singapore Pte Ltd., 2021)

\bibitem{Lindquist2012} Lindquist N C et al. \textit{Rep. Prog. Phys.} \textbf{75} 036501 (2012)

\bibitem{Qazi2016} Qazi U Y, Javaid R \textit{Advances in Nanoparticles} \textbf{5} 27 (2016) 

\bibitem{Kond-Leb_JRLR2021} Kondorskiy A D, Lebedev V S \textit{J. Russ. Laser Res.} \textbf{42} 697 (2021)

\bibitem{Miroshnichenko2022} Tribelsky M I, Miroshnichenko A E  \textit{Phys. Usp.} \textbf{65} 40 (2022); \textit{Usp. Fiz. Nauk.} \textbf{192} 45 (2022) 

\bibitem{Khlebtsov2022} Khlebtsov N G, Dykman L A, Khlebtsov B N \textit{Russ. Chem. Rev.} \textbf{91} RCR5058 (2022);
 \textit{Uspekhi Khimii} \textbf{91} RCR5058 (2022) 

\bibitem{Krasnok2013} Krasnok A E et al. \textit{Phys. Usp.} \textbf{56} 539 (2013); \textit{Usp. Fiz. Nauk.} \textbf{183} 561 (2013) 

\bibitem{Lepeshov2018} Lepeshov S I et al. \textit{Phys. Usp.} \textbf{61} 1035 (2018); \textit{Usp. Fiz. Nauk.} 
\textbf{188} 1137 (2018)


\bibitem{Barbillon2017} \textit{Nanoplasmonics: Fundamentals and Applications} (Ed. G Barbillon) (InTech Publisher, 2017)

\bibitem{Klimov2021} Klimov V V \textit{Phys. Usp.} \textbf{191} 1044 (2021); \textit{Usp. Fiz. Nauk.} \textbf{191} 1044 (2021) 

\bibitem{Diedenhofen2015} Diedenhofen S L et al. \textit{Light Sci. Appl.} \textbf{4} e234 (2015)

\bibitem{Tam2007} Tam F et al. \textit{Nano Lett.} \textbf{7} 49 (2007)

\bibitem{Liaw2009} Liaw J W et al. \textit{Opt Express} \textbf{17} 13532 (2009)

\bibitem{Ming2012} Ming T et al. \textit{J. Phys. Chem. Lett.} \textbf{3} 191 (2012)

\bibitem{Dong2015} Dong J et al. \textit{Nanophotonics} \textbf{4} 472 (2015)

\bibitem{LeeLee2020} Lee H et al. \textit{Nanophotonics} \textbf{9} 3089 (2020)

\bibitem{Kneipp1997} Kneipp K et al. \textit{Phys. Rev. Lett.} \textbf{78} 1667 (1997)

\bibitem{Ru2008} Le Ru E C, Etchegoin P G \textit{Principles of surface-enhanced Raman spectroscopy and related plasmonic effects} (Oxford: Elsevier, 2008)

\bibitem{Chen_2015} Chen S et al. \textit{J. Phys. Chem. C} \textbf{119} 5246 (2015)

\bibitem{Khlebtsov2023} Khlebtsov B N et al. \textit{Phys. Chem. Chem. Phys.} \textbf{25} 30903 (2023)

\bibitem{Arslanagic2015} Arslanagi{\'c} S, Ziolkowski R W \textit{Photonics Nanostructures: Fundam. Appl.} \textbf{13} 80 (2015) 

\bibitem{Smirnov2017} Smirnov B M \textit{Phys. Usp.} \textbf{60} 1236 (2017); \textit{Usp. Fiz. Nauk.} \textbf{187} 1329 (2017) 

\bibitem{LeeLawrie2021} Lee C et al. \textit{Chem. Rev.} \textbf{121} 4743 (2021)

\bibitem{Shapiro2006} Shapiro B I \textit{Russ. Chem. Rev.} \textbf{75} 433 (2006); \textit{Uspekhi Khimii} \textbf{75} 484 (2006) 

\bibitem{Wurthner2011} W{\"u}rthner F, Kaiser T E, Saha-M{\"o}ller C R \textit{Angew. Chem. Int. Ed. Engl.} \textbf{50} 3376 (2011)

\bibitem{J-Aggregates2012} \textit{J-Aggregates} Volume 2 (Ed T Kobayashi) (New Jersey: World Scientific, 2012) p 520

\bibitem{Bricks2018} Bricks J L et al. \textit{Methods and Applications in Fluorescence} \textbf{6} 012001 (2018) 

\bibitem{Hestand2018} Hestand N J, Spano F C \textit{Chem. Rev.} \textbf{118} 7069 (2018)

\bibitem{Otsuki2018} Otsuki J \textit{J. Mater. Chem. A} \textbf{6} 6710 (2018)

\bibitem{Hecht-Wurthner2021} Hecht M, W{\"u}rthner F \textit{Acc. Chem. Res.}, \textbf{54}, 642 (2021)

\bibitem{Ma2021} Ma S et al. \textit{Aggregate} \textbf{2} e96 (2021)

\bibitem{Shapiro_OE2018} Shapiro B I et al. \textit{Opt. Express} \textbf{26} 30324 (2018)

\bibitem{Shapiro_QE2018} Shapiro B I et al. \textit{Quantum Electron.} \textbf{48} 856 (2018); \textit{Kvantovaya Elektron.} \textbf{48} 856 (2018) 

\bibitem{Wiederrecht2008} Wiederrecht G P, Wurtz G A, Bouhelier A \textit{Chem. Phys. Lett.} \textbf{461} 171 (2008)

\bibitem{Fofang2008} Fofang N T et al. \textit{Nano Lett.} \textbf{8} 3481 (2008)

\bibitem{Lebedev2008} Lebedev V S et al. \textit{Colloids and Surfaces A: Physicochem. Eng. Aspects} \textbf{326} 204 (2008)

\bibitem{Lebedev2010} Lebedev V S et al. \textit{Quantum Electron.} \textbf{40} 246 (2010); \textit{Kvantovaya Elektron.} \textbf{40} 246 (2010) 

\bibitem{Antosiewicz2014} Antosiewicz T J, Apell S P, Shegai T \textit{ACS Photon.} \textbf{1} 454 (2014)

\bibitem{Shapiro2015} Shapiro B I et al. \textit{Quantum Electron.} \textbf{45} 1153 (2015); \textit{Kvantovaya Elektron.} \textbf{5} 1153 (2015) 

\bibitem{Todisco2018} Todisco F et al. \textit{ACS Photonics} \textbf{5} 143 (2018)

\bibitem{Song2019} Song T et al. \textit{Nanomaterials} \textbf{9}, 564 (2019)

\bibitem{Watanabe2006} Watanabe K et al. \textit{Chem. Rev.} \textbf{106} 4301 (2006)

\bibitem{Kravets2010a} Kravets V G et al. \textit{Phys. Rev. Lett.} \textbf{105} 246806 (2010)

\bibitem{Kravets2010b} Kravets V G et al. \textit{Nano Lett.} \textbf{10} 874 (2010)

\bibitem{Singh2024} Singh K et al. \textit{Microchem. J.} \textbf{197} 109888 (2024)

\bibitem{Walters2018} Walters C M et al. \textit{Adv. Mater.} \textbf{30} 1705381 (2018) 

\bibitem{Ralevich2018} Ralevi\'{c} U et al. \textit{Appl. Surf. Sci.} \textbf{434} 540 (2018)

\bibitem{A-Wang2015} Wang A X, Kong X \textit{Materials} \textbf{8} 3024 (2015)

\bibitem{Sorokin2015} Sorokin A V et al. \textit{J. Phys. Chem. C} \textbf{119} 2743 (2015)

\bibitem{Sorokin2020} Sorokin A V et al. \textit{J. Phys. Chem. C} \textbf{124} 10167 (2020)

\bibitem{L-Lu2002} Lu L et al. \textit{Langmuir} \textbf{18} 7706 (2002)

\bibitem{Hranisavljevic2002} Hranisavljevic J et al. \textit{J. Am. Chem. Soc} \textbf{124} 4536 (2002)

\bibitem{Jian-Zhang2007} Zhang J, Fu Y, Lakowicz J R \textit{J. Phys. Chem. C} \textbf{111} 50 (2007)

\bibitem{Akhavan2017} Akhavan S et al. \textit{ACS Nano} \textbf{11} 5430 (2017)

\bibitem{Kabbash2016} Kabbash M E et al. \textit{J Nanomaterials} \textbf{2016} 4819040 (2016)

\bibitem{Weeraddana2017} Weeraddana D et al. \textit{J. Chem. Phys.} \textbf{147} 074117 (2017)

\bibitem{Wiederrecht2004} Wiederrecht G P, Wurtz G A, Hranisavljevic J \textit{Nano Lett.} \textbf{4} 2121 (2004)

\bibitem{Bellessa2009}  Bellessa J et al. \textit{Phys. Rev. B} \textbf{80} 033303 (2009)

\bibitem{Yoshida2010} Yoshida A, Kometani N \textit{J. Phys. Chem. C} \textbf{114} 2867 (2010)

\bibitem{Leb-Medv2012} Lebedev V S, Medvedev A S \textit{Quantum Electron.} \textbf{42} 701 (2012); \textit{Kvantovaya Elektron.} \textbf{42} 701 (2012) 

\bibitem{Salomon2013} Salomon A et al. \textit{Chem. Phys. Chem.} \textbf{14} 1882 (2013)

\bibitem{Balci2013} Balci S \textit{Opt. Lett.} \textbf{38} 4498 (2013)

\bibitem{DeLacy2015} DeLacy B G et al. \textit{Nano Lett.} \textbf{15} 2588 (2015)

\bibitem{Zengin2015} Zengin G et al. \textit{Phys. Rev. Lett.} \textbf{114} 157401 (2015)

\bibitem{Kondorskiy2015} Kondorskiy A D et al. \textit{J. Russ. Laser Res.} \textbf{36} 175 (2015) 

\bibitem{Melnikau2022} Melnikau D et al. \textit{J. Lumin.} \textbf{242} 118557 (2022)

\bibitem{Bellessa2004} Bellessa J et al. \textit{Phys. Rev. Lett.} \textbf{93} 036404 (2004)

\bibitem{Symonds2008} Symonds C et al. \textit{New J. Phys.} \textbf{10} 065017 (2008)

\bibitem{Cade2009} Cade N I et al. \textit{Phys. Rev. B} \textbf{79} 241404 (2009) 

\bibitem{Bellessa2014} Bellessa J et al. \textit{Electronics} \textbf{3} 303 (2014)

\bibitem{Chmereva2016} Chmereva T M, Kucherenko M G, Kurmangaleev K S \textit{Opt. Spectrosc.} \textbf{120} 881 (2016); \textit{Opt. i Spektr.} \textbf{120} 941 (2016) 

\bibitem{Matsui2017} Matsui H, in \textit{Noble and Precious Metals -- Properties, Nanoscale Effects and Applications} (Eds. M. S. Seehra and A. D. Bristow) (IntechOpen 2017) ch. 5

\bibitem{Forn-Diaz2019} Forn-D\'{i}az P et al. \textit{Rev. Mod. Phys.} \textbf{91} 025005 (2019)

\bibitem{Bitton2019} Bitton O, Gupta S N, Haran G \textit{Nanophotonics} \textbf{8} 559 (2019)

\bibitem{Liu2021} Liu R et al. \textit{Phys Rev B} \textbf{103} 235430 (2021)

\bibitem{Kucherenko2022} Kucherenko M G, Nalbandyan V M, Chmereva T M \textit{Opt. Spectrosc.} \textbf{130} 593 (2022); \textit{Opt. i Spektr.} \textbf{130} 745 (2022) 

\bibitem{Kim-Barulin2023} Kim Y et al. \textit{Nanophotonics} \textbf{12} 413 (2023)

\bibitem{Hirai2023} Hirai K, Hutchison J A, Uji-i H \textit{Chem. Rev.} \textbf{123} 8099 (2023)

\bibitem{Jiang2019} Jiang P et al. \textit{Opt. Express} \textbf{27} 16613 (2019)

\bibitem{Tserkezis2020} Tserkezis C et al. \textit{Rep. Prog. Phys.} \textbf{83} 082401 (2020)

\bibitem{Deng2023} Deng X et al. \textit{Opt. Express} \textbf{31} 32082 (2023)

\bibitem{Tserkezis2023} Tserkezis C \textit{Phys. Rev. A} \textbf{107} 043707 (2023)

\bibitem{Nordlander2011} Manjavacas A, Garc{\'\i}a de Abajo F J, Nordlander P \textit{Nano Lett.} \textbf{11} 2318 (2011)

\bibitem{DeLacy2013} DeLacy B G et al. \textit{Opt. Express} \textbf{21} 19103 (2013)

\bibitem{Schlather2013} Schlather A E et al. \textit{Nano Lett.} \textbf{13} 3281 (2013)

\bibitem{Wurtz2007} Wurtz G A et al. \textit{Nano Lett.} \textbf{7} 1297 (2007)

\bibitem{Leb-Medv2013b} Lebedev V S, Medvedev A S \textit{Quantum Electron.} \textbf{43} 1065 (2013); \textit{Kvantovaya Elektron.} \textbf{43} 1065 (2013) 

\bibitem{Leb-Medv2013a} Lebedev V S, Medvedev A S \textit{J. Russ. Laser Res.} \textbf{34} 303 (2013) 

\bibitem{Moritaka2020} Moritaka S S et al. \text{Bull. Leb. Phys. Inst.} \textbf{47} 280 (2020); \textit{Kratkie Soobshcheniya po Fizike} \textbf{47} 41 (2020) 

\bibitem{Moritaka2023} Moritaka S S, Lebedev V S \text{Bull. Leb. Phys. Inst.} \textbf{50} 589 (2023); \textit{Kratkie Soobshcheniya po Fizike} \textbf{50} 112 (2023) 

\bibitem{H-Chen2012} Chen H et al. \textit{J. Phys. Chem. C} \textbf{116} 14088 (2012)

\bibitem{Thomas2018} Thomas R et al. \textit{ACS Nano} \textbf{12} 402 (2018)

\bibitem{Zengin2013} Zengin G et al. \textit{Sci. Rep.} \textbf{3} 3074 (2013)

\bibitem{Simon2016} Simon T et al. \textit{J. Phys. Chem. C} \textbf{120} 12226 (2016)

\bibitem{Nan2015} Nan F et al. \textit{Nano Lett.} \textbf{15} 2705 (2015)

\bibitem{Ni2010} Ni W et al. \textit{Nano Lett.} \textbf{10} 77 (2010)

\bibitem{Lekeufack2010} Lekeufack D D et al. \textit{Appl. Phys. Lett.} \textbf{96} 253107 (2010) 

\bibitem{Melnikau2013} Melnikau D et al. \textit{Nanoscale Res. Lett.} \textbf{8} 134 (2013)

\bibitem{Vasa2013} Vasa P et al. \textit{Nanoscale Res. Lett.} \textbf{8} 134 (2013) 

\bibitem{Vasa2013Rabi} Vasa P et al. \textit{Nature Photon.} \textbf{7} 128 (2013)

\bibitem{Vasa2018} Vasa P, Lienau C \textit{ACS Photon.} \textbf{5} 2 (2018)

\bibitem{Fain2019} Fain N, Ellenbogen T, Schwartz T \textit{Phys. Rev. B} \textbf{100} 235448 (2019)

\bibitem{Ates2020} Ates S et al. \textit{Opt. Lett.} \textbf{45} 5824 (2020)

\bibitem{Guo-Wu2021} Guo J et al. \textit{Nanoscale} \textbf{13} 15812 (2021)

\bibitem{Dey2023} Dey J, Virdi A, Chandra M \textit{Nanoscale} \textbf{15}, 17879 (2023)

\bibitem{Sukharev2017} Sukharev M, Nitzan A \textit{J. Phys.: Condensed Matter} \textbf{29} 443003 (2017) 

\bibitem{Manuel2019} Manuel A P et al. \textit{J. Mater. Chem. C} \textbf{7} 1821 (2019)

\bibitem{Kholmicheva2019} Kholmicheva N et al. \textit{Nanophotonics} \textbf{8} 613 (2019)

\bibitem{Vasa2020} Vasa P \textit{Adv. Phys. X} \textbf{5} 1749884 (2020)

\bibitem{He-Li2020} He Z et al. \textit{Appl. Sci.} \textbf{10} 1774 (2020)

\bibitem{Wei2021} Wei H et al. \textit{Adv. Funct. Mater.} \textbf{31} 2100889 (2021)

\bibitem{Barnes2015} T{\"o}rm{\"a} P, Barnes W L  \textit{Rep. Prog. Phys.} \textbf{78} 013901 (2015)

\bibitem{Cao2018} Cao E et al. \textit{Nanophotonics} \textbf{7} 145 (2018)

\bibitem{Bozhevolnyi2013} Han Z, Bozhevolnyi S I \textit{Rep. Prog. Phys.} \textbf{76} 016402 (2013)

\bibitem{Barnes2003} Barnes W L, Dereux A, Ebbesen T W \textit{Nature} \textbf{424} 824 (2003)

\bibitem{Zhang2012} Zhang J, Zhang L, Xu W \textit{J. Phys. D: Appl. Phys.} \textbf{45} 113001 (2012)

\bibitem{Raether1988} Raether H \textit{Surface Plasmons on Smooth and Rough Surfaces and on Gratings} (Berlin, Heidelberg: Springer, 1988)

\bibitem{Girard2000} Girard C, Joachim C, Gauthier S, \textit{Rep. Prog. Phys.} \textbf{63} 893 (2000)

\bibitem{Bohren1998} Bohren C F, Huffmann D R \textit{Absorption and Scattering of Light by Small Particles} (New York: John Wiley \& Sons, 1998)

\bibitem{Hohenau2007} Hohenau A, Leitner A, Aussenegg F R, in \textit{Surface Plasmon Nanophotonics} Chap. 2 (Eds M L Brongersma, P G Kik) (Springer, 2007)

\bibitem{Koch2001} Koch W, Holthausen M C \textit{A Chemist's Guide to Density Functional Theory} (Wiley VCH Verlag GmbH, 2001)

\bibitem{Chateau2015} Chateau D et al. \textit{Nanoscale} \textbf{7} 1934 (2015)

\bibitem{Cardinal2010} Cardinal M F et al. \textit{J. Phys. Chem. C} \textbf{114} 10417 (2010)

\bibitem{Melnikau2016} Melnikau D et al. \textit{J. Phys. Chem. Lett.} \textbf{7} 354 (2016)

\bibitem{Scarabelli2021} Scarabelli L, Liz-Marz{\'a}n L M  \textit{ACS Nano} \textbf{15} 18600 (2021)

\bibitem{Yin2016} Yin Z et al. \textit{RSC Adv.} \textbf{6} 86297 (2016) 

\bibitem{Swarnapali2015} Swarnapali A et al. \textit{Phys. Chem. Chem. Phys.} \textbf{17} 21133 (2015)

\bibitem{Gaponenko2010} Gaponenko S V \textit{Introduction to nanophotonics} (Cambridge University Press, 2010)

\bibitem{Novotny2012} Novotny L, Hecht B \textit{Principles of Nano-Optics} (Cambridge University Press, 2012)



\bibitem{Bigot1995} Bigot J Y et al. \textit{Phys. Rev. Lett.} \textbf{75} 4702 (1995)

\bibitem{Inouye1998} Inouye H et al. \textit{Phys. Rev. B} \textbf{57} 11334 (1998) 

\bibitem{Kreibig1995} Kreibig~U, Vollmer~M \textit{Optical Properties of Metal Clusters} (Berlin, Heidelberg: Springer, 1995) 

\bibitem{Johnson1972} Johnson~P B, Christy~R W \textit{Phys. Rev. B} \textbf{6} 4370 (1972)

\bibitem{Babar2015} Babar S, Weaver J H \textit{Appl. Opt.} \textbf{54} 477 (2015)

\bibitem{Palik1991} \textit{Handbook of Optical Constants of Solids II} (Ed. E D Palik) (San Diego: Academic, 1991)

\bibitem{Rakic1995} Raki{\'c} A D \textit{Appl. Opt.} \textbf{34} 4755 (1995)

\bibitem{Kondorskiy2023b} Kondorskiy A D, Mekshun A V \textit{Bull. Leb. Phys. Inst.} \textbf{50} 557 (2023;  \textit{
Kratkie Soobshcheniya po Fizike} \textbf{50}, No 12, 96 (2023)

\bibitem{Kondorskiy2023a} Kondorskiy A D, Mekshun A V \textit{J. Russ. Laser Res.} \textbf{44} 627 (2023) 

\bibitem{Fuchs1987} Fuchs R, Claro F \textit{Phys. Rev. B} \textbf{35} 3722 (1987) 

\bibitem{Ruppin1976} Ruppin R, Yatom H \textit{Phys. Status Solidi B} \textbf{74} 647 (1976)

\bibitem{Aleksandrov1999} Aleksandrov A F, Rukhadze A. A. \textit{Course on electrodynamics of plasmalike media} (Moscow: Mosk. Gos. Univ., 1999).

\bibitem{Kelly2003} Kelly K L et al. \textit{J. Phys. Chem. B} \textbf{107} 668 (2003)

\bibitem{Lam_JRLR2018} Kondorskiy A D, Lam N T, Lebedev V S \textit{J. Russ. Laser Res.} \textbf{39} 56 (2018)

\bibitem{Creighton1991} Creighton J A, Eadon D G \textit{J. Chem. Soc. Faraday Trans.} \textbf{87} 3881 (1991)

\bibitem{Kometani2001} Kometani N et al. \textit{Langmuir} \textbf{17} 578 (2001)

\bibitem{Mekshun2020}  Mekshun A V et al. \textit{Bull. Leb. Phys. Inst.} \textbf{47} 276 (2020); \textit{
Kratkie Soobshcheniya po Fizike}  \textbf{47}, No 9, 34 (2020) 

\bibitem{Payne2006} Payne E M et al. \textit{J. Phys. Chem. B} \textbf{110} 2150 (2006)

\bibitem{Khlebtsov2007} Khlebtsov B N, Melnikov A, Khlebtsov N G \textit{JQSRT} \textbf{107} 306 (2007)

\bibitem{Zenin2020} Zenin V A et al. \textit{ACS Photon.} \textbf{7} 1067 (2020)

\bibitem{Ray2021} Ray D, Kiselev A, Martin O J F \textit{Opt. Express} \textbf{29} 24056 (2021)

\bibitem{Frenkel1931} Frenkel J I \textit{Phys. Rev.} \textbf{37} 1276 (1931)

\bibitem{Davydov1971} Davydov A S \textit{Theory of Molecular Excitons} (New York, NY: Plenum, 1971); Translated from Russian: \textit{Teoriya molekulyarnyh excitonov} (Moscow: Nauka, 1968) 

\bibitem{Shapiro2008} Shapiro B I \textit{Nanotechnol. Russ.} \textbf{3} 139 (2008); \textit{Ros. nanotekhnologii} \textbf{3} 72 (2008)

\bibitem{Todisco2015} Todisco F et al. \textit{ACS Nano} \textbf{9} 9691 (2015)

\bibitem{Takeshima2020} Takeshima N, Sugawa K, Tahara H \textit{Nanoscale Res. Lett.} \textbf{15} 15 (2020) 

\bibitem{Asanuma2012} Asanuma H et al. \textit{J Photochem Photobiol C Photochem Rev} \textbf{13} 124 (2012)

\bibitem{Ciardelli2013} Ciardelli F, Ruggeri G, Pucci A \textit{Chem. Soc. Rev.} \textbf{42} 857 (2013) 

\bibitem{Shapiro-MDJH-spectra} Shapiro B I et al. \textit{Nanotechnol. Russ.} \textbf{5} 58 (2010); \textit{Ros. Nanotekhnologii} \textbf{5} 35 (2010)

\bibitem{Kobayashi1996} \textit{J-Aggregates} (Ed. T Kobayashi) (World Scientific, Singapore, 1996) 

\bibitem{Brixner2017} Brixner T et al. \textit{Adv. Energy Mater.} \textbf{7} 1700236 (2017)

\bibitem{Lee-Schenning-2009} Lee C C et al. \textit{Chem. Soc. Rev.} \textbf{38} 671 (2009)

\bibitem{Bardeen2014} Bardeen C J \textit{Annu. Rev. Phys. Chem.} \textbf{65} 127 (2014)

\bibitem{McRae1958} McRae E, Kasha M \textit{J. Chem. Phys.} \textbf{28} 721 (1958) 

\bibitem{Kasha1964} Kasha M \textit{Physical Processes in Radiation Biology} (New York: Academic Press, 1964)

\bibitem{Kasha1965} Kasha M, Rawls H R, El-Bayoumi M A \textit{Pure Appl. Chem.} \textbf{11} 371 (1965).

\bibitem{Didraga2004} Didraga C et al. \textit{J. Phys. Chem. B} \textbf{108} 14976 (2004)

\bibitem{Scheblykin2001} Scheblykin I G et al. \textit{J. Phys. Chem. B} \textbf{105} 4636 (2001)

\bibitem{MorLeb2023} Moritaka S S, Lebedev V S \textit{JETP Lett.} \textbf{118} 792 (2023);
\textit{Pis'ma v ZhETF} \textbf{118} 794 (2023) 

\bibitem{MorLeb2024} Moritaka S S, Lebedev V S \textit{J. Chem. Phys.} \textbf{160} 074901 (2024);  

\bibitem{Dicke1954} Dicke R H \textit{Phys. Rev.} \textbf{93} 99 (1954)

\bibitem{Gierschner2013a} Gierschner J et al. \textit{J Phys Chem Lett} \textbf{4} 2686 (2013)

\bibitem{Gierschner2013b} Gierschner J, Park S Y \textit{J. Mater. Chem. C} \textbf{1} 5818 (2013)

\bibitem{Bawendi2019} Chuang C et al. \textit{Chem.} \textbf{5} 3135 (2019)  

\bibitem{Deshmukh2019} Deshmukh A et al. \textit{J. Phys. Chem. C} \textbf{123} 18702 (2019)

\bibitem{Zhu2017} Zhu T, Wan Y. \textit{Acc. Chem. Res.} \textbf{50} 1725 (2017)

\bibitem{Levitz2018} Levitz A, Marmarchi F, Henary M \textit{Photochem. Photobiol. Sci.} \textbf{17} 1409 (2018)

\bibitem{Doria2018} Doria S et al. \textit{ACS Nano} \textbf{12} 4556 (2018)

\bibitem{Bogdanov1991} Bogdanov V L et al. \textit{JETP Lett.} \textbf{53} 105 (1991)

\bibitem{Wang1991} Wang Y \textit{J. Opt. Soc. Am. B} \textbf{8} 981 (1991)

\bibitem{Zhuravlev1992} Zhuravlev F A et al. \textit{JETP Lett.} \textbf{56} 264 (1992)

\bibitem{Shelkovnikov1993} Shelkovnikov V V et al. \textit{J. Struct. Chem.} \textbf{34} 909 (1993)

\bibitem{Gadonas1994} Gadonas R, Feller K-H, Pugzlys A \textit{Opt. Commun.} \textbf{112} 157 (1994)

\bibitem{Spano1994} Spano F C, Knoester J, in \textit{Advances in Magnetic and Optical Resonance} (Ed. W S Warren) (Academic Press, 1994) p. 117

\bibitem{Markov2000} Markov R V et al. \textit{Nonlinear Opt.} \textbf{25} 365 (2000)

\bibitem{Shelkovnikov2002} Shelkovnikov V V et al. \textit{Opt. Spectrosc.} \textbf{92} 884 (2002); \textit{Opt. i Spectr.} \textbf{92} 958 (2002)

\bibitem{Shelkovnikov2012} Shelkovnikov V V, Plekhanov A I, in \textit{Macro to Nano Spectroscopy} Chap. 16 (Ed. J Uddin) (Rijeka: IntechOpen, 2012) p. 317 

\bibitem{Lee2018} Lee Y U et al. \textit{Adv. Opt. Mater.} \textbf{6} 1701400 (2018)

\bibitem{Gerasimova2000} Gerasimova T N et al. \textit{Chem. Sustain. Dev.} \textbf{8} 109 (2000)

\bibitem{Markov2001} Markov R V et al. \textit{Quantum Electron.} \textbf{31} 1063 (2001); \textit{Kvantovaya Electron.} \textbf{31} 1063 (2001)

\bibitem{Markov2004} Markov R V et al. \textit{JETP} \textbf{99} 480 (2004); \textit{ZhETF} \textbf{126} 549 (2004) 

\bibitem{Sasaki2001} Sasaki F, Kano T, Kobayashi S \textit{Phys. Rev. B} \textbf{63} 205411 (2001)

\bibitem{Bednarz2001} Bednarz M, Knoester J \textit{J. Phys. Chem. B} \textbf{105}, 12913 (2001)

\bibitem{Kano2002} Kano H, Kobayashi T \textit{J. Chem. Phys.} \textit{116} 184 (2002)

\bibitem{Rehhagen2020} Rehhagen C et al. \textit{J. Phys. Chem. Lett.} \textbf{11} 6612 (2020)

\bibitem{Belko2022} Belko N V et al. \textit{J. Phys. Chem. C} \textbf{126} 7922 (2022)

\bibitem{Jumbo-Nogales2022} Jumbo-Nogales A et al. \textit{J. Phys. Chem. Lett.} \textbf{13} 10198 (2022)

\bibitem{Knoester1996} Knoester J, Spano F C, in \textit{J-Aggregates} (Ed. T. Kobayashi) (Singapore: World Scientific, 1996) p. 111

\bibitem{Nishimura2004} Nishimura K, Tokunaga E, Kobayashi T \textit{Chem. Phys. Lett.} \textbf{395} 114 (2004)

\bibitem{Dijkstra2008} Dijkstra A G, Jansen T C, Knoester J \textit{J. Chem. Phys.} \textbf{128} 164511 (2008)

\bibitem{Milota2009} Milota F et al. \textit{J. Chem. Phys.} \textbf{131} 054510 (2009)

\bibitem{Abramavicius2009} Abramavicius D et al. \textit{Chem. Rev.} \textbf{109} 2350 (2009)

\bibitem{Ginsberg2009} Ginsberg N S, Cheng Y-C, Fleming G R \textit{Acc. Chem. Res.} \textbf{42} 1352 (2009)

\bibitem{Bolzonello2016} Bolzonello L, Fassioli F, Collini E \textit{J. Phys. Chem. Lett.} \textbf{7} 4996 (2016)

\bibitem{Quenzel2022} Quenzel T et al. \textit{ACS Nano} \textbf{16} 4693 (2022)

\bibitem{Peruffo2023} Peruffo N, Mancin F, Collini E \textit{Adv. Opt. Mater.} \textbf{11} 2203010 (2023)

\bibitem{Russo2024} Russo M et al. \textit{Adv. Opt. Mater.} 2400821 (2024) [early view]

\bibitem{Fidder1993} Fidder H, Knoester J, Wiersma D A \textit{J. Chem. Phys.} \textbf{98} 6564 (1993)

\bibitem{Minoshima1994} Minoshima K et al. \textit{Chem. Phys. Lett.} \textbf{218} 67 (1994)

\bibitem{Gadonas1997} Gadonas R et al. \textit{J. Chem. Phys.} \textbf{106} 8374 (1997)

\bibitem{Hirschmann1988} Hirschmann R et al. \textit{J. Chem. Phys. Lett.} \textbf{151} 60 (1988)

\bibitem{Malyshev1996} Malyshev V A, Moreno P \textit{Phys. Rev. A} \textbf{53} 416 (1996)

\bibitem{Malyshev1998} Malyshev V A, Glaeske H, Feller K-H \textit{Phys. Rev. A} \textbf{58} 670 (1998)

\bibitem{Glaeske2002} Glaeske H, Malyshev V A, Feller K-H \textit{Phys. Rev. A} \textbf{65} 033821 (2002)

\bibitem{Klugkist2007} Klugkist J A, Malyshev V, Knoester J \textit{J. Chem. Phys.} \textbf{127} 164705 (2007)

\bibitem{Klugkist2008}  Klugkist J A, Malyshev V, Knoester J \textit{J. Chem. Phys.} \textbf{128} 084706 (2008)

\bibitem{Zabolotskiy2006} Zabolotskii A A \textit{Opt. Spectrosc.} \textbf{101} 606 (2006); \textit{Opt. i Spectr.}
 \textbf{101} 644 (2006)

\bibitem{Zabolotskiy2008} Zabolotskii A A \textit{JETP} \textbf{106} 404 (2008); \textit{ZhETF} \textbf{133} 466 (2008)

\bibitem{Nesterov2013} Nesterov L A et al. \textit{Opt. Spectrosc.} \textbf{115} 499 (2013); \textit{Opt. i Spectr.} \textbf{115} 572 (2013)

\bibitem{Zabolotskiy2016} Zabolotskii A A \textit{Optoelectron. Instrum. Data Process.} \textbf{52} 76 (2016)

\bibitem{Boyd2008} Boyd R W \textit{Nonlinear Optics} (3rd ed, Elsevier, 2008)

\bibitem{Wooten1972} Wooten F \textit{Optical Properties of Solids} (London: Academic Press, inc., 1972)

\bibitem{KML_OE2022} Kondorskiy A D, Moritaka S S, Lebedev V S \textit{Opt. Express} \textbf{30} 4600 (2022)

\bibitem{Kondorskiy2024} Kondorskiy A D \textit{Chinese Opt. Lett.} \textbf{22} 093602 (2024)

\bibitem{Grynberg2010} Grynberg G, Aspect A, Fabre C \textit{Introduction to Quantum Optics: From the Semi-Classical Approach to Quantized Light} (Cambridge University Press, 2010)

\bibitem{Meystre2007} Meystre P, Sargent M \textit{Elements of Quantum Optics} (Berlin, Heidelberg: Springer, 2007)

\bibitem{Abrikosov1963} Abrikosov A A, Gorkov L P, Dzyaloshinski I E \textit{Methods of Quantum Field Theory in Statistical Physics} (Englewood Cliffs: Prentice-Hall, inc., 1963); Translated from Russian: \textit{Metody Kvantovoi Teorii Polya v Statisticheskoi Fizike} (Moscow: Fizmatlit, 1962)

\bibitem{BBGKY2014} Campa A et al. \textit{Physics of Long-Range Interacting Systems} Chap. 7 (Oxford: Oxford Academic, 2014)

\bibitem{White2012} White A J, Galperin M \textit{Phys. Chem. Chem. Phys.} \textbf{14} 13809 (2012)

\bibitem{Zubarev1960} Zubarev D N \textit{Sov. Phys. Usp.} \textbf{3} 320 (1960); \textit{Usp. Fiz. Nauk} \textbf{71} 71 (1960)

\bibitem{Fuller1993} Fuller K A \textit{Opt. Lett.} \textbf{18} 257 (1993)

\bibitem{Aden1951} Aden A L, Kerker M \textit{J. Appl. Phys.} \textbf{22} 1242 (1951)

\bibitem{Guttler1952} G{\"u}ttler A \textit{Ann. Phys.} \textbf{11} 65 (1952)

\bibitem{Ruppin1968} Ruppin R, Englman R \textit{J. Phys. C} \textbf{1} 630 (1968)

\bibitem{Irimajiri1979} Irimajiri A, Hanai T, Inouye A \textit{J. Theor. Biolog.} \textbf{78} 251 (1979)

\bibitem{Bhandari1985} Bhandari R \textit{Appl. Opt.} \textbf{24} 1960 (1985)

\bibitem{Wu1991} Wu Z C,  Wang Y P \textit{Radio Sci.} \textbf{26} 1393 (1991)

\bibitem{Sinzig1994} Sinzig J, Quinten M \textit{Appl. Phys. A} \textbf{58} 157 (1994)

\bibitem{Stratton1948} Stratton J A \textit{Electromagnetic theory} (John Wiley \& Sons, Inc., Hoboken, 2007)

\bibitem{Barber1990} Barber P W, Hill S C \textit{Light Scattering by Particles: Computational Methods} (Singapore: World Scientific Publishing Co. Pte. Ltd., 1990)

\bibitem{Scaife1998} Scaife B K P \textit{Principles of Dielectrics} (Oxford: Oxford Science Publ, 1998)

\bibitem{Farafonov1993} Voshchinnikov N V, Farafonov V G \textit{Astrophys. Space Sci.} \textbf{204} 19 (1993)

\bibitem{Wang1982} Wang D S, Kerker M \textit{Phys. Rev. B} \textbf{25} 2433 (1982)

\bibitem{Ambjornsson2006} Ambj{\"o}rnsson T et al. \textit{Phys. Rev. B} \textbf{73} 085412 (2006)

\bibitem{Taflove2005} Taflove A, Hagnes S C \textit{Computational Electrodynamics: The Finite-Difference Time Domain Method} (Boston: Artech House, 2005)

\bibitem{Waterman1971} Waterman P C \textit{Phys Rev D} \textbf{3} 825 (1971)

\bibitem{Mishchenko2002} Mishchenko M I, Travis L D, Lacis A A \textit{Scattering, Absorption and Emission of Light by Small Particles} (Cambridge: Cambridge University Press, 2002)

\bibitem{Hafner1990} Hafner Ch \textit{The Generalized Multipole Technique for Computational Electromagnetics} (Boston: Artech House, 1990).

\bibitem{Moreno2002} Moreno E et al. \textit{JOSA A} \textbf{19} 101 (2002)

\bibitem{Yurkin2007} Yurkin M A, Hoekstra A G \textit{J. Quant. Spectrosc. Radiat. Transfer.} \textbf{106} 558 (2007)

\bibitem{Lee-Mal1995} Lee J K, Mal A K \textit{Appl. Math. Comput.} \textbf{67}, 135 (1995)

\bibitem{Nitzan1981} Nitzan A, Brus L E  \textit{J. Chem. Phys.} \textbf{75} 2205 (1981)

\bibitem{Gersten1981a}  Gersten J, Nitzan A \textit{J. Chem. Phys.} \textbf{75} 1139 (1981)

\bibitem{Gersten1981b} Gersten J, Nitzan A \textit{J. Chem. Phys.} \textbf{73} 3023 (1980)

\bibitem{Shah2013} Shah R A et al. \textit{Phys. Rev. B} \textbf{88} 075411 (2013)

\bibitem{Joe2006} Joe Y S, Satanin A M, Kim C S \textit{Phys. Scr.} \textbf{74} 259 (2006)

\bibitem{Miroshnichenko2010} Miroshnichenko A E, Flach S, Kivshar Y S \textit{Rev. Mod. Phys.} \textbf{82} 2257 (2010)

\bibitem{Krivenkov2019} Krivenkov V et al. \textit{Laser Photon. Rev.} \textbf{13} 1800176 (2019)

\bibitem{Fleischhauer2005} Fleischhauer M, Imamoglu A, Marangos J P \textit{Rev. Mod. Phys.} \textbf{77} 633 (2005)

\bibitem{Peng2014} Peng B et al. \textit{Nat. Commun.} \textbf{5} 5082 (2014)

\bibitem{Kivshar2017} Limonov M F et al. \textit{Nat. Photon.} \textbf{11} 543 (2017)

\bibitem{Fano-Springer} Gallinet B, in \textit{Fano Resonances in Optics and Microwaves: Physics and Applications} Chap. 5 (Eds E Kamenetskii, A Sadreev, A Miroshnichenko) (Cham: Springer, 2018) p. 110

\bibitem{Fano1961} Fano U \textit{Phys. Rev.} \textbf{124} 1866 (1961)

\bibitem{LL-QM} Landau L D, Lifshitz E M \textit{Quantum Mechanics: Non-Relativistic Theory} (Pergamon Press Ltd., 1991); Translated from: \textit{Kvantovaya Mekhanika: Nerelyativistskay Teoriya} (Moskva: Nauka, Fizmatlit, 1989)

\bibitem{Dong2010} Dong Z C et al. \textit{Nat. Photon.} \textbf{4} 50 (2010)

\bibitem{Jaynes-Cummings1963} Jaynes E T, Cummings F W \textit{Proc. IEEE} \textbf{51} 89 (1963) 

\bibitem{MorLeb-QE2024} Moritaka S S, Lebedev V S \textit{Bull. Lebedev Phys. Inst.} 51 S750 (2024); Moritaka S S, Lebedev V S \textit{Kvantovaya Elektron.} 54 362 (2024)

\bibitem{Gomez2014} Gomez D E \textit{J. Phys. Chem. C} \textbf{118} 23963 (2014)

\bibitem{Kond-Leb_QE2018} Kondorskiy A D, Lebedev V S \textit{Quantum Electron.} \textbf{48} 1035 (2018); \textit{Kvantovaya Electron.} \textbf{48} 1035 (2018) 

\bibitem{KondorMek2022} Kondorskiy A D, Mekshun A V \textit{Bull. Leb. Phys. Inst.} \textbf{49} 341 (2022);  \textit{
Kratkie Soobshcheniya po Fizike} \textbf{49}, No 10, 55 (2022)

\bibitem{Kond-Leb_OE2019} Kondorskiy A D, Lebedev V S \textit{Opt. Express} \textbf{27} 11783 (2019)

\bibitem{Wiederrecht2007} Wiederrecht G P, Hall, J E, Bouhelier A \textit{Phys. Rev. Lett.} \textbf{98} 083001 (2007)

\bibitem{Petoukhoff2020} Petoukhoff C E, Dani K M, O'Carroll D M \textit{Polymers} \textbf{12} 2141 (2020) 

\bibitem{Faucheaux2014} Faucheaux J A, Fu J, Jain P K \textit{J Phys Chem C} \textbf{118} 2710 (2014)

\bibitem{Chen2012} Chen H et al. \textit{J. Phys. Chem. C} \textbf{116} 14088 (2012)

\bibitem{Fleischmann1974} Fleischmann M, Hendra P J, McQuillan A J \textit{Chem. Phys. Lett.} \textbf{26} 163 (1974) 

\bibitem{Campion1998} Campion A, Kambhampati P \textit{Chem. Soc. Rev.} \textbf{27} 241 (1998) 

\bibitem{Cui2015} Cui L et al. \textit{Sci. Rep.} \textbf{5} 11920 (2015)

\bibitem{Pockrand1979} Pockrand I et al. \textit{J. Chem. Phys.} \textbf{70} 3401 (1979)

\bibitem{Pompa2006} Pompa P P et al. \textit{Nat. Nanotechnol.} \textbf{1} 126 (2006)

\bibitem{Yeh2008} Yeh D-M, Shi Y, Chen M \textit{Nanotechnology} \textbf{19} 345201 (2008)

\bibitem{Dulkeith2002} Dulkeith E et al. \textit{Phys. Rev. Lett.} \textbf{89} 203002 (2002) 

\bibitem{Liu2015} Liu X et al. \textit{Nat. Photon.} \textbf{9} 30 (2015)

\bibitem{Liu2016} Liu W et al. \textit{Nano Lett.} \textbf{16} 1262 (2016)

\bibitem{Pelton2019} Pelton M, Storm S D, Lenga H \textit{Nanoscale} \textbf{11} 14540 (2019)

\bibitem{Khitrova2006} Khitrova G. et al. \textit{Nat. Phys.} \textbf{2} 81 (2006)

\bibitem{Wu2021} Wu F et al. \textit{ACS Nano} \textbf{15} 2292 (2021)

\bibitem{Petoukhoff2019} Petoukhoff C E, Dani K M, O'Carroll D M, in: \textit{JSAP-OSA Joint Symposia 2019 Abstracts, OSA Technical Digest} (Optica Publishing Group, 2019), p. 18p\_E208\_13

\bibitem{Xiong2019} Xiong X et al. \textit{Nanophotonics} \textbf{9} 0333 (2019) 

\bibitem{Balci2019} Balci F M, Sarisozen S, Balci N P S \textit{J. Phys. Chem. C} \textbf{123} 26571 (2019)

\bibitem{Schwartz2011} Schwartz T et al. \textit{Phys. Rev. Lett.} \textbf{106}, 196405 (2011)

\bibitem{Sato2001} Sato T et al. \textit{Chem. Lett.}  \textbf{30} 402 (2001)

\bibitem{Wurtz2003} Wurtz G A, Hranisavljevic J, Wiederrecht G P \textit{J. Microsc.} \textbf{210} 340 (2003)

\bibitem{Uwada2007} Uwada T et al. \textit{J. Phys. Chem. C} \textbf{111} 1549 (2007)

\bibitem{Yoshida2009a} Yoshida A, Yonezawa Y, Kometani N \textit{Langmuir} \textbf{25} 6683 (2009)

\bibitem{Vujacic2012} Vuja\v{c}i\'{c} A et al. \textit{J. Phys. Chem. C} \textbf{116} 4655 (2012)

\bibitem{Laban2014} Laban B et al. \textit{J. Phys. Chem. C} \textbf{118} 23393 (2014)

\bibitem{Laban2015} Laban B, Vodnik V, Vasi\'{c} V \textit{Nanospectroscopy} \textbf{1} 5460 (2015)

\bibitem{Gulen2010} G{\"u}len D \textit{J. Phys. Chem. C} \textbf{114} 13825 (2010)

\bibitem{Balci2016} Balci S et al. \textit{ACS Photon.} \textbf{3} 2010 (2016)

\bibitem{Lam-Kond-Leb2019} Lam N T, Kondorskiy A D, Lebedev V S \textit{Bull. Leb. Phys. Inst.} \textbf{46} 390 (2019)

\bibitem{Yoshida2009b} Yoshida A, Uchida N, Kometani N \textit{Langmuir} \textbf{25} 11802 (2009)

\bibitem{Solomon2020} Solomon M L et al. \textit{Acc. Chem. Res.} \textbf{53} 588 (2020) 

\bibitem{ZhuWu2021} Zhu J et al. \textit{Nano Lett.} \textbf{21} 3573 (2021)

\bibitem{He-Guo2023} He C et al. \textit{Nano Lett.} \textbf{23} 9428 (2023) 

\bibitem{Kumar2023} Kumar M et al. \textit{ACS Appl. Nano Mater.} \textbf{6} 13894 (2023)

\bibitem{Cheng-Yang2023} Cheng Q et al. \textit{Nano Lett.} \textbf{23} 11376 (2023)

\bibitem{Gunnarsson2005} Gunnarsson L et al. \textit{J. Phys. Chem. B} \textbf{109} 1079 (2005)

\bibitem{Scherer1984} Scherer P, Fischer S F \textit{Chem. Phys.} \textbf{86} 269 (1984)

\bibitem{Misawa1994} Misawa K et al. \textit{J. Lumin.} \textbf{60-61} 812 (1994)

\bibitem{Tani2007} Tani T et al. \textit{J. Lumin.} \textbf{122-123} 244 (2007)

\bibitem{Tani2012} Obara Y et al. \textit{Int. J. Mol. Sci.} \textbf{13} 5851 (2012)

\bibitem{Haverkort2014} Haverkort F, Stradomska A, Knoester J \textit{J. Phys. Chem. B} \textbf{118} 8877 (2014)

\bibitem{Pugzlys2006} Pugzlys A et al. \textit{Int. J. Photoenergy} \textbf{2006} 029623 (2006)

\bibitem{Joannopoulos2010} Qiu C et al. \textit{Laser Photonics Rev.} \textbf{4} 268 (2010)

\bibitem{Roth1973} Roth J, Dignam M J \textit{J. Opt. Soc. Am.} \textbf{63} 308 (1973)

\bibitem{Ru2018a} Augui\'{e} B, Le Ru E C \textit{J. Phys. Chem. C} \textbf{122} 19110 (2018)

\bibitem{Ru2018b} Tang C, Augui\'{e} B, Le Ru E C \textit{ACS Photonics} \textbf{5} 5002 (2018)

\bibitem{Tang2021} Tang C, Augui\'{e} B, Le Ru E C \textit{Phys. Rev. A} \textbf{104} 033502 (2021)

\bibitem{Ru2019} Augui\'{e} B, Darby B L, Le Ru E C \textit{Nanoscale} \textbf{11} 12177 (2019)

\bibitem{Tang2020} Tang C, Augui\'{e} B, Le Ru E C \textit{Phys. Rev. B} \textbf{103} 085436 (2021)

\bibitem{Kalsin2006} Kalsin A M et al. \textit{Science} \textbf{312} 420 (2006)

\bibitem{Kowalczyk2012} Kowalczyk B et al. \textit{Nat. Mater.} \textbf{11} 227 (2012)

\bibitem{Stete2023} Stete F et al. \textit{ACS Photon.} \textbf{10} 2511 (2023)

\end{thebibliography}
\end{document}